\documentclass[pdfusetitle,aps,reprint,prd,twocolumn,superscriptaddress,nofootinbib]{revtex4-1}
\pdfoutput=1

\usepackage{amssymb}
\usepackage{amsmath}
\usepackage{epsfig}
\usepackage{breakurl}
\usepackage{float}
\usepackage{multirow}
\usepackage{array}
\usepackage{bm}
\usepackage{color}
\usepackage{tikz}
\usepackage[utf8]{inputenc}
\usepackage{braket}
\usepackage{graphicx}
\usepackage{footmisc}
\usepackage{tensor}
\usepackage[normalem]{ulem}

\usepackage[T1]{fontenc}
\usepackage{array}
\usepackage{booktabs}
\usepackage{multirow}
\usepackage{amstext}
\usepackage{epsfig}
\usepackage{graphicx}
\usepackage{caption}
\usepackage{subcaption}
\makeatletter

\providecommand{\tabularnewline}{\\}

\makeatother

\usepackage{babel}

\usepackage[force]{feynmp-auto}

\xdefinecolor{mylinkcolor}{rgb}{0,0,0.7}
\usepackage[
	bookmarksnumbered, bookmarksopen, bookmarksopenlevel=2,
	breaklinks=true,
	colorlinks=true, filecolor=mylinkcolor, citecolor=mylinkcolor,
	linkcolor=mylinkcolor, urlcolor=mylinkcolor, menucolor=mylinkcolor,
]{hyperref}
\usepackage{bookmark}

\usepackage{cleveref}
\crefformat{section}{\S#2#1#3} 
\crefformat{subsection}{\S#2#1#3}
\crefformat{subsubsection}{\S#2#1#3}

\usepackage{colortbl}
\definecolor{blue2}{cmyk}{1, 0.1, 0.1, 0}

\definecolor{pyBlue}{RGB}{31, 119, 180}
\definecolor{pyRed}{RGB}{214, 39, 40}
\definecolor{pyGreen}{RGB}{44, 160, 44}
\definecolor{pyBlue2}{RGB}{0, 111, 237}
\definecolor{pyRed2}{RGB}{224, 52, 36}

\definecolor{summersky}{cmyk}{0.71,0.33,0,0.5}
\definecolor{flamingo}{cmyk}{0,0.51,0.71,0.5}
\definecolor{rp}{cmyk}{0.2, 1, 0.6, 0}
\definecolor{pacificblue}{cmyk}{0.95,0.3,0, 0.5}
\definecolor{gray60}{cmyk}{0.4,0.4,0,0.8}

\renewcommand{\vec}[1]{\mathbf{#1}}

\usetikzlibrary{decorations.markings}
\usepackage{cleveref}
\crefformat{section}{\S#2#1#3}
\crefformat{subsection}{\S#2#1#3}
\crefformat{subsubsection}{\S#2#1#3}

\makeatletter
\def\simgt{\mathrel{\lower$\frac{5}{2}$pt\vbox{\lineskip=0pt\baselineskip=0pt
           \hbox{$>$}\hbox{$\sim$}}}}
\def\simlt{\mathrel{\lower$\frac{5}{2}$pt\vbox{\lineskip=0pt\baselineskip=0pt
           \hbox{$<$}\hbox{$\sim$}}}}

\makeatother

\def\spa#1.#2{\left\langle#1\,#2\right\rangle}
\def\spb#1.#2{\left[#1\,#2\right]}
\def\sand#1.#2.#3{%
\left\langle#1{\vphantom1}\right|{#2}\left|#3\right]}%
\def\sandmp#1.#2.#3{%
\left\langle#1{\vphantom1}\right|{#2}\left|#3\right]}%
\def\sandpm#1.#2.#3{%
\left[#1{\vphantom1}\right|{#2}\left|#3\right\rangle}%
\def\sandmm#1.#2.#3{%
\left\langle#1{\vphantom1}\right|{#2}\left|#3\right\rangle}%
\def\sandpp#1.#2.#3{%
\left[#1{\vphantom1}\right|{#2}\left|#3\right]}%

\renewcommand{\imath}{\mathrm{i}}

\newcommand{\be}{\begin{equation}}
\newcommand{\ee}{\end{equation}}

\def\S{{\mathbb S}}

\allowdisplaybreaks

\begin{document}

\title{Absorptive Effects in Black Hole Scattering}

\author{Yilber Fabian Bautista}
\email{yilber-fabian.bautista-chivata@ipht.fr}
\affiliation{Institut de Physique Théorique, CEA, Université Paris–Saclay,
F–91191 Gif-sur-Yvette cedex, France}

\author{Yu-Tin Huang}
\email{yutin@phys.ntu.edu.tw}
\affiliation{Department of Physics and Center for Theoretical Physics, National Taiwan University, Taipei
10617, Taiwan}
\affiliation{Physics Division, National Center for Theoretical Sciences, Taipei 10617, Taiwan}

\author{Jung-Wook Kim}
\email{jung-wook.kim@aei.mpg.de}
\affiliation{Max Planck Institute for Gravitational Physics (Albert Einstein Institute),
Am M\"uhlenberg  1, D-14476 Potsdam, Germany}

\begin{abstract}
In this paper we define absorptive Compton amplitudes, which captures the absorption factor for waves of spin-weight-$s$ scattering in black hole perturbation theory. At the leading order, in the $GM\omega$ expansion, such amplitudes are purely imaginary and expressible as contact terms. Equipped with these amplitudes we compute the mass change in black hole scattering events via Kosower-Maybee-O'Connell formalism, where the rest mass of Schwarzschild/Kerr black hole is modified due to absorption of gravitational, electromagnetic, or scalar fields sourced by other compact object. We reproduced the power loss previously computed in the post-Newtonian expansion. The results presented here hold for similar mass ratios and generic spin orientation, while keeping the Kerr spin parameter to lie in the physical region $\chi\le1$.

\end{abstract}

\maketitle
\tableofcontents

\section{Introduction}
Characterizing the dynamics involving compact objects in general relativity (GR) has received tremendous attention in the last decade due to its importance for analyzing data collected in gravitational wave detectors~\cite{LIGOScientific:2016aoc}. In general, studying the dynamical evolution of such gravitating objects is a very difficult problem as the non-linearity of the gravitational field induces dynamical changes of the sources themselves, and those  at the same time change the structure of the spacetime. Fortunately, typical systems composed of compact objects admit a hierarchical separation of scales~\cite{Goldberger:2004jt,Porto:2016pyg}, which allows modeling compact objects as point particles, with the finite size effects introduced in an controlled, systematic manner either via effective multipole moments associated to the compositeness of the compact object~\cite{Porto:2007qi,Goldberger:2020fot,Goldberger:2005cd,Goldberger:2012kf,Goldberger:2020wbx,Saketh:2022xjb,Saketh:2024juq,Ivanov:2024sds}, mass-changing amplitudes~\cite{Kim:2020dif,Aoude:2023fdm} and hidden sectors~\cite{Jones:2023ugm}, or by solving directly the field equations given the initial state of the source in the region where the finite-size effects become manifest~\cite{Poisson:2004cw,Maselli:2017cmm,Datta:2019epe,Brown:2007jx,Tagoshi:1997jy,Endlich:2015mke,Chatziioannou:2016kem,Taracchini:2013wfa,Chia:2020yla,deCesare:2023rmg}. Object's finite size imprints on a gravitational wave signal can be divided into tree categories: tidal deformations, induced spin multipole moments, and tidal heating  and absorption. In this paper we will be concerned with the latter.

The simplest system involving dynamical evolution of gravitating compact objects is given by a plane wave scattering off black holes (BHs) of GR. According to Einstein's theory, the incident wave can be absorbed by the BH as long as its wavelength is comparable or smaller than the object's size. From the point of view of an asymptotic observer, the fluxes of energy and angular momentum entering the BH's horizon  are seen  as induced mass and current quadrupole (and higher) moments on the BH~\cite{Poisson:2004cw}. Indeed, due to the no-hair theorem, the physical effects the absorbed waves have on the BH is simply to change its macroscopic properties such as its mass and angular momentum.

From a more microscopic perspective, absorption is expected to be excitations of low-lying BH internal modes during the wave scattering process~\cite{Goldberger:2020wbx}; the states describing the excitation of such modes are hidden behind the BH horizon and therefore inaccessible from outside the BH. Therefore, for an asymptotic observer, the Einsteinian elastic evolution\footnote{By elastic we mean the binary system does not emit energy and angular momentum towards infinity.} of the wave + BH system seems non-unitary, with the non-unitarity caused by the absorption of energy and angular momentum by the BH.

While models of absorptive effects in  binary black hole systems have been focused mostly on the bounded orbit scenario~\cite{Porto:2007qi,Goldberger:2020fot,Saketh:2022xjb}, and aligned spins in the case of Kerr BH binaries, a few results are available for the scattering scenario in the case of Schwarzschild BHs~\cite{Goldberger:2020wbx,Jones:2023ugm} (the worldline actions modeling absorptive effects presented in  Refs.~\cite{Goldberger:2020fot,Saketh:2022xjb}, are in principle usable for both, bounded or unbounded scenarios, but in those references the authors restricted to solve the equations of motion (EOM) for BHs in closed orbits). In this work we expand on studying the absorptive dynamics for spinning  BHs in   hyperbolic encounters, while allowing for generic BH spin orientations.  

In the low frequency approximation  (or simply the post-Minkowskian (PM) approximation $\epsilon:=2GM\omega\ll1$), existing effective models of BH absorptive effects in the literature typically rely on a spectral expansion of the two-point correlation functions of quadrupole moments (or effective operators), with the  effective coefficients in the expansion matched to the BH absorption cross section~\cite{Porto:2007qi,Goldberger:2020fot,Saketh:2022xjb,Goldberger:2020fot,Saketh:2022xjb,Aoude:2023fdm, Jones:2023ugm, Chen:2023qzo,Vidal:2024inh}. At face value this appears to be taking a detour; one introduces an asantz for the spectral function fixed against black hole perturbation theory (BPHT), and recycles it to compute other classical observables. It would be desirable to identify directly the relevant information from the solution in BHPT itself.
Here we bypass the need for a spectral  decomposition and show that at leading PM-order, the absorptive effects in the elastic Compton amplitude  obtained from the solutions to the Teukolsky equation, 
can be  isolated in what we call the leading order ``absorptive Compton amplitude''. Such an amplitude is then recycled to compute leading PM absorptive  observables in the binary black hole problem for hyperbolic encounters.

Our approach is somewhat a hybrid between the UV gravitational approach (see for instance Ref.~\cite{Poisson:2004cw}) and the modern amplitudes approach. That is, we do not rely on an effective description to model the absorptive effects by a single BH but we directly take the UV solutions as dictated by the Teukolsky equation in the wave + BH scattering problem, and then incorporate absorptive effects in the binary-BH problem via the in-in Kosower-Maybee-O'Connell (KMOC) formalism~\cite{Kosower:2018adc}, closely following recent worldline~\cite{Goldberger:2020wbx} and on-shell~\cite{Jones:2023ugm} computations done for scattering of Schwarzschild BHs. In addition, we note that at the leading PM order, the computation of absorptive observables is greatly simplified by noticing that it is equivalent to computing the triangle leading singularity (LS) for 1-loop two-body observables. Furthermore, we also clarify how to obtain absorptive Schwarzschild observables as the spinless limit of the Kerr counterpart; this limit was previously known to be discontinuous in the PM expansion~\cite{Poisson:2004cw}. We remark that such a limit is well defined when Kerr BH spin parameter lies in the physical region $\chi\le1$ and non-rational functions of $\chi$ in the Teukolsky solutions are kept, which are the famous polygamma contributions~\cite{Bautista:2022wjf} characterizing the finite-size linear response of the Kerr BH to small wave perturbations~\cite{Bautista:2024agp}.

This paper is organized as follows: In Section~\ref{sec:ComptonsBHPT} we discuss how to isolate the leading PM abortive contributions to the elastic Compton amplitude obtained from BHPT analysis, and provide explicit covariant amplitudes for Schwarzschild BHs in subsection~\ref{sec:schwarzschild} and Kerr BHs in subsection~\ref{sec:Kerramplitudes}, for waves of generic spin-weight-$s$ scattering off the BH. The leading PM absorptive contribution is controlled by the leading $\ell = |s|$ harmonic in the absorption factor. In subsection~\ref{sec:spectralIntegral} we show the leading PM absorptive Compton amplitudes can be obtained from the effective model of inaccessible reaction channels excited by absorption in the framework of mass changing three-point amplitudes. In particular, we show that the imaginary part of the spectral gluing of two mass-changing three-point amplitudes recovers the absorptive Compton amplitude, while the real part provide additional constraints on the near-threshold  ``off-shell'' spectral function of the BH. In section~\ref{sec:mass change} we use absorptive Compton amplitudes to compute the change in mass for a BH in binary scattering event based on KMOC formalism and triangle LS computation, providing explicit results for Schwarzschild BH absorbing spin-weight-$s$ waves in subsection~\ref{sec:schchangemass} and Kerr BH in subsection~\ref{sec:kerr_mass_change}. We also discuss the aligned spin, Schwarzschild, and non-relativistic limits  of the Kerr case, finding agreement with reported results in overlapping regions of validity. In section~\ref{sec:conclusions} we conclude with a discussion and future directions.  We provide a set of appendices: In Appendix~\ref{app:cross_section} we review inelastic scattering processes in GR and quantum mechanics,  Appendix~\ref{app:etafromF} provides the technical details to obtain the absorption factor in the BHPT computation via the Nekrasov-Shatashvili function, in Appendix~\ref{app:polvec} we include  useful expressions for polarization vectors and discuss the forward limit of the elastic Compton amplitude, and in Appendix~\ref{app:forde} we review the extraction of one loop integral coefficients via the Forde's method and comment on the connection of the triangle coefficient with the triangle LS.

\textbf{Note}: While this work was  prepared for submission, the work of  Ref.~\cite{aoude2024inelasticcoupledchanneleikonalscattering} appeared, which address Schwarzschild BH absorptive effects  using a similar multi-channels scattering  language.   

\section{Absorptive Compton Amplitudes from BHPT}\label{sec:ComptonsBHPT}

In this section we isolate the leading order---in a post-Minkowskian sense---absorptive contributions to the BH Compton amplitude obtained  directly in the framework of BHPT. 

Consider an incoming plane wave of momentum  $k_2^\mu$, energy $\omega$, 
and spin-weight\footnote{These spin-weight values correspond to scalar, neutrino, photon, gravitino and gravitational waves, respectively. In the case of half-integer spin perturbations, we refer to them as waves in the sense of Chandrasekhar (see e.g. Chapter 10 of Ref.~\cite{Chandrasekhar:1985kt}).} $s = \{0,-1/2,-1,-3/2,-2\}$, scattering off a Kerr BH with momentum is $p^\mu = M u^\mu$ and spin vector $\vec{a}$. The angle formed by the incoming wave and the BH's spin direction is $\gamma$, while the scattered wave has outgoing  momentum $k_3^\mu$ parametrized by the $(\theta,\phi)$-direction on the celestial sphere.  
Such a scattering process is traditionally studied in the framework of linear BHPT where the dynamical evolution of the perturbations is dictated by the so-called Teukolsky master equation (TME)~\cite{PhysRevLett.29.1114}.  TME is a second order partial differential equation obeyed by the Teukolsky scalar ${}_s\psi$. Nontrivially, this scalar admits separation of angular and radial perturbation equations, which are respectively known as the angular Teukolsky equation (ATE) and the radial Teukolsky equation (RTE).  Both the ATE and the RTE are ordinary differential equations of Confluent Heun type~\cite{Batic_2007,Bonelli:2022ten,Aminov:2020yma} and do not possess closed form solutions in terms of traditional mathematical functions. Therefore, analytic solutions to the TME can be obtained only under certain perturbative expansions such as e.g. the PM-approximation and the spin multipole expansion (SME), as we discuss below. 

A formal solution to the RTE imposing physical boundary conditions of purely incoming waves at the black hole outer horizon  can be written. The radial functions, ${}_s R_{\ell m}(r)$, have the asymptotic $r\to \infty$ behavior
\begin{equation}
{}_s R_{\ell m}(r) =\frac{{}_s B_{\ell m}^\text{inc}}{r}e^{-i \omega r^*}+ \frac{{}_s B_{\ell m}^\text{ref}}{r^{2s+1}}e^{-i \omega r^*}\,.
\end{equation}

The ratio of the incoming and reflection coefficients define a partial wave scattering operator (see e.g. Refs.~\cite{futterman88,Ivanov:2022qqt,Bautista:2023sdf,Saketh:2023bul}):
\begin{align}
    \frac{{}_s B_{\ell m}^{\text{ref}}}{{}_s B_{\ell m}^{\text{inc}}}:= {}_s\mathsf{S}_{\ell m}^P=  {}_s\eta_{\ell m}e^{2i{}_{s}\delta_{\ell m }^{P}} \,. \label{eq:connectionformula}
\end{align}
We refer to $ {}_s\mathsf{S}^P_{\ell m}$ as the \textit{elastic-channel} ($2\to2$) partial wave scattering matrix obtained by computing the total real elastic phase shift ${}_{s}\delta_{\ell m }^{P}={}_{s}\delta_{\ell m }^{\text{FZ},P}+{}_{s}\delta_{\ell m }^{\text{NZ}}$---where the near-zone and far-zone components are individually purely real---and the real factor ${}_s\eta_{\ell m}$, accounting for inelastic contributions in the scattering process. Such inelastic contributions are in general expected to be present in BH scattering scenarios as waves with sufficiently short wavelengths can be partially absorbed by the BH. Absorptive effects therefore open additional \textit{reaction-scattering-channels} inaccessible for an  observer outside the BH, 
and the elastic-channel partial wave scattering matrix of Eq.\eqref{eq:connectionformula} is non-unitary but rather satisfies the inequality $| {}_s\mathsf{S}^P_{\ell m}| = {}_s\eta_{\ell m} < 1$. 

\subsection{Elastic scattering in the presence of inelastic processes}\label{sec:isolation}
Even in  the presence of reaction-channels, the elastic-channel partial wave scattering matrix $ {}_s\mathsf{S}^P_{\ell m}$, introduced in Eq.\eqref{eq:connectionformula}, defines an elastic two-to-two (wave + BH $\to $ wave + BH) scattering amplitude.\footnote{Following the terminology of scattering in quantum mechanics, elastic scattering is defined as the scattering process where the nature of the scattered particles is unaltered.} However, since $ {}_s\mathsf{S}^P_{\ell m}$ is non-unitary, the  simultaneous excitation of other reaction-channels in the wave + BH scattering process is needed to restore unitarity, which we address in Section~\ref{sec:spectralIntegral}; see also e.g. Ref.~\cite{Landau1981Quantum} and  Appendix~\ref{app:cross_section}. For waves of generic spin-weights scattering off the BH, we can define two elastic helicity amplitudes describing the helicity preserving and helicity reversing scenarios. We call these elastic amplitudes the spin-weight-$s$ Compton amplitudes, which are given respectively by the partial wave sums~\cite{futterman88,Dolan:2008kf}:
\begin{widetext}
\begin{align}
    {}_s f(\Omega)&=N\sum_{\ell=|s|}^{\infty} \sum_{m=-\ell}^{\ell}{}_{s}S_{\ell m}(\gamma,0;a\omega){}_{s}S_{\ell m}(\theta,\phi;a\omega)\sum_{P=\pm1}\left({}_s\mathsf{S}_{\ell m}^P-1\right), \label{Eq:fKerrGeneric}\\
     {}_s g(\Omega)&=N\sum_{\ell=|s|}^{\infty} \sum_{m=-\ell}^{\ell}{}_{s}S_{\ell m}(\gamma,0;a\omega){}_{s}S_{\ell m}(\pi-\theta,\phi;a\omega)\sum_{P=\pm1}P(-)^{\ell}\left({}_s\mathsf{S}_{\ell m}^P{-}1\right) \label{Eq:gKerrGeneric}\,,
\end{align}
\end{widetext}
where $N=\frac{(-1)^{|s|}2\pi}{i\omega}$, $\Omega$ stands for angular variables $(\gamma,\theta,\phi)$, and $P$ is the parity label. The sum over parity $P = \pm 1$ appears as consequence of changing from the parity basis to the helicity basis.
The only parity-dependent contribution to the phase shift is in the far-zone phase shift, and can be traced to the phase of the Teukolsky-Starobinsky constant, which has an imaginary part only in the $s=-2$ case. 
This implies ${}_{s}\delta_{\ell m }^{\text{FZ},P}={}_{s}\delta_{\ell m }^{\text{FZ}}$ for $s\ne2$, therefore the helicity reversing amplitude $ {}_s g(\Omega)$ vanishes for waves of spin-weight $s\ne2$ due to the linear in $P$ factor inside the asymmetric $P$-sum in Eq.\eqref{Eq:gKerrGeneric}. Finally, 
the partial wave amplitudes defined above recover the Compton amplitudes in the Schwarzschild limit ($a\to0$). In this case, the spin-weighted spheroidal harmonics ${}_{s}S_{\ell m}(\theta,\phi;a\omega)$ reduce to the spin-weighted spherical harmonics  ${}_{s}Y_{\ell m}(\theta,\phi)$.

As mention above,  knowledge of all the non-diagonal reaction-channels is needed to restore unitarity of the scattering matrix. However, at leading order in the PM expansion we can identify the piece in the elastic amplitude given in Eq.\eqref{Eq:fKerrGeneric}, responsible for the  violation of the unitarity of condition of $ {}_s\mathsf{S}^P_{\ell m}$; we refer to this piece as the leading order  \textit{absorptive contribution to the elastic amplitude}, and to isolate it we proceed as follows: Let us further  split the total elastic amplitude in Eq.\eqref{Eq:fKerrGeneric} as
\begin{equation}
   {}_sf(\Omega)= {}_sf_{1}(\Omega)  {+}{}_sf_{2}(\Omega) \,,
\end{equation}
where
\begin{equation}\label{eq:elastic_amplitude}
\begin{split}
     \frac{{}_sf_{1}(\Omega)}{N}{=}\sum_{\ell,m,P} {}_{s}S_{\ell m}(\gamma;a\omega){}_{s}S_{\ell m}(\theta,\phi;a\omega)
     ({}_s\eta_{\ell m}{-}1){}_s e^{2i{}_{s}\delta_{\ell m }^{P}},
    \end{split}
\end{equation}
and
\begin{equation}
\label{eq:absortive_amplitude_def}
    \begin{split}
       {}_sf_{2}(\Omega)
     &=   {}_sf(\Omega) -  {}_sf_{1}(\Omega)\,.
    \end{split}
\end{equation}
The key observation is the scaling of different contributions to the phase-shift and absorption factor in the PM expansion, recall $\epsilon=2GM\omega$,
\begin{subequations}\label{PM exp}
\begin{align}
{}_s\delta_{\ell m}^{\text{FZ},P}&\sim O(\epsilon)\,, \label{PM exp FZ}\\
{}_s\delta_{|s| m}^{\text{NZ}}&\sim \epsilon^{2|s|{+}2}(1+\log \epsilon+\cdots)+\cdots\,,\label{PM exp NZ}\\
{}_s\eta_{|s| m}&\sim 1 + \epsilon^{2|s|{+}1} + \epsilon^{2|s|{+}2} \label{PM exp eta} \\
&\phantom{\sim asdf} + \epsilon^{2|s|{+}3}(1+\log(\epsilon) + \cdots) + \cdots \,. \nonumber 
\end{align}
\end{subequations}
In other words, the leading PM contribution of the near-zone comes from the $\ell = |s|$ harmonic. Isolating the identity from the scattering matrix,
\begin{equation}\label{eq:SlmDef}
{}_s\mathsf{S}^P_{\ell m}=1+i \,{}_s\mathsf{T}^P_{\ell m}={}_s\eta_{\ell m}\,{}_se^{2i{}_{s}\delta_{\ell m }^{P}}\,,
\end{equation}
we see from Eq.\eqref{PM exp} that absorption effects first appear at $\epsilon^{2|s|{+}1}$ order, which become the leading contribution to the imaginary part of ${}_s\mathsf{T}^P_{\ell m}$. This contribution is precisely the leading term in ${}_sf_{\text{1}}(\Omega)$. We will show shortly that the imaginary leading terms suggests that they are polynomial contact terms in the elastic amplitude.

At leading PM order we can isolate purely absorptive pieces in the BH Compton amplitudes as
\begin{equation}
\label{eq:absortive_amplitude_def}
    \begin{split}
    \boxed{
       {}_sf^{\text{LO}}_{\text{A}}(\Omega)
     {=} {}_sf_{\text{1}}(\Omega)
     \Big|_{\text{L0}}
       {=}N\sum_{\ell,m}{}_s\mathcal{S}_{\ell m}^2\left({}_s\eta_{\ell m}{-}1\right){+}O(\epsilon^{2|s|{+}3}) \,,}
    \end{split}
\end{equation}
which is the main object of study in this work. Notice that the identification of ${}_s f_1(\Omega)$ with pure  absorptive effects is valid only at the leading PM-order, as in general ${}_s f_1(\Omega)$ as defined in Eq.~\eqref{eq:elastic_amplitude} mixes absorptive and conservative far-zone effects at subleading PM orders.

To simplify the expressions,  we have denoted in  Eq.~\eqref{eq:absortive_amplitude_def} the two copies of the spheroidal harmonics as ${}_s\mathcal{S}^2_{\ell m} = {}_{s}S_{\ell m}(\gamma,0;a\omega){}_{s}S_{\ell m}(\theta,\phi;a\omega)$.  
In principle, there are $O(\epsilon^{2|s|+2})$ terms due to interference between ${}_s\eta_{|s| m}$ [Eq.\eqref{PM exp eta}] and ${}_s\delta^{\text{FZ}}_{\ell m}$ [Eq.\eqref{PM exp FZ}], which are of iteration type and contribute to the real part of the Compton amplitude. The iteration contributions are expected to be absent in observables, as prescribed for instance by the KMOC formalism~\cite{Kosower:2018adc}. We therefore expect the absorptive amplitude in Eq.\eqref{eq:absortive_amplitude_def} to be exact up to corrections of order  $O(\epsilon^{2|s|{+}3})$. This observation will be crucial for recovering Schwarzschild observables from that of Kerr in the $\chi\to 0 $ limit. 

The leading PM-separation of the absorptive contributions in the helicity reversing amplitude in Eq.\eqref{Eq:gKerrGeneric} can be done analogously. It is not hard to show that due to the linear $P$ factor in Eq.\eqref{Eq:gKerrGeneric}, the purely absorptive piece in ${}_s g(\omega)$ vanishes for wave perturbations of generic spin-weight $s$ at leading PM order, including the gravitational case. Vanishing of such amplitude has been associated to a hidden self-duality symmetry of the inelastic contributions to the BH Compton amplitudes~\cite{Jones:2023ugm,Chen:2023qzo}. 

As for the   real contributions in the elastic Compton amplitudes, up to order $\epsilon^{2|s|{+}2}$, ${}_sf_{\text{2}}(\Omega)$   as  given by  Eq.\eqref{eq:absortive_amplitude_def}  can be interpreted as a purely conservative contribution to the Compton amplitude; indeed, for $s=-2$ perturbations,  this piece  was     recently used in Ref.~\cite{Bautista:2024agp} to study the conservative dynamics of the gravitational  spinning two-body problem up to six order in the SME,  i.e. the expansion in powers of  $(a\omega)$.

As a consequence of the separation discussed above, at leading PM order the inelastic cross section accounting for  all reaction channels excited in the BH + wave scattering process follows from the imaginary part of the forward   inelastic amplitude in the visible \textit{input channel} (See also Appendix~\ref{app:cross_section}).
\begin{equation}\label{eq:cross_section_LO_absoprtion}
{}_s \sigma_{\text{inelastic}}^{\text{LO}} = \frac{4\pi}{\omega} \text{Im}[{}_s f_A^{\text{LO, Forward}}(\theta,\phi)]\,.
\end{equation}
Finally, in light of the  discussion for the remaining sections,  it  is   convenient to  write the PM-expansion of the absorption factor for the leading $\ell = |s|$ harmonic,  ${}_s \eta_{|s| m}$, as
\begin{equation}\label{eq:beta-splittingeta}
{}_s \eta_{|s| m}= 1{+}\sum_{n=0} ^1  \epsilon^{2|s|{+}1{+}n} {}_s\beta_{|s|
 m}^{(2|s|{+}1{+}n)}\,.
\end{equation}

\subsection{Leading order absorptive Compton amplitudes for Schwarzschild}\label{sec:schwarzschild}
As a warm up we start by studying  absorptive effects in the Compton amplitudes describing scattering of waves off the Schwarzschild BH. Notice that  due to the spherical symmetry of the BH background, only the angular momentum number $\ell$ is important in the absorptive factor ${}_s\eta_{\ell m}$. We refer the reader to Appendix~\ref{app:etafromF} for details on how to compute the absorption factor ${}_s\eta_{\ell m}$ within the framework of  BHPT and its modern  connection to CFT and supersymmetric gauge theories~\cite{Bautista:2023sdf,Bonelli:2022ten,Aminov:2020yma,Bianchi:2021mft}.
\begin{table*}[th]
\caption{Leading order absorptive Compton amplitudes for waves of different spin-weights $s$, scattering off Schwarzschild black hole as obtained from  Eq.\eqref{eq:absortive_amplitude_def}. The scalar amplitudes are given in Eq.\eqref{eq:scalaramplitudes}. The leading order absorption cross section is obtained from the leading order absorptive amplitudes in the forward limit, as indicated in Eq.\eqref{eq:cross_section_LO_absoprtion}. Alternatively, it can also be obtained directly from Eq.\eqref{eq:absoption_cross_Section_all}. }
\label{tab:summary_amplitudes_schw}
\vspace{0.2cm}
\setlength\extrarowheight{8pt}
\begin{ruledtabular}
\begin{centering}
\begin{tabular}{c c c c}
spin-weight $s$ & $_{s}\beta_{|s|}^{(2|s|{+}1{+}n)}$ & ${}_s f_{\text{A,{Schw.}}}^{\text{{LO}}}$ & ${}_s\sigma_{\text{inelastic}}^{\text{LO},\,\text{Schw.}}$. 
\tabularnewline
\hline  
$0$ & $n=0$: $ -2$, $n=1$: $-2\pi$  & ${}_0 A^{(0)} (i) 2 GM \epsilon  (\pi  \epsilon {+}1)$ & $16 \pi  G^2 M^2 (1{+}2 \pi  G M \omega )$ \tabularnewline
\hline 
$-\frac{1}{2}$ &  $n=0$: $ -\frac{1}{8}$, $n=1$: $-\frac{\pi}{8}$ & ${}_{-\frac{1}{2}}A^{(0)}\frac{(i)  }{4  }GM\epsilon  (\pi  \epsilon {+}1)e^{\frac{i}{2}\phi}$ & $2  \pi  G^2 M^2 (1{+}2 \pi  G M \omega )$\tabularnewline
\hline 
$-1$ & $ n=0$: $0$, $n=1$: $-\frac{2}{9}$ & $_{-1}A^{(0)}\frac{(i)}{3}2 GM \epsilon^{3}e^{i\phi}$ & $\frac{64}{3}  \pi  G^4 M^4 \omega ^2$ \tabularnewline
\hline 
$-\frac{{3}}{2}$ & $n=0:$ $-\frac{1}{128}$, $n=1:$ $-\frac{\pi}{128}$ & 

$ {}_{-\frac{3}{2}}A^{(0)} \frac{(i) }{32 } GM\epsilon ^3 (\pi  \epsilon {+}1) e^{\frac{3i}{2}\phi}$
& $\pi  G^4 M^4 \omega ^2 (1{+}2 \pi  G M \omega )$
\tabularnewline
\hline 
$-2$ & $ n=0$: $0$, $n=1$:  $-\frac{2}{225}$ & $_{-2}A^{(0)}\frac{(i)}{45}2 GM\epsilon^{5}e^{2i\phi}$ & $\frac{256}{45} \pi  G^6 M^6 \omega ^4$\tabularnewline
\end{tabular}
\par\end{centering}
\end{ruledtabular}
\end{table*}

For wave perturbations of spin-weight   $s=-2$, 
the  leading absorptive contribution comes from the $\ell=2$ harmonic. In such a case, the $\beta$-coefficients entering in Eq.\eqref{eq:beta-splittingeta} are non-zero starting at $n=1$. We obtain 
\begin{equation} 
{}_{-2}\beta_{2}^{(6)} =-\frac{2}{225} \,.
\end{equation}
The sum in Eq.\eqref{eq:absortive_amplitude_def} can now be easily performed. We can set $\gamma=0$ without loss of generality;\footnote{Alternatively, one can keep $\gamma$ generic. In such a case, all possible values of $m$ contribute to the sum in Eq.\eqref{eq:absortive_amplitude_def} as a consequence of the non-spherical nature of the plane wave. After summing over $m$, any dependence on $\gamma$ drops out from the final result.} this localizes the sum in the azimuthal number $m$ to the $m=l=|s|$ harmonic. The $\ell=m=2$ spin-weighted spherical harmonic is
\begin{equation}
{}_{-2}Y_{2,2}(\theta,\phi) = \frac{1}{2} \sqrt{\frac{5}{\pi }} e^{2 i \phi } \cos ^4\left(\frac{\theta }{2}\right)\,,
\end{equation}
therefore  Eq.\eqref{eq:absortive_amplitude_def} gives
\begin{equation}\label{eq:schwgravity}
{}_{-2}f_{A,\text{ Schw.}}^{\text{LO}}(\Omega) = \frac{i \epsilon ^6 e^{2 i \phi } \cos ^4\left(\frac{\theta }{2}\right)}{45 \omega} \,.
\end{equation}

The amplitude  in Eq.\eqref{eq:schwgravity} can be written in a more covariant manner using  the scalar helicity factor  
\begin{equation}
\label{eq:compt_scalar_hcl}
    {}_{-2} A^{(0)}= \cos^{4}(\theta/2)= \frac{\langle 2|u|3]^{4}}{(2\omega)^{4}}  
\end{equation}
where $u^\mu = (1,0,0,0)$ is the velocity of the black hole in its rest  frame, and we used the spinor parametrization of Ref.~\cite{Bautista:2022wjf}: 
\begin{equation}\label{eq:spinhel}
    \begin{split}
      |\lambda\rangle =& \sqrt{2\omega}\left(-e^{-i\phi/2} \sin\theta/2, e^{i\phi/2} \cos\theta/2 \right),\\
      [\lambda| =& \sqrt{2\omega}\left(-e^{i\phi/2} \sin\theta/2, e^{-i\phi/2} \cos\theta/2 \right)\,.
    \end{split}
\end{equation}
We can construct momentum matrices for the momenta of incoming ($k_2$) and outgoing ($k_3$) waves from these spinors as $(k_{\alpha \dot{\alpha}})_i = (|\lambda\rangle_{\alpha})_i ([\lambda|_{\dot \alpha})_i$, where we set $\phi=\theta=0$ for the former and $\phi=0$ for the latter.

\begin{figure}[H]
    \centering
\includegraphics[width=0.6\linewidth]{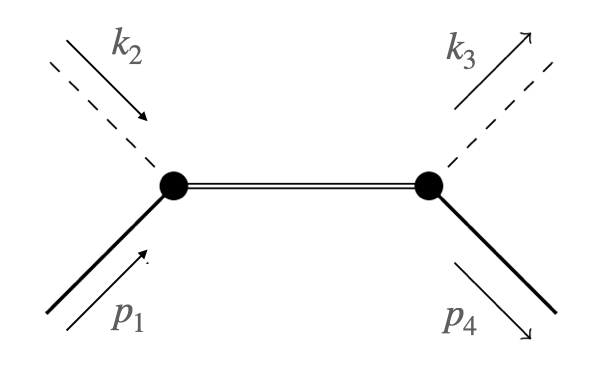}
    \caption{Schematic representation for the ``off-shell'' propagation of absorptive degrees of freedom in the elastic Compton amplitudes.  }
    \label{fig:tree-gluing}
\end{figure}

With this parametrization at hand, Eq.\eqref{eq:schwgravity} becomes\footnote{The phase $e^{i|s|\phi}$ is due to normalization of spin-weighted spheroidal harmonics, and can be reabsorbed by the helicity factors of Eq.\eqref{eq:compt_scalar_hcl} via a little group transformation for the massless spinors $|2\rangle,|3\rangle$ and $[2|,[3|$ (see e.g. Ref.~\cite{Aoude:2023fdm}). We chose to keep it explicit in this section but we will drop it in Section \ref{sec:mass change}. } 
\begin{equation}\label{eq:ampgrav}
{}_{-2}f_{A,\text{ Schw.}}^{\text{LO}}={}_{-2} A^{(0)}\frac{(i) }{45} \frac{\epsilon^6}{\omega}e^{2i\phi} \,.
\end{equation}
Finally, the absorption cross section is obtained from the imaginary part of the forward limit of this amplitude, as indicated in Eq.\eqref{eq:cross_section_LO_absoprtion}. 
For scalar waves, the forward limit is simply obtained by sending $\theta\to0$, which aligns the momenta of the incoming and outgoing wave. For waves of non-zero spin-weight however, the forward limit also requires alignment of polarization in the incoming and outgoing waves. Since in the BHPT computation the polarization of the incoming wave is fixed by $\phi=0$, as discussed below Eq.\eqref{eq:spinhel}, the polarization of the outgoing wave also needs to be set to $\phi=0$ in the forward limit (see Appendix~\ref{app:polvec}). In summary, the forward limit of the absorptive  amplitudes involving spin-weight-$s$ waves is taken via $(\theta,\phi)\to (0,0)$. In this limit, the the helicity factor   ${}_{-2} A^{(0)}\to1$, whereas the 
$e^{2i\phi}$ phase in Eq.\eqref{eq:ampgrav} vanishes. This produces the well-known absorption cross section~\cite{Starobinskil:1974nkd}:
\begin{equation}\label{eq:abs_sch_s2}
{}_{-2}\sigma_{\text{inelastic}}^{\text{LO, Schw.}} = \frac{256}{45}\pi (G M)^6 \omega^4\,.
\end{equation}
In a similar manner,  absorptive Compton amplitudes for wave perturbations of spin-weight $s=0,-1/2,-1,-3/2$ can be obtained from the Teukolsky solutions. 
The  scalar helicity factor for generic spin-weight ${}_sA^{(0)}$  is  generalized  to 
\begin{equation}\label{eq:scalaramplitudes}
{}_sA^{(0)} =\cos^{2|s|}(\theta/2)= \frac{\langle 2|u|3]^{2|s|}}{(2\omega)^{2|s|}} \,,
\end{equation}
and the amplitude picks up a phase proportional to $e^{i|s|\phi}$ from the spin-weight-$s$ spherical harmonics.
In Table~\ref{tab:summary_amplitudes_schw} we summarize the leading order contribution to the absorptive amplitudes for wave perturbations of different spin-weight values off the Schwarzschild BH. Notice for genetic spin-weights,   the   leading $\ell=|s|$  absorptive cross section for Schwarzschild can be put in a unified form 
\begin{equation}
{}_{|s|}\sigma_{\text{inelastic}}^{\text{LO, Schw.}} {=}\frac{\pi(2|s|{+}1)}{-\omega^2} \sum_{n=0}^{1}\epsilon^{2|s|{+}1{+}n}{}_s\beta_{|s|}^{(2|s|{+}1{+}n)} {+}O(\epsilon^{2|s|{+}3}) \,.
\end{equation}
Interestingly, we have obtained absorptive Compton amplitudes without   relying on an EFT spectral decomposition, and more importantly, we have landed on amplitudes without a $t$-channel pole. 
In subsection~\ref{sec:Kerramplitudes} we show that Kerr BH Compton amplitudes share this feature. The non-existence of the $t$-channel pole is not surprising since at leading PM order, the absorptive Compton amplitude can also be obtained from the light mode spectral integration~\cite{Jones:2023ugm,Chen:2023qzo} by gluing two mass-changing three-point amplitudes~\cite{Aoude:2023fdm}, where graviton exchange does not occur as indicated in Fig.~\ref{fig:tree-gluing}. We expand on this in Section~\ref{sec:spectralIntegral}.

\subsection{Leading order absorptive Compton amplitudes for Kerr}\label{sec:Kerramplitudes}
Let us now  proceed with the extraction of the leading absorptive contribution to Compton amplitudes for Kerr. Since Kerr's spin breaks spherical symmetry, one has to consider the contributions to the partial waves from all azimuthal numbers ($m$).  In addition, since finding   explicit formulae for the  Heun spin-weighted spheroidal harmonics  to  all orders in the BH  spin is an open problem,   to get explicit analytic expressions for the absoprtive Kerr Compton amplitudes we  perform  an additional SME, in powers of $(a\omega)$. In particular, in this subsection we obtain explicit results for the  absorptive Compton amplitudes up to order $(a\omega)^{2|s|{+}1}$, which only needs $\ell=|s|$ contributions in the absorption factor ${}_s\eta_{\ell m}$.  This is also the order at which transcendental functions of the Kerr spin parameter $\chi$ start to appear, and more importantly, the order that Schwarzschild results can be recovered as the $a\to0$ limit of Kerr.

Similar to the Schwarzschild case, we  discuss in detail how to extract the LO absorptive amplitude for  $s=-2$ perturbations and provide directly  the  results for wave perturbations of  other spin-weights.   For gravitational wave scattering, the $\beta$-coefficients entering   in Eq.\eqref{eq:beta-splittingeta} are now more complicated as both azimuthal numbers $m$ and the black hole spin parameter $\chi=a/(GM)$ enter the absorption factor. Explicitly, we get contributions to  Eq.\eqref{eq:beta-splittingeta} for $n=0$ and $n=1$, 
given respectively by
\begin{equation}\label{eq:modes_spin45}
\begin{split}
    {}_{{-}2}\beta_{2,m}^{(5)}&{=}\frac{m \chi }{900}  \left(\left(m^2-4\right) \chi ^2+4\right) \left(\left(m^2-1\right) \chi ^2+1\right)\,,\\
    {}_{{-}2}\beta_{2,m}^{(6)}{=}& \frac{1}{8100}\Big[ 45 \pi  \kappa ^2 m^3 \chi ^3
    {+}36 \kappa ^4 (\pi  m \chi -1)-9 \pi  m \chi\\
    &
      \times \left(4 \kappa ^4{+}m^4 \chi ^4{+}5 \kappa ^2 m^2 \chi ^2\right) \coth \left(\frac{\pi  m \chi }{\kappa }\right){+} \\
    &   \left(\chi ^2 \left(9 \pi  m^3 \chi {-}20 m^2{+}5 \left(m^4{-}5 m^2{+}4\right) \chi ^2{+}95\right)\right.\\
    &\left.
    {-}115\right)m^2 \chi ^2\Big]\,.
 \end{split}
\end{equation}
We used  $\kappa=\sqrt{1-\chi^2}$.
This is also the PM order  considered in the analysis of Ref.~\cite{Saketh:2022xjb}. The $ {}_{-2}\beta_{2,m}^{(6)}$ for $m = 0$ is divergent due to the factor  $\coth{\Big(\frac{\pi m\chi}{\kappa}}\Big)$. To evaluate this mode one has take the physical limit $m\to0$, which  produces a  finite answer 
\begin{equation}
     {}_{-2}\beta_{2,\lim_{m=0}}^{(6)} = -\frac{1}{225}\kappa^4(1{+}\kappa)\,.
\end{equation}
Indeed,  in the Schwarzschild limit $\kappa\to1$, we have  ${}_{-2}\beta_{\ell=2,\lim_{m=0}}^{(6)} \to -\frac{2}{225}$ which recovers the coefficient reported  in the $s=-2$ row of Table~\ref{tab:summary_amplitudes_schw}. Importantly, the Schwarzschild  limit can only  be  obtained when the transcendental functions of the Kerr spin parameter $\chi$ are included; we return to this discussion at the end of this subsection. 

In order to obtain a more compact  expression for $  {}_{-2}f^{\text{LO}}_{A,\text{Kerr}}$, it is convenient to match the partial wave sum in Eq.\eqref{eq:absortive_amplitude_def} to a covariant amplitude ansatz. Details on the matching procedure can be found in Ref.~\cite{Bautista:2021wfy}. We use the customary spin basis
\begin{equation}\label{eq:basisspinwave}
\{k_2{\cdot} a,k_3{\cdot} a,w{\cdot} a, a\omega\}\,,\quad \text{with }\quad w^\mu = \frac{2u\cdot k_2}{\langle2|u|3]} \langle2|\sigma^\mu|3]\,.
\end{equation}
The spin-weighed spheroidal harmonics need to be expanded up to order $(a\omega)^1$ and the sums in Eq.\eqref{eq:absortive_amplitude_def} receive contributions from the  $\ell=2,3$ modes. An ansatz fully capturing the non-trivial BHPT solution has the form ($p_1{+}k_2=k_3{+}p_4$)
\begin{equation}\label{eq:ansatzA45}
\begin{split}
   \frac{  {}_{-2}f^{\text{LO}}_{A,\text{Kerr}}}{{}_{-2}A^{(0)}\xi} &= G M \Big[\Big(c_{4,1}z{+}\frac{ c_{4,2}}{\xi} (w{\cdot} a)^2 \Big)(a \omega) (w{\cdot} a){+}(w{\cdot} a)^2\times\\
    &\Big(
    \frac{(w{\cdot} a)^2}{\xi} 
( c_{5,1}w{\cdot} a {+}c_{5,2}a\omega){+}
z( c_{5,3}w{\cdot} a{+}c_{5,4}a\omega)
\Big)\\
&
 {+}\Big( c_{5,7}z{+}\frac{c_{5,8}}{\xi} (w{\cdot} a)^2  \Big)(a \omega) (w{\cdot} a)(k_2 {+}k_3){\cdot} a\\
&
 {+}z^2 ( c_{5,5}w{\cdot} a{+}c_{5,6}a\omega)
\xi 
 \Big]e^{2i\phi}\,.
\end{split}
\end{equation}
We have introduced the optical parameter 
\begin{equation}
    \xi^{-1} = \frac{M^2 t}{(s-M^2)^2} = -\sin^2(\theta/2)\,,
\end{equation}
and to ensure the cancellation of the spurious pole $\xi=-1$ at order $(a\omega)^5$, one needs to impose  $c_{5,5}=-c_{5,1}{+}c_{5,3}$.
The helicity factor ${}_{-2}A^{(0)}$ is given explicitly in Eq.\eqref{eq:compt_scalar_hcl}.

Importantly, all contributions in Eq.\eqref{eq:ansatzA45} are contact terms, i.e. they are polynomial in the Mandelstams. The $t$-channel pole due to the factor of $\xi^{-1}$ in the l.h.s. of Eq.\eqref{eq:ansatzA45} is canceled either by factors of $\xi^{-1}$ or the combination $z=(k_2\cdot a-w \cdot a)(k_3\cdot a-w\cdot a)$ in the r.h.s., whereas the $s$-  and $u$- channel poles are canceled by a copy of either $w\cdot a$, or $\omega a$, with $\omega = (s-M^2)/(2M)$.

Via a matching procedure we identify the free coefficients:
\begin{align}
c_{4,1}= &\frac{ 4 i \left(9 \chi^2{+}8\right) }{45 \chi^3}\,,\,\,\,
    c_{4,2}= \frac{32 i \left(3 \chi^2{+}1\right) }{45 \chi^3}\,,\label{eq:c41and2} \\
     c_{5,1}=&\frac{64 i \pi  \left(3 \chi ^2{+}1\right)}{45 \chi ^4} \,,\,\,\, c_{5,3}=\frac{8 i \pi  \left(33 \chi ^2{+}16\right)}{45 \chi ^4}\,,\\
  c_{5,2}=& \frac{32 i}{135 \chi ^5} \Big(3 \pi  \left(3 \chi ^2{+}1\right) \chi  \tanh \left(\frac{\pi  \chi }{\kappa }\right) \left(\coth ^2\left(\frac{\pi  \chi }{\kappa }\right){+}1\right)\nonumber\\
  &\qquad\qquad {+}6 \chi ^4{+}35 \chi ^2{+}3\Big)\,,\label{eq:c52}\\
  c_{5,4}= &\frac{8 i}{135 \chi ^5} \left(9 \pi  \left(7 \chi ^2{+}4\right) \chi  \coth \left(\frac{2 \pi  \chi }{\kappa }\right)\right.\nonumber\\
  &\left.{+}3 \pi  \left(4-3 \chi ^2\right) \chi  \, \text{csch}\left(\frac{2 \pi  \chi }{\kappa }\right){+}9 \chi ^4{+}157 \chi ^2{+}24\right)\,,\\
  c_{5,6}=&\frac{4 i }{135 \chi ^5}\Big(3 \pi  \chi  \tanh \left(\frac{\pi  \chi }{\kappa }\right) \left(5 \coth ^2\left(\frac{\pi  \chi }{\kappa }\right){+}3 \chi ^2{+}1\right)\nonumber\\
  &{+}34 \chi ^2{+}24{+}9 \kappa^5\Big)\,,\\
  c_{5,7}=& -\frac{8 i \left(9 \chi ^2{+}8\right)}{135 \chi ^3}\,,\,
    c_{5,8}= -\frac{64 i \left(3 \chi ^2{+}1\right)}{135 \chi ^3}\,.\label{eq:last-coeff}
\end{align}

It is interesting to make connection of the results obtained from these expressions with those of Ref.~\cite{Bautista:2022wjf} obtained in the super-extremal limit. For the latter,  the only terms that produce an amplitude with tree-level scaling are
\begin{align}
     c_{5,2}|_{\chi\to\infty}=&\pm\frac{64}{15}\,,\,\,c_{5,4}|_{\chi\to\infty}= \pm\frac{16}{5}\,,\,\,  c_{5,6}|_{\chi\to\infty}=\pm \frac{4}{15}\,,
\end{align}
where the sign indicates the branch choice in the analytic continuation of $\kappa=\sqrt{1-\chi^2}$. These analytically continued coefficients agree with the  coefficients reported in Ref.~\cite{Bautista:2022wjf}  labeled by an $\eta$-tag. We remark that the super-extremal limit is only considered for comparison and we will keep the Kerr spin parameter to lie in the physical region $\chi\le1$ for the rest of this work. 

Returning to the discussion, we now comment on how to obtain the inelastic cross section for gravitational wave scattering off Kerr. It is  obtained via the forward limit of the absorptive amplitude Eq.\eqref{eq:ansatzA45},  as prescribed by  Eq.\eqref{eq:cross_section_LO_absoprtion}. For this we use  the parametrization for the spin-operators given in Eq.\eqref{eq:basisspinwave} as

\begin{subequations}
\begin{align}
    k_2\cdot a=& -a \omega  \cos\gamma \\
    k_3 \cdot a=& a \omega  (\sin\gamma  \sin \theta  \cos \phi -\cos\gamma  \cos \theta )\\
    w\cdot a = &-a \omega  \left(\cos\gamma -e^{i \phi } \sin\gamma  \tan \left(\frac{\theta }{2}\right)\right)\,,\label{eq:wdota}
\end{align} 
\end{subequations}
and arrive at
\begin{equation}\label{eq:crosssectGr}
\begin{split}
\frac{{}_{-2}\sigma_{\text{inelastic}}^{\text{LO}\,,\text{Kerr}}}{(GM)^2} &= \frac{1}{4} i \pi  \chi ^4 \epsilon ^3 \left(c_{4,1} \sin\gamma  \sin (2 \gamma ){+}2 c_{4,2} \cos ^3(\gamma )\right)\\
&-\frac{1}{32} i \pi  \chi ^5 \epsilon ^4 \left(8 \left(c_{5,6} \sin ^4\gamma {+}c_{5,2} \cos ^4\gamma \right.\right.\\
&\left.\left.{+}2 c_{5,8} \cos ^4\gamma {+}\sin ^2\gamma  \cos\gamma  \left(\left(c_{5,4}{+}2 c_{5,7}\right) \cos\gamma \right.\right.\right.\\
&\left.\left.\left.-c_{5,3}\right)\right)-4 c_{5,1} (\cos\gamma {+}\cos (3 \gamma ))\right)\,.
\end{split}
\end{equation}
We have checked that by replacing the coefficients (\ref{eq:c41and2}-\ref{eq:last-coeff}) into Eq.\eqref{eq:crosssectGr} one gets the same result as that obtained by using directly Eq.\eqref{eq:absoption_cross_Section_all}. Therefore  both, operators proportional to $|a| \omega$, and those proportional to $w\cdot a$ in the Compton amplitude in  Eq.\eqref{eq:ansatzA45} contribute to the absorption cross section.  In Ref.~\cite{Bautista:2022wjf}, the $|a|\omega$ operators were viewed as remnants of analytically continuing absorptive effects to the super-extremal limit. In this work we learn that in the physical region $\chi\le1$, both   $|a|\omega$ and $w\cdot a$ operators contribute to absorption.

In the Schwarzschild limit ($\chi\to 0,\kappa\to1$) the absorption cross section in Eq.\eqref{eq:crosssectGr} reduces to that reported in Table~\ref{tab:summary_amplitudes_schw} for gravitational waves. 
The relevant coefficients are $c_{5,i}$ for $i=2,4,6$, with all of them having rational and transcendental functions in $\chi$. 
These transcendental contributions come directly from ${}_{{-}2}\beta_{2,m}^{(6)}$ in Eq.\eqref{eq:modes_spin45}. 
For instance,  typical hyperbolic functions appearing  in these coefficients have  the small spin expansion
\begin{align}\label{eq:smallspin}
\tanh\left( \frac{\pi\chi}{\kappa}\right)& = \pi  \chi + \frac{1}{6} \pi  \left(3-2 \pi ^2\right) \chi ^3 + O\left(\chi ^4\right)\,,\\
\coth\left(\frac{\pi\chi}{\kappa}\right) &= \frac{1}{\pi  \chi } + \left(\frac{\pi }{3}-\frac{1}{2 \pi }\right) \chi + O\left(\chi ^2\right) \,.
\end{align}

This translates into the small-spin expansion of the relevant coefficients
\begin{align}
c_{5,2}=&\frac{32 i(\alpha+1 ) }{45 \chi ^5} {+} O\left(\chi ^{-4}\right)\,,\\
c_{5,4}=&\frac{64 i (\alpha+1) }{45 \chi^5}{+} O\left(\chi ^{-4}\right)\,,\\
c_{5,6}=& \frac{4 i (5 \alpha+11)}{45 \chi^5}{+} O\left(\chi ^{-4}\right)\,,
\end{align}
where we introduced an auxiliary parameter $\alpha=1$ to track contributions from the hyperbolic functions. 
 The final contribution to the cross section in Eq.\eqref{eq:crosssectGr} is of order $O(a^0)$. 
The remaining coefficients scale as $O(\chi^{\ge-4})$ and do not contribute to the Schwarzschild cross section. In summary, the Schwarzschild observable is contained in the $(a\omega)^5$ terms in the SME of the absorptive Compton amplitude in Eq.\eqref{eq:ansatzA45}, and to obtain the correct result we have to keep the Kerr spin parameter $\chi\le1$.

This thus clarifies how to recover absorptive Schwarzschild observables from the small spin limit of Kerr observables, which was reported as an issue in Ref.~\cite{Poisson:2004cw}. In Section~\ref{sec:mass change} we  return to this  discussion in  the context of  the Kerr binary black hole scattering problem.

A second interesting  limiting case  of the absorptive cross section in Eq.\eqref{eq:crosssectGr} is given by  the polar and anti-polar scattering scenarios; these are obtained by setting $\gamma\to0$ and $\gamma\to \pi $ in Eq.\eqref{eq:crosssectGr}, respectively.
In such cases,  the absorption cross section drastically simplifies to 
\begin{equation}
\begin{split}
\frac{{}_{-2}\sigma_{\text{inelastic}}^{\text{LO}\,,\text{Kerr}}|_{\text{polar}}}{(GM)^2}=&\pm
\frac{1}{2} i \pi  \chi ^4 \epsilon ^3 c_{4,2}\\
&\pm\frac{1}{4} i \pi  \chi ^5 \epsilon ^4 \left(c_{5,1}\mp c_{5,2}\mp 2 c_{5,8}\right)\,,
\end{split}
\end{equation}
with the upper/lower sign for the polar/anti-polar case. After substituting the coefficients $c_{4,2}$ of Eq.\eqref{eq:c41and2}, the first line agrees with the result reported in  Eq.(26) in Ref.~\cite{Porto:2007qi}.\footnote{We believe  Eq.(26) in this reference is missing a multiplicative factor of $a_*$. This is most likely  a typo.  } As we show in subsection~\ref{sec:kerr_mass_change}, the polar scattering result is enough 
to recover two-body absorptive observables in the align spin limit, at least for the first few orders in the post-Newtonian (PN)  expansion.  

Finally, in the equatorial limit ($\gamma\to\pi/2$), Eq.\eqref{eq:crosssectGr} becomes 
\begin{equation}
\frac{{}_{-2}\sigma_{\text{inelastic}}^{\text{LO}\,,\text{Kerr}}|_{\text{ecuatorial}}}{(GM)^2}=-\frac{1}{4} i \pi  \chi ^5 \epsilon ^4 c_{5,6}
\end{equation}
which as $\chi\to0$ and with $c_{5,6}\to \frac{64 i }{45 \chi^5}$ recovers the Schwarzschild cross section reported in Table~\ref{tab:summary_amplitudes_schw}, as expected.

The leading PM absorptive Kerr Compton amplitudes and cross sections for waves of spin-weights $s=0,-1/2,-1,-3/2$ can be obtained analogously. We summarize our findings in Table~\ref{tab:summary_amplitudes_kerr}.

It is interesting to understand  the behavior of the absorption cross sections computed above as a function of the energy of the perturbing wave. As it is well known, for a given $\ell$-mode, different superradiant $m$-modes can be exited if the energy of the wave satisfies  
\be\label{eq:superradiant_cond}
\omega-\omega_{c,m}\le 0\,,\quad \omega_{c,m} = \frac{m\chi}{2GM(1+\sqrt{1-\chi^2})}\,,
\ee
where the critical supper-radiant frequency $\omega_{c,m}$, becomes maximal for  $m=|\ell|$. When a super-radiant $m-$mode is exited, the absorption cross section for that given mode becomes negative as energy and angular momentum have been  taken away from the BH. However, when   the azimuthal sum in Eq. \eqref{eq:absoption_cross_Section_all} is performed, supper-radiant effects in the total absorption cross section are expected to average to zero, see Ref.~ \cite{Porto:2007qi}.

\begin{figure}[H]
\begin{center}
    \begin{tabular}{c}
        \begin{subfigure}{0.45\textwidth}
            \centering
            \includegraphics[width=\textwidth]{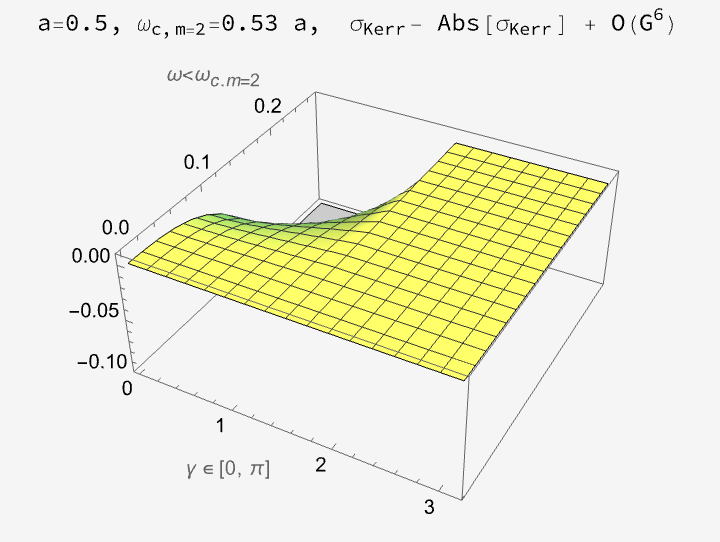}
        \end{subfigure} \\
        \begin{subfigure}{0.45\textwidth}
            \centering
            \includegraphics[width=\textwidth]{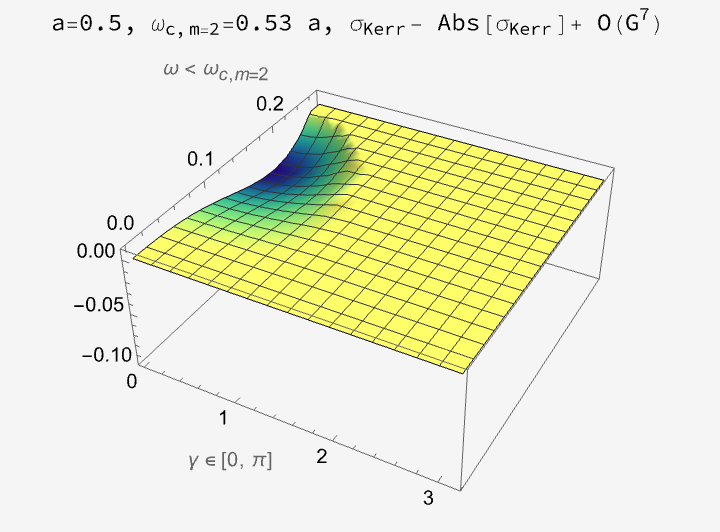}
        \end{subfigure} 
    \end{tabular}
   \end{center} 
\caption{Absolute difference for the  gravitational ($s=-2$) absorptive cross section and its absolute value, up to $O(G^5)$ (top) and $O(G^6)$ (bottom) as function of the wave's   incoming orientation  $\gamma$, and the frequency $\omega$ inside the superradiant region as defined by the inequality in Eq. \eqref{eq:superradiant_cond}. We  have used $GM=1$, and $m=2$, and $a=0.5$ as the value for the Kerr spin parameter. Notice in the Schwarzchild limit ($\omega\to0$), the abosrtive cross section is always possitive.  }
    \label{fig:superrad}
\end{figure}

To illustrate this phenomena and  specializing  to gravitational wave absorption, in
 Fig. \ref{fig:superrad} we have plotted the absolute difference between the  Kerr gravitational absorptive cross section given in Eq. \eqref{eq:crosssectGr} and its absolute value,  as a function of both the  direction of the incoming wave $\gamma$, and the frequency of the wave inside the super-radiance region; in the top plot we included the leading $G^5$ contributions whereas in the bottom one we included up to the  $G^6$ order. We see in the top plot  apparent negative values for the total absorption cross section which appears inside the super-radiance bound. These negative values are present  as a consequence of the truncation of the PM-expansion as can be seen from the  bottom figure, and are expected to totally disappear for the all orders in $G$ result.

In a related front, the absorptive cross section given in Eq.\eqref{eq:crosssectGr}  (and its analogs for wave perturbations of other spin-weight shown in Table \ref{tab:summary_amplitudes_kerr}) predicts an  asymmetry of absorption when  gravitons in a given circular polarization   state  are absorbed more than  those in the other polarization state  depending on the incidence angle, leading to a net  polarizing effect on the outgoing flux of gravitons;\footnote{The authors would like to thank the anonymous referee for bringing this point to attention.} this is the equivalent of the \textit{spin-induced polarization}  which appeared  in  the conservative sector as given for instance in Eq. $(3.18)$ in Ref.~\cite{Bautista:2021wfy}, but now in the absorptive sector. For the latter, 
the absorption cross section for positive helicity gravitons ${}_{+2} \sigma$ can be obtained from that of negative helicity gravitons ${}_{-2} \sigma$ by time reversal ($\omega \to - \omega$) or parity inversion ($\gamma \to \pi - \gamma$). The latter is used to obtain the absorptive cross section ${}_{+2} \sigma_{\text{inelastic}}^{\text{LO}\,,\text{Kerr}}$, since absorption is a manifestly time-asymmetric process.

The net spin-induced polarization  can then be quantified by  ${}_sP$, defined by the ratio \cite{PhysRevD.16.237,Dolan:2008kf,Barbieri:2005kp,Bautista:2021wfy}\footnote{Since the LO helicity reversing amplitude in the absorptive sector vanishes for any spin-weight-$s$, here we define the spin-induce polarization with respect to the absorption cross section computed in  the helicity states rather than the circular polarization states as done in Refs.~\cite{PhysRevD.16.237,Dolan:2008kf,Barbieri:2005kp,Bautista:2021wfy}. },
\begin{align}
   {}_s P &:= \frac{\left[ {}_{-s} \sigma_{\text{inelastic}}^{\text{LO}\,,\text{Kerr}} \right] - \left[ {}_{+s} \sigma_{\text{inelastic}}^{\text{LO}\,,\text{Kerr}} \right]}{\left[ {}_{-s} \sigma_{\text{inelastic}}^{\text{LO}\,,\text{Kerr}} \right] + \left[ {}_{+s} \sigma_{\text{inelastic}}^{\text{LO}\,,\text{Kerr}} \right]} \,,
\end{align}
which vanishes for Schwarzschild BHs. For instance, for integer spin-weight  $s=-2,-1,0$ perturbations, in the small spin expansion we have 
\begin{equation}
{}_{s} P = \chi  \cos (\gamma ) \left(\frac{1}{\epsilon }+\pi \right)\frac{s}{2}+O(\chi^2)\,.
\end{equation}
The positivity of the cross sections ${}_{\pm s} \sigma \ge 0$ imply the bound $|P| \le 1$, where saturation of the bound $P \sim \pm 1$ implies maximal polarizing effects from absorption; only one circular polarization is absorbed, while the other is not. Note that saturation of the bound does not imply total absorption of one polarization.

\begin{figure}[H]
   \centering
    \begin{tabular}{c}
        \begin{subfigure}{0.45\textwidth}
            \centering
            \includegraphics[width=\textwidth]{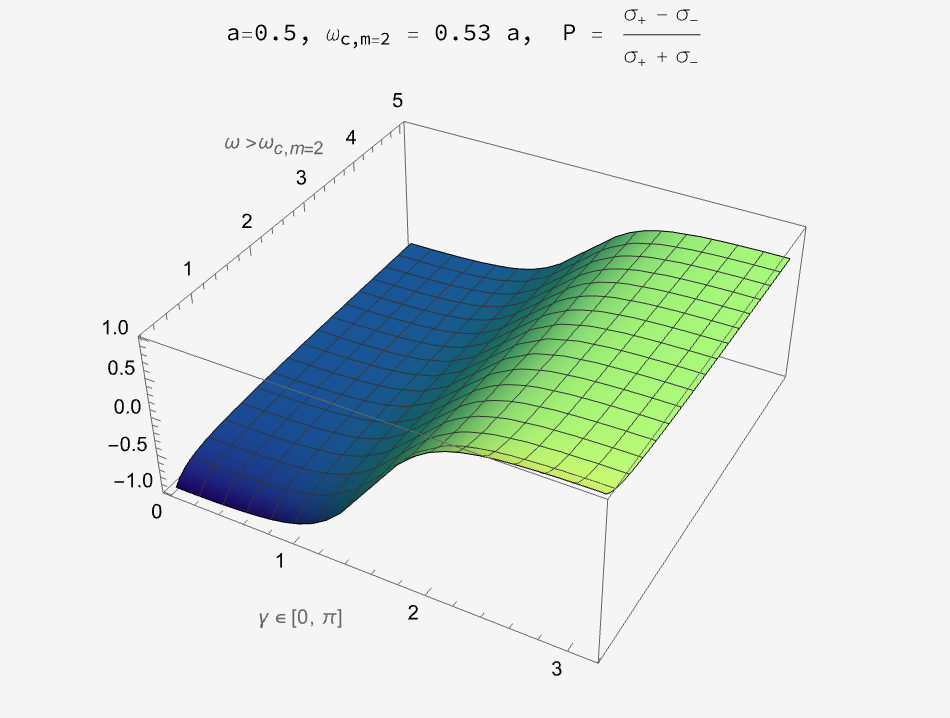}
        \end{subfigure} \\
        \begin{subfigure}{0.45\textwidth}
            \centering
            \includegraphics[width=\textwidth]{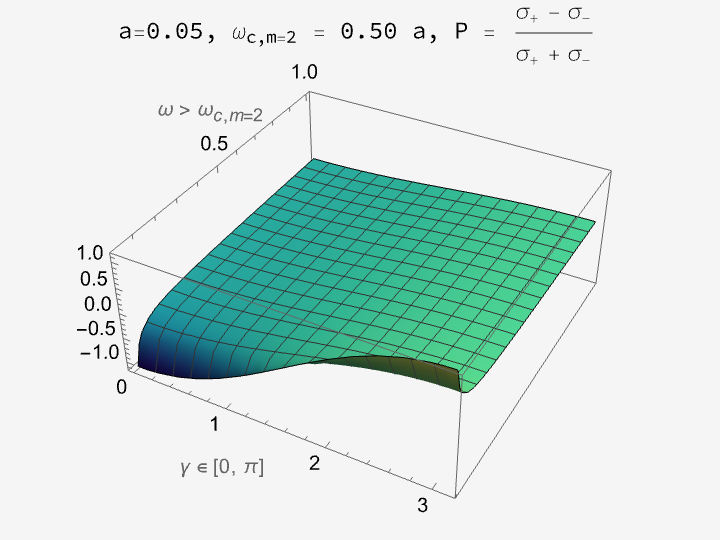}
        \end{subfigure} 
    \end{tabular}
    \caption{Spin-induced polarization ${}_{-2}P$ for gravitational scattering for fast-spinning ($a = 0.5$, upper) and slow-spinning ($a = 0.05$, lower) Kerr black holes ($GM=1$). Fast-spinning black holes have stronger polarizing effect than slow-spinning black holes. The polarizing effect becomes stronger as graviton's frequency approaches the superradiance threshold (towards the front) and as the incidence angle approaches the polar axis ($\gamma \to 0$ or $\gamma \to \pi$).}
    \label{fig:polarization}
\end{figure}

The spin-induced polarization ${}_{-2}P$ for gravitational perturbations of Kerr is plotted in Fig.~\ref{fig:polarization} for fast-spinning ($a = 0.5$) and slow-spinning ($a = 0.05$) black holes for various incidence angles $\gamma \in [0 , \pi]$ and frequencies above the critical frequency $\omega \ge \omega_{c,m = 2}$, where we set $GM=1$. The lower cutoff $\omega_{c,m=2}$ was introduced because the absorptive cross section $\sigma_{\text{inelastic}}^{\text{LO}\,,\text{Kerr}}$ is not perturbatively well-controlled due to PM truncation errors; see the discussions around Eq.\eqref{eq:superradiant_cond}. It is evident from Fig.~\ref{fig:polarization} that polarizing effect is stronger for faster-spinning black holes, polar incidence angles ($\gamma = 0$ or $\gamma = \pi$), and graviton frequencies near the superradiance threshold $\omega \gtrsim \omega_{c,m=2}$.

\begin{table*}
\caption{Leading order absorptive Compton amplitudes and cross section for 
 spin-weight $s=0,-1/2,-1,-3/2$ wave perturbations off Kerr as obtained from BHPT analysis. The coefficients $d_i,e_i$ and $n_i$ entering in each   amplitude ansatz, are obtained directly from the Teukolsky solutions. 
 We have checked that in the Schwarzschild limit we recover the results presented in subsection~\ref{sec:schwarzschild}.  }\label{tab:summary_amplitudes_kerr}
\setlength\extrarowheight{11pt}
\begin{ruledtabular}
\begin{centering}
\begin{tabular}{c||c||l}
\multirow{4}{*}{$0$} & \multirow{2}{*}{$_{0}\beta_{0}^{(1{+}n)}$} & $n=0:$ $-(1{+}\kappa)$\tabularnewline
 &  & $n=1:$ $-\pi(1{+}\kappa)$\tabularnewline
\cline{2-3} 
 & $_{0}f_{A,\text{Kerr}}^{\text{LO}}$ & ${}_{0}A^{(0)} i(\kappa{+}1)(1{+}\epsilon\pi) GM \epsilon \xi$\tabularnewline
\cline{2-3} \cline{3-3} 
 & $_{0}\sigma_{\text{inelastic}}^{\text{LO}\,,\text{Kerr}}$ & $8\pi G^2M^2(\kappa{+}1)(\pi\epsilon{+}1)$\tabularnewline
\hline 
\hline
\multirow{6}{*}{$-\frac{1}{2}$} & \multirow{2}{*}{$_{-\frac{1}{2}}\beta_{\frac{1}{2}m}^{(2{+}n)}$} & $n=0:$ $-\frac{1}{8}\left(\kappa^{2}{+}4m^{2}\chi^{2}\right)$\tabularnewline
 &  & $n=1:$ $\frac{1}{72}\left(-9\pi\left(\kappa{+}4m^{2}\chi^{2}\right){+}9\pi\left(\kappa{+}4m^{2}\chi^{2}\right)\tanh\left(\frac{\pi m\chi}{\kappa}\right){+}8m\chi\left(-4m^{2}\chi^{2}{+}\chi^{2}{+}8\right)\right)$
 \tabularnewline
\cline{2-3} \cline{3-3} 
 & $_{-\frac{1}{2}}f_{A,\text{Kerr}}^{\text{LO}}$ & $_{-\frac{1}{2}}A^{(0)}\xi GM\Big[ d_{1}\frac{a\omega}{\xi}{+}d_{2}\left(z{+}\frac{(w\cdot a)^{2}}{\xi}\right){+}\frac{a\omega}{\xi}\left(d_{3}\left(k_{2}\cdot a{+}k_{3}\cdot a\right){+}d_{4}w\cdot a\right)\Big]e^{\frac{i}{2}\phi}$\tabularnewline
\cline{2-3} \cline{3-3} 
 & $d_{i}$ & $d_{1}=-\frac{i}{2\chi}\,,$ $d_{2}=-\frac{i\pi}{\chi^{2}},$
$d_{3}=\frac{i}{6\chi}\,,$ $d_{4}=-\frac{i\left(3\pi\tanh\left(\frac{\pi\chi}{2\kappa}\right){+}11\chi\right)}{3\chi^{2}}$\tabularnewline
\cline{2-3} \cline{3-3} 
 & $_{-\frac{1}{2}}\sigma_{\text{inelastic}}^{\text{LO}\,,\text{Kerr}}$ & $(GM)^2\pi\Big[-2i \chi^{2}\epsilon\left(d_{2}-\left(2d_{3}{+}d_{4}\right)\cos(\gamma)\right)-4i \chi d_{1}\Big]$\tabularnewline
\cline{2-3} \cline{3-3} 
 & $_{-\frac{1}{2}}\sigma_{\text{inelastic}}^{\text{LO}\,,\text{Kerr}}|_{\text{polar}}$ & $(GM)^2\pi\Big[-2i \chi^{2}\epsilon\left(d_{2}\mp\left(2d_{3}{+}d_{4}\right)\right)-4i \chi d_{1}\Big]$\tabularnewline
\hline 
\hline
\multirow{7}{*}{$-1$} & \multirow{3}{*}{$_{-1}\beta_{1m}^{(3{+}n)}$} & $n=0:$ $\frac{1}{9}m\chi\left(\left(m^{2}-1\right)\chi^{2}{+}1\right)\,$\tabularnewline
 &  & $n=1:$ 
 $\begin{array}{l}
\frac{1}{36}\Big[4\pi m\chi\left(-\left(\left(m^{2}-1\right)\chi^{2}\right)-1\right)\coth\left(\frac{\pi m\chi}{\kappa}\right)\\
{+}\chi\left(4\pi m\left(\left(m^{2}-1\right)\chi^{2}{+}1\right){+}\chi\left(3m^{4}\chi^{2}-3m^{2}\left(\chi^{2}{+}3\right){+}4\right)\right)-4\Big]
\end{array}
$
\tabularnewline
\cline{2-3} 
 & $_{-1}f_{A,\text{Kerr}}^{\text{LO}}$ & 
  $_{-1}A^{(0)}\xi GM\Big[
\frac{e_{1}(a\omega)w{\cdot}a}{\xi}{+}z\left(e_{2}w{\cdot}a{+}e_{3}(a\omega)\right)
{+}\frac{(w{\cdot}a)^{2}\left(e_{2}w{\cdot}a{+}e_{4}(a\omega)\right)}{\xi}{+}\frac{e_{5}(a\omega)w{\cdot}a\left(k_{2}{\cdot}a{+}k_{3}{\cdot}a\right)}{\xi}\Big]e^{i\phi}
$
\tabularnewline
\cline{2-3} \cline{3-3} 
 & $e_{i}$ & $e_{1}=i\frac{4}{3\chi}\,,$ $e_{2}=\frac{8i\pi}{3\chi^{2}}\,,$
$e_{3}=\frac{2i\left(2\kappa^{3}{+}2\pi\chi\coth\left(\frac{\pi\chi}{\kappa}\right){+}\chi^{2}{+}4\right)}{3\chi^{3}}\,,$
$e_{4}=\frac{4i\left(2\pi\chi\coth\left(\frac{\pi\chi}{\kappa}\right){+}3\chi^{2}{+}2\right)}{3\chi^{3}}\,,$
$e_{5}=-\frac{2i}{3\chi}$ \tabularnewline
\cline{2-3} \cline{3-3} 
 & $_{-1}\sigma_{\text{inelastic}}^{\text{LO}\,,\text{Kerr}}$ & $(GM)^2\pi\Big[2i \chi^{2}e_{1}\epsilon\cos(\gamma)-i\chi^{3}\epsilon^{2}\left(e_{3}\sin^{2}(\gamma){+}\left(e_{4}{+}2e_{5}\right)\cos^{2}(\gamma)-e_{2}\cos(\gamma)\right)\Big]$\tabularnewline
\cline{2-3} \cline{3-3} 
 & $_{-1}\sigma_{\text{inelastic}}^{\text{LO}\,,\text{Kerr}}|_{\text{polar}}$ & $(GM)^2\pi\Big[\pm2i \chi^{2}e_{1}\epsilon-i\chi^{3}\epsilon^{2}\left(\left(e_{4}{+}2e_{5}\right)\mp e_{2}\right)\Big]$\tabularnewline
\hline 
\hline
\multirow{12}{*}{$-\frac{3}{2}$}& \multirow{3}{*}{$_{-\frac{3}{2}}\beta_{\frac{3}{2}m}^{(4{+}n)}$} & $n=0:$ $-\frac{\left(\kappa^{2}{+}4m^{2}\chi^{2}\right)\left(9\kappa^{2}{+}4m^{2}\chi^{2}\right)}{1152}$\tabularnewline
 &  & $n=1:$ $
 \begin{array}{l}
 \frac{1}{28800}\left(-225\pi\kappa^{4}-16m^{3}\chi^{3}(m\chi(16m\chi{+}25\pi)-100)-144\kappa^{4}m\chi\right.\\
\left.{+}25\pi\left(\kappa^{2}{+}4m^{2}\chi^{2}\right)\left(9\kappa^{2}{+}4m^{2}\chi^{2}\right)\tanh\left(\frac{\pi m\chi}{\kappa}\right)-40\kappa^{2}m\chi(m\chi(16m\chi{+}25\pi)-50)\right)
 \end{array}$\tabularnewline
\cline{2-3} 
 & $_{-\frac{3}{2}}f_{A,\text{Kerr}}^{\text{LO}}$ & $
\begin{array}{ll}
 _{-\frac{3}{2}}A^{(0)}\xi GM \Big[&\left(n_{1}z{+}\frac{n_{2}(w{\cdot}a)^{2}}{\xi}\right)a\omega{+}\Big(n_{3}z^2\xi{+}\frac{(w{\cdot}a)^{2}\left(n_{4}(w{\cdot}a)^{2}{+}a\omega\left(n_{5}(k_2{+}k_3)\cdot a{+}n_{6}w{\cdot}a\right)\right)}{\xi}\\
 &{+}z\left(n_{7}(w{\cdot}a)^{2}{+}a\omega\left(n_{8}(k_2{+}k_3)\cdot a{+}n_{9}w{\cdot}a\right)\right)\Big)\Big]e^{\frac{3i}{2}\phi}
 \end{array}
 $\tabularnewline
\cline{2-3} 
 & $n_{i}$ & $\begin{array}{l}
 n_{1}=-\frac{i\left(4\chi^{2}{+}3\right)}{12\chi^{3}}\,,n_{2}=-\frac{i\left(8\chi^{2}{+}1\right)}{4\chi^{3}}\,,n_{3}=-\frac{\left(4\chi^{2}{+}3\right)i\pi}{6\chi^{4}}\,,n_{4}=-\frac{\left(8\chi^{2}{+}1\right)i\pi}{2\chi^{4}}\,,\\
 n_{5}=\frac{3i\left(8\chi^{2}{+}1\right)}{20\chi^{3}}\,,n_{6}=-\frac{i\left(15\pi\left(8\chi^{2}{+}1\right)\tanh\left(\frac{3\pi\chi}{2\kappa}\right){+}\chi\left(88\chi^{2}{+}191\right)\right)}{30\chi^{4}}\,,n_{7}= n_{3}{+}n_{4}\,,n_{8}=\frac{i\left(4\chi^{2}{+}3\right)}{20\chi^{3}}\,,\\
 n_{9}=-\frac{i\left(5\pi\left(32\chi^{2}{+}9\right)\tanh\left(\frac{3\pi\chi}{2\kappa}\right){+}5\pi\left(8\chi^{2}-9\right)\sinh\left(\frac{\pi\chi}{2\kappa}\right)\text{sech}\left(\frac{3\pi\chi}{2\kappa}\right){+}56\chi^{3}{+}382\chi\right)}{60\chi^{4}}\,
 \end{array}
 $\tabularnewline
\cline{2-3} 
 & $_{-\frac{3}{2}}\sigma_{\text{inelastic}}^{\text{LO}\,,\text{Kerr}}$ & $
 \begin{array}{ll}
(GM)^2\pi\Big[& -\frac{1}{8}i \chi^{4}\epsilon^{3}\left(4n_{3}\sin^{4}(\gamma){+}n_{7}\sin^{2}(2\gamma){+}4n_{4}\cos^{4}(\gamma)-8n_{5}\cos^{3}(\gamma)-\left(3n_{6}{+}2n_{8}{+}n_{9}\right)\cos(\gamma)\right.\\
 &\left.{+}\left(-n_{6}{+}2n_{8}{+}n_{9}\right)\cos(3\gamma)\right)-i \chi^{3}\epsilon^{2}\left(n_{1}\sin^{2}(\gamma){+}n_{2}\cos^{2}(\gamma)\right)\Big]
 \end{array}
 $\tabularnewline
\cline{2-3}
 & $ _{-\frac{3}{2}}\sigma_{\text{inelastic}}^{\text{LO}\,,\text{Kerr}}|_{\text{polar}}$ & $(GM)^2\pi\Big[-\frac{1}{8}i \chi^{4}\epsilon^{3}\left(4n_{4}\mp8n_{5}\mp\left(3n_{6}{+}2n_{8}{+}n_{9}\right)\pm\left(-n_{6}{+}2n_{8}{+}n_{9}\right)\right)-i \chi^{3}\epsilon^{2}n_{2}\Big]$\tabularnewline
\end{tabular}
\end{centering}
\end{ruledtabular}
\end{table*}

\subsection{ Absorptive Compton amplitudes   from spectral integration. }\label{sec:spectralIntegral}

Finding a prescription that separates the absorptive contributions to the  elastic scattering operator $ {}_s\mathsf{S}_{\ell m}^P$ to all orders in the PM expansion is in general a very hard task. 
The difficulties are twofold; firstly, elastic and inelastic contributions at a fixed PM order mix at higher PM orders due to interference, and secondly, the knowledge of all possible reaction-channels excited in the BH + wave scattering process is necessary. 
In other words, we  need to  supplement the diagonal (input-channel) $1+1'\to 1+1'$ elements ${}_s\mathsf{S}^P_{\ell m}$ by all possible non-diagonal processes,  e.g. $1+1'\to \tilde{1}+\tilde{1'}$ scatterings, $1+1'\to 1''$ excitations, and so on (see chapter XVIII in Ref.~\cite{Landau1981Quantum} and  also Appendix~\ref{app:cross_section}).
Unfortunately, in the classical BHPT  computation  such  non-diagonal elements  are inaccessible since the BH-states describing the  energy and angular momentum absorbed by the BH cannot be accessed from an observer outside the BH. 
This problem is circumvented in classical GR computations by restoring unitarity (by hand) through flux balancing laws at the BH horizon, which is enough to restore unitarity at the level of observables.

Several proposals for EFT parametrizations of BH reaction-channels can be found in the literature. At leading PM order the channels are expected to be described by on-shell mass-changing three-point amplitudes~\cite{Kim:2020dif,Aoude:2023fdm}, or by amplitudes involving   operators parametrizing hidden sectors coupled to gravity~\cite{Jones:2023ugm}. In both of these  approaches, the introduction of an EFT spectral function to model the invisible sector is necessary. The  problem of restoring unitarity can be rephrased as ``what is the spectral density for a black hole?''

Finding such a spectral density is a non-trivial task and in this section we  show that the absorptive amplitudes of Eq.\eqref{eq:absortive_amplitude_def} can be reproduced from a spectral integral, which takes the ``on-shell'' near-threshold part of  the  spectral density as the main input. ``Off-shell'' contributions are also necessary to ensure the vanishing of the BH's Love numbers, as  we  discuss  below.  Other prescriptions considered  in the literature for modeling absorptive effects are based on writing ans\"atze for the spectral decomposition of two-point functions of hidden-sector operators localized on the compact objects~\cite{Goldberger:2020fot,Saketh:2022xjb,Porto:2007qi,Ivanov:2024sds}.

Following the treatment of  Refs.~\cite{Aoude:2023fdm,Jones:2023ugm,Chen:2023qzo}, we proceed to derive the absorptive Compton amplitudes for Schwarzschild\footnote{The case for Kerr is analogous and we do not discuss it here.} BHs using the EFT mass-changing three-point amplitudes, where a BH of mass $M_1$ and momentum $p_1^\mu$ absorbs a graviton  (or a massless quanta of helicity $h<2$) with momentum $k^\mu$ and transitions into an excited state $X$ of momentum $p_2^\mu$,  mass $M_2$ and spin $s_2$, where the details of the ``microstate'' is inaccessible to a distant observer. Such an amplitude is uniquely determined by kinematics and takes the form
\begin{equation}\label{eq:masschanging3pt}
A_3(p_2,s_2|p_1,k,h) {=} g_{s_2}^{|h|} (\bar{\omega})M_1^{1-2s_2} \langle \bar{\textbf{2}}k\rangle^{s_2{-}h}[ \bar{\textbf{2}}k]^{s_2{+}h}\,,
\end{equation}
where $\bar{\omega}=\frac{2p_1\cdot k}{M_1^2}$ is dimensionless. In the classical limit the amplitude is proportional to spin-weighted spherical harmonics~\cite{Aoude:2023fdm} (see Ref.~\cite{Chen:2023qzo} for discussions on the Kerr case).

The ``Wilson coefficients'' $g_{s_2}^{|h|} (\bar{\omega})$ and the spectral density $\rho_l(\mu^2)$ parametrize our ignorance associated with the possible set of states $X$. However, we can impose constraints by matching to the BH absorption cross-section. For example, the inclusive absorption probability for the leading $\ell=|h|$ harmonic leads to the constraint~\cite{Aoude:2023fdm}:
\begin{equation}\label{eq:probab}
{}_h P_{|h|} =
\rho_{|h|}(M_1^2)\Big[\frac{ M_1^2 \bar{w}^{2|h|+1}}{  4(2|h|+1)}\Big|  g_{|h|}^{|h|}(\bar{\omega})\Big|^2\Big] \,.
\end{equation}
Using the PM decomposition of Eq.\eqref{eq:beta-splittingeta}, we can relate the l.h.s. to the $\beta$-coefficients as 
\begin{equation}\label{eq:matching}
{}_h P_{|h|} = 1-{}_{h}\eta_{h,m}^2 =-\sum_{n=0}^{1}\epsilon^{2|h|+1+n}{}_h\beta_{|h|}^{(2|h|{+}1{+}n)} \,.
\end{equation}
Combining the two, we can partially determine the BH spectral function. 
Note that due to the ``squared'' nature of the matching  procedure one  can only account for the product of the spectral density and the modulus square of the EFT coefficients, and information about their phases cannot be obtained from the matching.

The mass-changing three-point amplitude also induces absorptive effects for the Compton amplitude. This process is captured via a tree-level exchange diagram as shown in Fig.~\ref{fig:tree-gluing}. Schematically, such a contribution is given by,
\begin{equation}\label{eq:4pt}
\int d \mu^2  \sum_{\{b\}}  \rho_{\{b\}}(\mu^2) \frac{{}_s A_{\{b\}}^L \times {}_s A_{\{b\}}^R}{s-\mu^2+i0}+(s\leftrightarrow u)\,.
\end{equation}
Importantly, the above integral requires the knowledge of the ``off-shell'' spectral function $\rho_{\{b\}}(\mu^2)$ where $\mu^2\neq 2M_1\omega+M_1^2$. 
This information \emph{cannot} be obtained from the matching procedure of the absorptive cross section, since the matching only involves the on-shell three-point amplitude and hence is only relevant for strict threshold kinematics $\mu^2=2M_1\omega+M_1^2$. 

We expect the absorptive effect contained in Eq.\eqref{eq:4pt} to be the leading contribution to the imaginary part of the elastic scattering matrix  $({}_s\mathsf{S}^P_{\ell m}{-}1)$. To isolate absorptive terms from the amplitude in Eq.\eqref{eq:4pt}, as discussed near Eq.\eqref{eq:SlmDef}, we need to remove the phase associated to the spin-weighted harmonics. Since the three-point amplitudes entering in Eq.\eqref{eq:4pt} are spin-weighted spherical harmonics, we take them outside of the spectral integral and the imaginary part of $\mathsf{T}^P_{\ell m}$ arises from taking  
\begin{equation}\label{eq: Im}
\text{Im}\left[\int d\mu^2 \frac{1}{s-\mu^2+i0}\right] = \pi\,.
\end{equation}

Thus, although the gluing of the three-point amplitudes involves an integration over a continuous spectrum, the imaginary part is a simple polynomial contact term, i.e. on the support of Eqs.(\ref{eq:probab}-\ref{eq:matching})  
\begin{align}\label{eq: ImComp}
& \sum_{\{b\}}  \rho_{\{b\}}(\mu^2) {}_s A_{\{b\}}^L  {}_s A_{\{b\}}^R\bigg|_{\mu^2=2M_1\omega{+}M_1^2}={}_s f_{A}^{\text{LO}} \,.
\end{align}
In other words, when parametrizing the momenta of the incoming and outgoing gravitons by the spinors of Eq.\eqref{eq:spinhel}, we recover the absorptive amplitude for Schwarzschild, defined in Eq.\eqref{eq:absortive_amplitude_def}, at leading PM order for the leading  $l=|h|$ harmonic. 

In summary, we learned that at leading PM order, cutting the propagating   line in Fig.~\ref{fig:tree-gluing} is equivalent to making an insertion of the absorptive contributions to the Compton amplitude. This observation will be useful in the two-body context, and allows us to obtain absorptive two-body observables from the triangle LS, rather than box LSs as na\"ively expected from the KMOC formalism.

Finally, the real part of the spectral integral in Eq.\eqref{eq:4pt} requires the knowledge of off-shell spectral functions; we comment on these issues in  Section~\ref{sec:conclusions}.

\section{Mass Change for a Black Hole in a Binary Scattering }\label{sec:mass change}
\subsection{Absorptive binary  observables from KMOC formalism}\label{sec:KMOC}

In this section we compute the mass change of a Schwarzschild or a Kerr BH with initial mass $m_1$ and incoming momentum $p_1^\mu = m_1 u_1^\mu$, due to the absorption of massless fields of spin-weight-$s$ sourced by another compact body of mass $m_2$ and initial incoming momentum $p_2^\mu = m_2 u_2^\mu$, in a massive $2\to2$ scattering process. For mass-changing Kerr, we denote its spin as $a_1$ and the dimensionless spin parameter as $\chi = a_1/(Gm_1) = S/(Gm_1^2)$.
The binary system's impact parameter is $b^\mu$.

We will utilize the  KMOC formalism~\cite{Kosower:2018adc}, where change in the expectation value of an observable $\mathcal{O}$ is given by,  
\be
\langle \Delta \mathcal{O}\rangle = i\langle \text{in}|[T,\mathcal{O}]|\text{in}\rangle+  \langle \text{in}|T^\dagger [\mathcal{O},T]|\text{in}\rangle\,.
 \ee
The two contributions here are conventionally termed ``virtual'' and ``real'' kernels respectively. In Ref.~\cite{Jones:2023ugm}, it was shown that for $\mathcal{O}=\mathbb{P}_1^\mu$ at leading order in the two-body PM expansion, the transverse (with respect to $p_1^\mu$) part of the virtual kernel cancels against the real kernel, while the longitudinal piece is determined by the real kernel. Moreover, the absorptive contribution to the impulse $\Delta \mathbb{P}^\mu_1$ is proportional to the mass-shift, $\langle\Delta \mathbb{P}^\mu_1\rangle=(\Delta m_1) \hat{u}^\mu_1$, where $\hat{u}_1$ is the dual velocity vector satisfying $\hat{u}_i \cdot u_j=\delta_{ij}$ for $i,j = 1,2$. This suggests that at leading order, we can simply compute $\langle\Delta \mathbb{P}^2_1\rangle$ as the contribution to the impulse.

To do so, it is more convenient to work with the following version of the KMOC formula,
\begin{align} \label{eq:KMOC_new}
\begin{aligned}
    \langle \Delta \mathcal{O} \rangle &= \langle \text{in} | S^\dagger \mathcal{O} S - \mathcal{O} |\text{in}\rangle\,,
    \\ &= i \langle \text{in}| \mathcal{O} T - T^\dagger \mathcal{O} |\text{in}\rangle + \langle \text{in}|T^\dagger \mathcal{O} T |\text{in}\rangle \,.
\end{aligned}
\end{align}

If the wavepacket state $| \text{in} \rangle$ consists of $\mathcal{O}$-eigenstates with a common eigenvalue, we can pull out $\mathcal{O}$ as the eigenvalue $\mathcal{O}_{\text{in}} = \langle \text{in} | \mathcal{O} | \text{in} \rangle$ and write Eq.\eqref{eq:KMOC_new} as
\begin{align} \label{eq:KMOC_mod}
    \langle \Delta \mathcal{O} \rangle &= - \mathcal{O}_{\text{in}} \frac{\langle \text{in}| T |\text{in}\rangle - \langle \text{in}| T^\dagger |\text{in}\rangle}{i} + \langle \text{in}|T^\dagger \mathcal{O} T |\text{in}\rangle \nonumber\,,
    \\ &= \langle \text{in}|T^\dagger \left( \mathcal{O} - \mathcal{O}_{\text{in}} \right) T |\text{in}\rangle \,,
\end{align}
where unitarity of the $S$-matrix was used to obtain the second line and the $\mathcal{O}_{\text{in}} $-insertion is understood to be multiplied by the identity operator. Setting $\mathcal{O} = \mathbb{P}_1^2$, 
\begin{align} \label{eq:mass_gain_op}
    \mathcal{O} - \mathcal{O}_{\text{in}} &= \frac{\mathbb{P}_1^2 - m_1^2}{2 m_1} = \Delta m_1 \left[ 1 + O \left( \frac{\Delta m_1}{m_1} \right) \right] \,,
\end{align}
we can compute the mass change  at leading order from Eq.\eqref{eq:KMOC_mod}. In a nutshell, by computing $\langle \Delta\mathbb{P}_1^2\rangle$, we extract the longitudinal piece of the impulse for $\mathbb{P}_1^\mu$. We remark that Hawking's area theorem is trivially satisfied when positivity of the operator Eq.\eqref{eq:mass_gain_op} is assumed, i.e. when we only consider absorption effects.

\begin{figure*}
    \centering
    \includegraphics[width=0.95\textwidth]{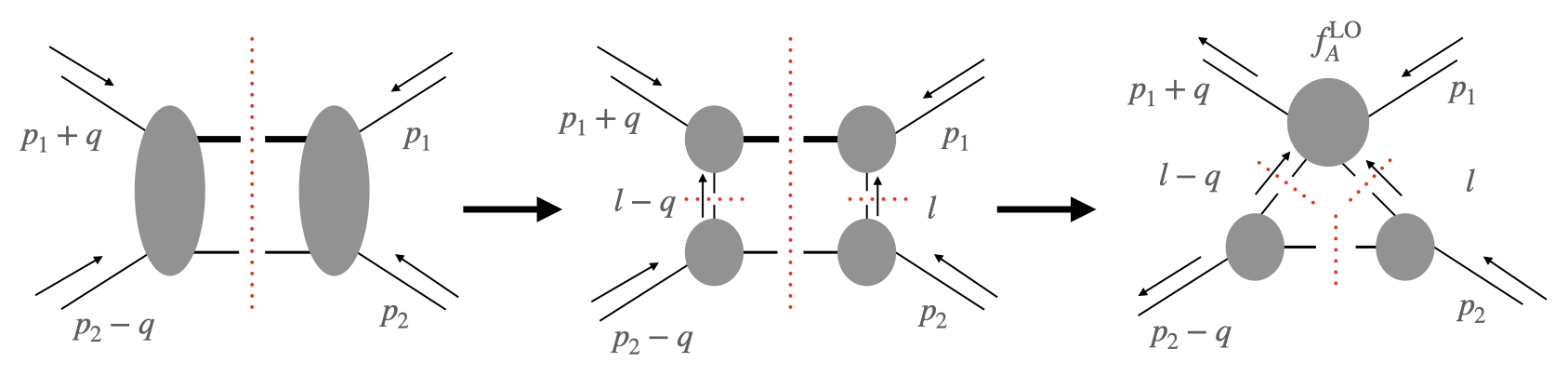}
    \caption{ KMOC representation for  the BH mass-change observable and the triangle reduction sequence. Left: The product of two four-point amplitudes in the original definition. Middle: We isolate classical terms by extracting the none-analytic terms in each four-point. Right: Extracting the imaginary part from the spectral integral collapses into contact terms. }
    \label{fig:cutting}
\end{figure*}

Note that in Eq. (\ref{eq:KMOC_mod}), one is essentially inserting an operator in a unitarity cut. To isolate the absorptive effect, one inserts a spectral projector $\mathfrak{P}_{(\mu^2,m_2^2)}$ 
which projects onto  the intermediate two-particle states with masses $\mu^2$ and $m_2^2$, where $\mu^2$ is the mass of the hidden states associated with the absorptive process. The spectral projector $\mathfrak{P}_{(\mu^2,m_2^2)}$ contains the information of the spectral function $\rho_\ell(\mu^2)$ discussed earlier in the gluing of three-points into the Compton amplitude. The leading order mass change can then be written as
\begin{widetext}
\begin{align}\label{eq: Triangle}
\begin{aligned}
\langle \Delta m_1 \rangle_{\text{LO}} &= \int d\mu^2 \, \langle \text{in} | T^\dagger \mathfrak{P}_{(\mu^2,m_2^2)} \frac{\mathbb{P}_1^2 - m_1^2}{2 m_1} T | \text{in} \rangle = \int d\mu^2 \, \frac{\mu^2 - m_1^2}{2 m_1} \langle \text{in} | T^\dagger \mathfrak{P}_{(\mu^2,m_2^2)} T | \text{in} \rangle \, \\
&\simeq \int \hat{d}^{4}q \, \frac{ \hat{\delta}(q\cdot u_1)\hat{\delta}(q\cdot u_2)}{4 m_1 m_2} e^{iq\cdot b} \int_{0}^\infty d\mathsf{s} \,\mathsf{s}\rho_\ell(\mathsf{s})  \\
&\phantom{=asdfasdfasdf} \times \int d \Phi_{r_1} d \Phi_{r_2} \, \mathcal{A}^*_4(p_1 + q,p_2 - q ; r_1 , r_2) \mathcal{A}_4(p_1,p_2; r_1 , r_2 ) \, \hat{\delta}^4 (p_1 + p_2 - r_1 - r_2) \,,
\end{aligned}
\end{align}
\end{widetext}
where $\mathsf{s}\equiv \frac{\mu^2{-}m^2_1}{2m_1}$, $r_1^\mu$($r_2^\mu$) are the momenta of the particle with mass $\mu^2$ ($m_2^2$), $d\Phi_{r_{1,2}}$ are the Lorentz invariant phase space (LIPS) measure associated to the cut momentum $r_{1,2}^\mu$, and $\mathcal{A}_4$ is the BH + BH$\rightarrow$ BH + X scattering amplitude. The $\simeq$ in the second line denotes that we have neglected the terms that trivialise in the classical limit, such as the integration over wavefunctions of the wavepacket $\psi (p_1) \psi^\ast (p_1 {+} q)$ or the positive energy condition $\Theta (p_1^0 {+} q^0)$. The expression can be diagrammatically expressed as the left diagram of Fig.~\ref{fig:cutting}. Re-parameterising the cut momenta by $r_1^\mu = p_1^\mu + l^\mu$, the LIPS integral in the classical limit reduces to
\begin{widetext}
\begin{align}\label{eq:LIPS_red}
\begin{aligned}
&\int d \Phi_{r_1} d \Phi_{r_2} \, \mathcal{A}^*_4(p_1 + q,p_2 - q ; r_1 , r_2) \mathcal{A}_4(p_1,p_2; r_1 , r_2 ) \, \hat{\delta}^4 (p_1 + p_2 - r_1 - r_2)
\\ &\simeq \int \hat{d}^4 l \, \frac{\hat{\delta} (u_1 \cdot l - s) \hat{\delta} (u_2 \cdot l)}{4 m_1 m_2} 
\, \mathcal{A}^*_4(p_1 + q,p_2 - q ; p_1 + l , p_2 - l) \mathcal{A}_4(p_1,p_2; p_1 + l , p_2 - l ) \,.
\end{aligned}
\end{align}
\end{widetext}
Since the time component of the ``loop momentum'' $l^\mu$ is constrained by the delta constraints, we have simplified the positive energy conditions to $\Theta(p_1^0 {+} l^0) \Theta(p_2^0 {-} l^0) = 1$.

As discussed in ref.~\cite{Jones:2023ugm}, classical dynamics is governed by non-analytic terms in $\mathcal{A}_4$, which at leading order arises from pole contributions corresponding to the exchange of one massless mediator. We can write the integrand of Eq.\eqref{eq:LIPS_red} as
\begin{align} \label{eq:LIPS_red2}
    \int \hat{d}^4 l \, \frac{\hat{\delta} (u_1 \cdot l - s) \hat{\delta} (u_2 \cdot l)}{4 m_1 m_2 \, l^2 (l-q)^2} \times \mathcal{N} (p_1, p_2, l, q, \mathsf{s}) \,,
\end{align}
to make the poles manifest, where $\mathcal{N}$ is the numerator of the ``loop'' integrand. Importantly, the only terms relevant in $\mathcal{N}$ are those non-vanishing on the support of $l^2=(l-q)^2=0$.  Terms that vanish correspond to pinch contributions and do not affect the classical dynamics. Thus, we isolate the classical terms by imposing the cut conditions on $\mathcal{N}$, as illustrated in the middle of Fig.~\ref{fig:cutting}. On the locus of cut constraints the numerator $\mathcal{N}$ can be written as products of 3pt amplitudes and $\mathsf{s}\rho_\ell(\mathsf{s})$,
\begin{align}
    \mathcal{N} &= \mathsf{s} \,\mathcal{F} \, A^*_3(p_2{-}q ; l {-} q , p_2 {-} l) A_3(p_2, {-}l; p_2{-}l) \,, \nonumber
    \\ \mathcal{F} &= \rho_\ell(\mathsf{s}) \, A^*_3(p_1{+}q , l {-} q ; p_1 {+} l) \, A_3(p_1, l ; p_1 {+} l ) \,.
\end{align}

Na\"ively, this would correspond to computing the leading singularity of a box integral, since the two cut conditions are already present in Eqs. (\ref{eq:LIPS_red}-\ref{eq:LIPS_red2}). However, we may exchange the order of integration in Eq.\eqref{eq: Triangle} and evaluate the spectral integral $\int d\mathsf{s}$ first, which localizes to $\mathsf{s} = u_1 \cdot l$ by removing one of the delta constraints. Then $\mathcal{F}$ can be identified  exactly with the imaginary part of the absorptive Compton amplitude $f_{A}^{\text{LO}}$, as discussed around Eq.\eqref{eq: ImComp}
. In the two-body computation, this is illustrated in the rightmost diagram of  Fig.~\ref{fig:cutting}. The resulting integral has a triangle topology,
\begin{align}\label{eq:Triangle3}
    \frac{{\rm Cut}_{\rm tria}[(u_1 \cdot l)  \times  A_3\times   A_3 \times  f_{A}^{\text{LO}}]}{4 m_1 m_2} \times \int \hat{d}^4 l \, \frac{\hat{\delta} (u_2 \cdot l)}{l^2 (l-q)^2} \,,
\end{align}
where ${\rm Cut}_{\rm tria}$ extracts the triangle LS of the numerator $\mathcal{N}|_{\mathsf{s} = u_1 \cdot l}$.
This follows from the fact that computation of the triangle LS can be identified as integral reduction for the triangle integral 
as demonstrated in Appendix~\ref{app:forde}. Finally using that~\cite{Jones:2023ugm} 
\begin{equation}\label{eq:tria_integral}
\int \hat{d}^{4} l \, \frac{\hat{\delta}(u_2 \cdot l)}{l^2(l-q)^2}=\frac{1}{8\sqrt{-q^2}}\,,
\end{equation}
we arrive at      
\begin{align}\label{eq:changeinmass}
\langle \Delta m_1 \rangle_{\text{LO}} &\simeq \frac{1}{16 m_1^2 m_2^2}\int \hat{d}^4q \,\hat{\delta}(q \cdot u_1) \hat{\delta}(q \cdot u_2) \frac{e^{iq{\cdot} b} }{8\sqrt{-q^2}} \nonumber \\
&\phantom{\simeq} \times {\rm Cut}_{\rm tria} \left[(u_1 \cdot l)  \times  A_3\times   A_3 \times  f_{A}^{\text{LO}}\right] \,\nonumber \\
&\equiv\frac{1}{16 m_1^2 m_2^2}\int \hat{d}^4q \,\hat{\delta}(q \cdot u_1) \hat{\delta}(q \cdot u_2) e^{iq{\cdot} b} \text{Im}[\mathcal{M}_4]\,,
\end{align}
where the last line defines $\text{Im}[\mathcal{M}_4]$. Note that we have implicitly used crossing symmetry and converted $A_3^\ast$ to $A_3$.
We now turn to extracting the triangle LS for spin-$s$ massless mediators.

\subsubsection{The real kernel and triangle leading singularity}

We have learned that the change in mass  in a binary BH scattering  process is  controlled by the   absorptive contribution to the $2\to2$ elastic amplitude~\cite{Jones:2023ugm}, and at leading PM order, such contributions are determined from the  triangle cut   depicted in the rightmost diagram of Fig.~\ref{fig:cutting}, where the the leading  absorptive effects are fully captured by the absorptive Compton amplitudes studied in Section~\ref{sec:ComptonsBHPT}.

\begin{figure}
    \centering
    \includegraphics[width=0.6\linewidth]{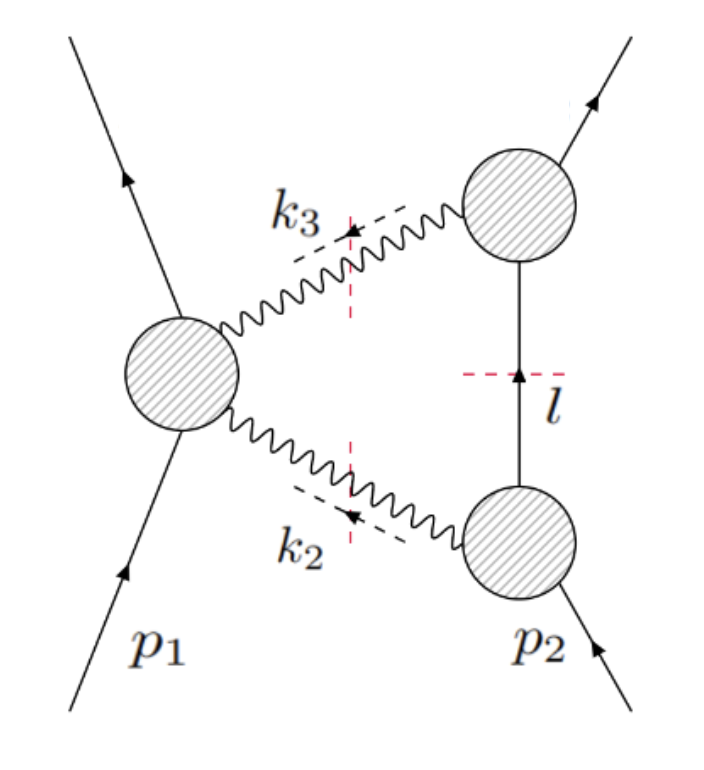}
    \caption{Triangle leading singularity configuration~\cite{Cachazo:2017jef}.  }
    \label{fig:trisngle-cut}
\end{figure}

To evaluate the triangle LS and mass-change as prescribed by Eq.\eqref{eq:changeinmass}, let us relabel the loop momentum in Fig.~\ref{fig:cutting} to match the conventions of the original LS and holomorphic classical limit (HCL) parametrization of Refs.~\cite{Guevara:2017csg,Cachazo:2017jef,Chung:2018kqs,Bautista:2023szu}. In the remaining parts of this section we follow the loop momentum labeling of Fig.~\ref{fig:trisngle-cut}. 

We use helicity amplitudes to compute the triangle-cut in Eq.\eqref{eq:Triangle3}, i.e. we compute the triangle LS as
\begin{align}
\Big[(u_1\cdot k_2){}_s A_{4,\text{abs.}}^{+-}\times{}_s  A_3^-\times  {}_sA_3^+\Big]_{\rm Cut-conditions} \,.
\end{align}
This is enough to compute the mass change since the helicity-reversing absorptive Compton vanishes at leading PM order.
The helicity-preserving absorptive Compton, ${}_s A_{4,\text{abs.}}^{+-}$, is given by
\begin{align}
{}_s A_{4,\text{abs.}}^{+-}={}_sf_{A,\text{Schw./Kerr}}^{\text{LO}}|_{k_3\to-k_3} \,,
\end{align}
due to all-incoming convention for the cut computation, in contrast to the in-out convention for the momenta in the BHPT analysis.

The mass-preserving three-point amplitudes involving one massless particle of spin-weight $s=\{0,-1,-2\}$ can written in a compact form:\footnote{Half-integer spin three-point couplings are forbidden due to the spin-statistics theorem.}
\begin{align}
{}_sA_3(p_2,k^-,p_3) =& {}_sQ^- \left( \frac{[k|p_2|\zeta \rangle}{\langle k\zeta\rangle} \right)^{2|s|}\,,\\
{}_sA_3(p_2,k^+,p_3) = & {}_sQ^+ \left( \frac{\langle k|p_2|\zeta ]}{[k\zeta]} \right)^{2|s|}\,,
\end{align}
where $|\zeta\rangle, |\zeta]$ are reference spinors, and $Q_s^{\pm}$ the spin-weight-$s$ coupling constant; for example, ${}_{-2}Q^{\pm} =- \frac{\sqrt{32\pi G}}{2}$ for gravitons.

Our strategy in the following is to provide an explicit parametrization of the three-point and Compton amplitudes  that allows to evaluate the triangle LS as a contour integral, while allowing generic spin orientations for the BH. We proceed by introducing the LS and  HCL  parametrization of Ref.~\cite{Guevara:2017csg,Cachazo:2017jef,Chung:2018kqs}, where a  generalization for the LS-HCL construction for BHs with generic spin orientation was introduced in Appendix~A of Ref.~\cite{Bautista:2023szu}.
The labelling of momenta can be found in Fig.~\ref{fig:trisngle-cut}. 
The parametrization for the massless momenta $k_2,k_3$ are
\be\label{eq:spin_op}
\begin{split}
k_2^\mu&{=}\frac{|q|\sqrt{z^2{-}1}(m_2p_1^\mu{-}m_1\sigma p_2^\mu){+}m_1m_2\sigma v q^\mu{+}iz \mathcal{E}^\mu}{2m_1m_2\sigma v}\,,\\
k_3&{=}{-}\frac{|q|\sqrt{z^2{-}1}(m_2p_1^\mu{-}m_1\sigma p_2^\mu){-}m_1m_2\sigma v q^\mu{+}iz \mathcal{E}^\mu}{2m_1m_2\sigma v}\,,\\
w^\mu&{=}{-}\frac{|q|\sqrt{z^2{-}1}\big(m_2p_1^\mu z{+}m_1\sigma p_2^\mu(v-z)\big){+}i(z^2-1)\mathcal{E}^\mu}{2m_1m_2(1{-}vz)\sigma}\,,
\end{split}
\ee
where the momentum transfer is $q^\mu = k_2^\mu + k_3^\mu$.
The null vector $w^\mu$ is built from the spinors of the exchanged massless  momenta and provides an extra handle needed for computing spin effects. It is defined as $w^\mu= \frac{2u_1\cdot k_2}{u_1\cdot \epsilon_2} \epsilon_2^\mu$, where 
$
\epsilon_2^\mu=\frac{ \langle 2|\sigma^\mu|3]}{\sqrt{2}[32]}$. We have also defined $\sigma= u_1\cdot u_2 = \frac{1}{\sqrt{1-v^2}}$ to denote the relativistic Lorentz factor, and the vector $\mathcal{E}^\mu=\epsilon^{\alpha\beta\gamma\mu} q_\alpha p_{1\beta}p_{2\gamma}$. These expressions contain the leading in $|q|=\sqrt{-q^2}$ contributions to the classical amplitude. The parameter $z$ was introduced to simplify the expression, and the relation $z=\frac{1+y^2}{2y}$ needs to be inserted when computing the LS.

On the triangle cut, the product $(u_1\cdot k_2)$ evaluates to 
\begin{equation}
u_1\cdot k_2 = \omega_c(y) =  \frac{|q| \sqrt{\sigma ^2-1} \left(y^2-1\right)}{2 y} \,.
\end{equation}
In other words, the expectation value of the mass-change can be thought of as the expectation value of the absorbed energy $\omega_c$ carried by the massless mediator. 

Using this  LS construction,  Eq.\eqref{eq:Triangle3}  is  reduced to the  contour integral 
\begin{equation}\label{LS} 
\frac{1}{8\sqrt{-q^2}} \int_{\Gamma_{\rm{LS}}} \frac{dy}{2\pi y} \omega_c(y) {}_sA_{4,\text{abs.}}^{+-}(\{A\}) {}_s A_3^-(\{B\})  {}_sA_3^+(\{C\}) ,
\end{equation}
where we have used Eq.\eqref{eq:tria_integral}. The contour $\Gamma_{\text{LS}}$ computes the residue at $y = 0$ minus that at $y= \infty$~\cite{Cachazo:2017jef}.\footnote{The correct prescription is to take the average (see Appendix~\ref{app:forde}), but we can compensate the missing factor of $\frac{1}{2}$ by only evaluating the $(k_2^+,k_3^-)$ helicity configuration.} We also included the momenta for the three-point and four-point amplitudes, which follow the conventions of Fig.~\ref{fig:trisngle-cut}: $\{A\}=\{p_1,{-}(p_1+q),k_2^{+},k_3^{-}\}$, $\{B\}=\{p_2,{-}l,{-}k_2^-\}$ and $\{C\}=\{-(p_2-q),l,{-}k_3^+ \}$.

Thus, to compute the change in mass in the scattering process we simply evaluate the LS integral [Eq.\eqref{LS}], 
and substitute the result into the mass changing formula [Eq.\eqref{eq:changeinmass}].
In the remainder of this section we evaluate the change in mass in the  BH-BH scattering  for    different BH binary components, as well as   different spin-weight mediators.

\subsection{Schwarzschild black hole mass change}\label{sec:schchangemass}

\begin{table*}[th]
\caption{LS coefficients  for electromagnetic ($s=-1$) and gravitational ($s=-2$) exchange in  Kerr BH + Schwarzschild BH scattering.  }\label{tab:summary_coeffs}
\setlength\extrarowheight{11pt}
\begin{ruledtabular}
\begin{centering}
\begin{tabular}{c||c||l}
\multirow{1}{*}{$-1$} &\multirow{1}{*}{${}_{-1}A_{j,i}$}  &
 $\begin{array}{l}
 {}_{-1}A_{2,1}= \frac{i e_1 G m_2^2 Q_e^2 \sigma }{64m_1^2}\,,\,  {}_{-1}A_{3,1}=-\frac{G m_2^2 Q_e^2 \left(\sigma ^2-1\right) \left(e_3 \left(5 \sigma ^2-4\right)+3 e_4-2 e_5\right)}{512m_1^3}\,,\,  {}_{-1}A_{3,2}=-\frac{5 e_2 G m_2^2 Q_e^2 \sigma  \left(\sigma ^2-1\right)}{512m_1^3}\\
 {}_{-1}A_{3,3}=-\frac{G m_2^2 Q_e^2 \left(e_3 \left(7-5 \sigma ^2\right)+e_4 \left(5 \sigma ^2-7\right)+2 \left(5 e_5 \sigma ^2+e_5\right)\right)}{512m_1^3}\,,\, {}_{-1}A_{3,4}=-\frac{Gm_2^2 Q_e^2 \left(\sigma ^2-1\right) (3 e_3-3 e_4+2 e_5)}{512m_1^3}
\end{array}
$

 \tabularnewline
\hline 
\hline
\multirow{2}{*}{$-2$}& \multirow{1}{*}{${}_{-2}A_{j,i}$} & 
  $
 \begin{array}{l}
 {}_{-2}A_{4,1}= \frac{i G^2 m_2^4  \sigma  \left(\left(7 \sigma ^2-6\right) c_{4,1}+3 c_{4,2}\right)}{256 m_1^4}\,,
 {}_{-2}A_{4,2}= -\frac{21 i G^2 m_2^4  \sigma  \left(c_{4,1}-c_{4,2}\right)}{256 m_1^4}\,,
 {}_{-2}A_{4,3}= \frac{3 i G^2 m_2^4 \sigma  \left(c_{4,1}-c_{4,2}\right)}{256 m_1^4}
 \\
 {}_{-2}A_{5,1}= -\frac{\pi  G^2  m_2^4 \left(\sigma ^2-1\right) \left(21 \sigma ^4 c_{5,6}-28 \sigma ^2 c_{5,6}-14 \sigma ^2 c_{5,7}+\left(7 \sigma ^2-6\right) c_{5,4}+5 c_{5,2}+8 c_{5,6}+4 c_{5,7}-2 c_{5,8}\right)}{1024 m_1^5}\\
  {}_{-2}A_{5,2}=
  \frac{7 \pi  G^2  m_2^4 \sigma  \left(\sigma ^2-1\right) \left(\left(3 \sigma ^2-7\right) c_{5,1}+\left(4-3 \sigma ^2\right) c_{5,3}\right)}{1024 m_1^5}\,,{}_{-2}A_{5,4}=-\frac{7 \pi  G^2  m_2^4 \sigma  \left(3 \sigma ^2-5\right) \left(2 c_{5,1}-c_{5,3}\right)}{1024 m_1^5} \\
   {}_{-2}A_{5,3}=\frac{\pi  G^2  m_2^4 \left(\left(46-42 \sigma ^2\right) c_{5,2}-7 \left(3 \sigma ^4-10 \sigma ^2+7\right) c_{5,4}+2 \left(\left(9-21 \sigma ^4\right) c_{5,7}+\left(21 \sigma ^4-49 \sigma ^2+26\right) c_{5,6}-8 c_{5,8}\right)\right)}{1024 m_1^5}\\
   {}_{-2}A_{5,5}= -\frac{7 \pi  G^2  m_2^4 \left(3 \sigma ^2 c_{5,6}-6 \sigma ^2 c_{5,7}+6 \sigma ^2 c_{5,8}+\left(3 \sigma ^2-7\right) c_{5,2}+\left(7-3 \sigma ^2\right) c_{5,4}-7 c_{5,6}-2 c_{5,7}+2 c_{5,8}\right)}{1024 m_1^5}\\
    {}_{-2}A_{5,6}=\frac{\pi  G^2  m_2^4 \left(\sigma ^2-1\right) \left(\left(7 \sigma ^2-11\right) c_{5,4}-2 \left(\left(7 \sigma ^2-6\right) c_{5,6}+\left(7 \sigma ^2-3\right) c_{5,7}+2 c_{5,8}\right)+10 c_{5,2}\right)}{1024 m_1^5}\\
   {}_{-2}A_{5,7}=  \frac{21 \pi  G^2  m_2^4 \sigma  \left(\sigma ^2-1\right) \left(2 c_{5,1}-c_{5,3}\right)}{1024 m_1^5}\,, {}_{-2}A_{5,9}=-\frac{\pi  G^2  m_2^4 \left(\sigma ^2-1\right) \left(5 c_{5,2}-5 c_{5,4}+5 c_{5,6}+2 c_{5,7}-2 c_{5,8}\right)}{1024 m_1^5}\\
   {}_{-2}A_{5,8}= \frac{\pi  G^2  m_2^4 \left(21 \sigma ^2 c_{5,6}+\left(21 \sigma ^2-23\right) c_{5,2}+\left(23-21 \sigma ^2\right) c_{5,4}-23 c_{5,6}-8 c_{5,7}+8 c_{5,8}\right)}{512 m_1^5}
 \end{array}$

 \tabularnewline

\end{tabular}
\end{centering}
\end{ruledtabular}
\end{table*}

As a warm up, we study the change in mass in a Schwarzschild BH-BH scattering scenario using absorptive Compton amplitudes in Table~\ref{tab:summary_amplitudes_schw},
obtained from the solutions to the  Teukoslky equations in the $a\to0$ limit. The computation serves as a check against reported PM results in the literature.

From the LS parametrization in Eq.\eqref{eq:spin_op}, the product of three-point amplitudes to be included in Eq.\eqref{LS}  takes the simple form 
\begin{equation}
{}_sA_3^-\times {}_sA_3^+ =  {}_sQ^+ {}_sQ^-  m_2^{2|s|}\,.
\end{equation}
Similarly, the wave-scattering PM parameter is
\begin{equation}
\epsilon = 2 G m_1 \omega_c =
\frac{G m_1 |q| \sqrt{\sigma ^2-1} \left(y^2-1\right)}{2 y}\,.
\end{equation}
Finally, we also introduce the LS parametrization of the   helicity factor for the spin-weight-$s$ absorptive Compton amplitudes
\begin{equation}
\langle 2|u_1|3] = \frac{|q| \left(\sqrt{\sigma ^2-1}+\sqrt{\sigma ^2-1} y^2-2 \sigma  y\right)}{4 y}\,,
\end{equation}
 which then produces  
\begin{equation}\label{eq:scalaramplitudes}
\begin{split}
{}_sA^{(0)} =& \frac{\langle 2|u_1|3]^{2|s|}}{(2\omega_c)^{2|s|}} \\
=&2^{2s}\left(\sigma ^2{-}1\right)^s \left(y^2{-}1\right)^{2 s} \left(\sqrt{\sigma ^2{-}1} \left(y^2{+}1\right){-}2 \sigma  y\right)^{{-}2 s}\,.
\end{split}
\end{equation}

These are the building blocks for evaluating the LS integral Eq.\eqref{LS} involving only Schwarzschild BHs.

Specializing to the case of gravitons (spin-weight $s=-2$)
and using the  LS parametrization introduced above, the  LS integral in Eq.\eqref{LS}  produces 
\begin{equation}
{}_{-2}\text{Im}[\mathcal{M}_4^{\text{Schw.}}] = \frac{\pi  G^7}{720}   m_1^5 m_2^4 |q|^5 \left(1{-}\sigma ^2\right) \left(21 \sigma ^4{-}14 \sigma ^2{+}1\right)\,.
\end{equation}
When inserted in the KMOC mass-change formula [Eq.\eqref{eq:changeinmass}], the mass change is obtained as
\begin{equation}\label{eq:deltam1schgrav}
{}_{-2}\Delta m_1^{\text{Schw.}} 
{=} \frac{5  G^7\pi m_1^6 m_2^2 \sqrt{\sigma ^2
{-}1} \left(21 \sigma ^4{-}14 \sigma ^2{+}1\right)}{16 |b|^{7}}\,,
\end{equation}
where $|b|$ is the magnitude of the impact parameter.
Eq.\eqref{eq:deltam1schgrav} recovers the results reported in Refs.~\cite{Goldberger:2020wbx,Jones:2023ugm}.

The mass-change can also be evaluated for scalar and photon exchange between the BHs.   We  recover the change in mass expressions reported in Eq.(3.36) and Eq.(3.38) of Ref.~\cite{Jones:2023ugm}.

\subsection{Kerr black hole mass change}\label{sec:kerr_mass_change}

We now move to the mass-change analysis for Kerr BH + Schwarzschild BH scattering. In particular, we obtain new results for the change in mass of the Kerr BH with generic spin orientations, which we validate by comparing against existing results in the PN literature for the aligned spin configuration. We also discuss in detail how to recover the Schwarzschild results presented in subsection~\ref{sec:schchangemass}, as  the spinless limit of the Kerr observables.

In order to obtain compact expressions, we expand the result of the  evaluation of the LS integral in terms of the following spin operators:
\begin{equation}
\begin{split}
  \mathcal{H}{=}   \{o_1{=}\varepsilon(q,u_1,u_2,S),o_2{=}q{\cdot} S,o_3{=}|q|u_2{\cdot} S,o_4 {=} |q||S|\}\,.
    \end{split}
\end{equation}
At the $i$-th order in BH SME, the LS can be written in a symbolic form
\begin{equation}\label{eq:cons}
{}_s \text{Im}[\mathcal{M}_4]^{i,\text{Kerr}} =\frac{\pi}{\sqrt{-q^2}}\sum_{j}{}_sA_{i,j}\mathcal{H}^{\otimes^i}_j \,.
\end{equation}
where the ${}_sA_{i,j}$  coefficients are determined by explicit evaluation of the contour integral in Eq.\eqref{LS}.

\subsubsection{Graviton exchange}
The ${}_{_-2}A_{i,j}$ coefficients in Eq.\eqref{eq:cons} for graviton exchange can be obtained from the absorptive Compton amplitude [Eq.\eqref{eq:ansatzA45}] written in the SME. This implies that the leading order for absorptive effects is the fourth order in the SME. The basis of spin operators [Eq.\eqref{eq:basisspinwave}] can be evaluated using the LS parametrization [Eq.\eqref{eq:spin_op}], whereas the optical parameter is given by
\begin{equation}\label{eq:optical}
 \xi =\frac{2(u_1\cdot k_2)^2}{k_2\cdot k_3} =- \sigma^2 v^2 \frac{(1-y^2)^2}{4y}\,.
\end{equation}

 At fourth order in the SME,  the triangle LS [Eq.\eqref{LS}] evaluates to
\be
{}_{-2} \text{Im}[\mathcal{M}_4]^{4,\text{Kerr}}  = 
o_1 o_4 \left(o_2^2 {}_{-2}A_{4,3}{+}o_4^2 {}_{-2}A_{4,1}{+}o_3^2 {}_{-2}A_{4,2}\right) \,.
\ee
For fifth order in the SME,
\be\label{eq:LSspin5gr}
\begin{split}
{}_{-2} \text{Im}[\mathcal{M}_4]^{5,\text{Kerr}}   = o_4 \Big(&o_2^4 {}_{-2}A_{5,9}+o_3^2 o_2^2 {}_{-2}A_{5,8}+o_4^4 {}_{-2}A_{5,1}\\
&+o_3 o_4^3 {}_{-2}A_{5,2}+o_3^4 {}_{-2}{}_{-2}A_{5,5}\\
&+o_4^2 \left(o_2^2 {}_{-2}A_{5,6}+o_3^2 {}_{-2}A_{5,3}\right)\\
&+o_4 \left(o_3^3 {}_{-2}A_{5,4}+o_2^2 o_3 {}_{-2}A_{5,7}\right)\Big)\,.
\end{split}
\ee
The   coefficients  ${}_{s}A_{i,j}$   are reported in Table~\ref{tab:summary_coeffs}.

We take the center of momentum (CoM) frame to simplify comparison with the results available in the PN literature.
The momenta of the BHs are parametrized as (see e.g. Ref.~\cite{Bern:2020buy} for details):
\begin{equation}
p_1{=}{-}(E_1,\boldsymbol{p})\,,\,\,p_2{=}{-}(E_2,-\boldsymbol{p})\,,\,\,q{=}(0,\boldsymbol{q})\,,\,\,\boldsymbol{p}{\cdot}\boldsymbol{q}{=}\frac{\boldsymbol{q}^2}{2}\,,
\end{equation}
Here, $\boldsymbol{p}$ is the asymptotic incoming three-momentum and $\boldsymbol{q}$ is the three-momentum transfer in the scattering process. The total energy is $E=E_1+E_2$, whereas the Lorentz factor $\sigma^2 = \frac{\boldsymbol{p}^2E^2}{m_1^2m_2^2}$.

The covariant spin operators can analogously be mapped to  their CoM representations via  
\be\label{eq:spin_kin_com}
\begin{split}
o_1 &= \frac{E}{m_1 m_2}\,\boldsymbol{S}\cdot(\boldsymbol{p}\times\boldsymbol{q})\,,\,\,\,
o_2=\boldsymbol{q}\cdot\boldsymbol{S}+O(\boldsymbol{q}^2)\,,\\
o_3&=-\frac{|\boldsymbol{q}|E}{m_1m_2} \boldsymbol{p}\cdot \boldsymbol{S}+O(\boldsymbol{q}^2)
\,,\,\,o_4=|\boldsymbol{S}||\boldsymbol{q}|\,.
\end{split}
\ee

The remaining Fourier transform in the KMOC integral [Eq.\eqref{eq:changeinmass}] can be easily evaluated to give:
\begin{widetext}
\be
\begin{split}\label{eq:changemasskerrgravspin4}
{}_{-2}\Delta m_1^{4,\text{Kerr}} =\frac{15  E |\boldsymbol{S}| \boldsymbol{ b}{\cdot}\left(\boldsymbol{p}\times \boldsymbol{S}\right) }{\boldsymbol{ b}^9 m_1^2 m_2^5 \sqrt{\sigma ^2-1}}\left(3 \boldsymbol{ b}^2 E^2 {}_{-2}A_{4,2} \left(\boldsymbol{p}\cdot \boldsymbol{S}\right){}^2+m_1^2 m_2^2 \left({}_{-2}A_{4,3} \left(7 \left(\boldsymbol{ b}\cdot \boldsymbol{S}_{\perp}\right){}^2-\boldsymbol{ b}^2 \boldsymbol{S}_{\perp}^2\right)+3 \boldsymbol{ b}^2 \boldsymbol{S}^2 {}_{-2}A_{4,1}\right)\right)\,,
\end{split}
\ee
\end{widetext}
where $\boldsymbol{S}_{\perp} = \boldsymbol{S}-\boldsymbol{p}\frac{\boldsymbol{p}\cdot \boldsymbol{S}}{\boldsymbol{p}^2}$ is the transverse component of $\boldsymbol{S}$.

The  $A$-coefficients generally contain negative powers of the Kerr spin parameter $\chi$. In particular, the mass change [Eq.\eqref{eq:changemasskerrgravspin4}] has coefficients that scale as $O(\chi^{-3})$ and higher in $\chi$. Therefore, although the norm of the multipole operators $o_i$ scales as $|o_i|\sim O(\chi)$, the net scaling of Eq.\eqref{eq:changemasskerrgravspin4} starts linear in $\chi$ (see for instance Eq.\eqref{eq:changemasskerrspin1gr} for the aligned spin configuration). 
This means the small-$\chi$ expansion and the SME are not the same. 
To correctly capture absorptive effects for BHs with generic spin orientations, one needs to keep at least to fourth order in the SME for the Compton amplitude, which is different from the scaling of $\chi$ in the observable.

Next we specialize to the aligned spin configuration,
\begin{equation}\label{eq:aligned}
\begin{split}
\boldsymbol{ b}\cdot(\boldsymbol{p}\times \boldsymbol{S}){=} \frac{\sqrt{\sigma ^2-1} |\boldsymbol{S}| |\boldsymbol{b}|}{E(m_1 m_2)^{-1}},\,\,\boldsymbol{S}_{\perp}^2{=}\boldsymbol{S}^2,\, 
\boldsymbol{p}\cdot \boldsymbol{S}{=} \boldsymbol{b}\cdot \boldsymbol{S}_{\perp}{=}0\,.
\end{split}
\end{equation}
The change in mass [Eq.\eqref{eq:changemasskerrgravspin4}] becomes
\be\label{eq:changemasskerrspin1gr}
\begin{split}
&{}_{-2}\Delta m_1^{4,\text{Kerr}}\Big|_{\text{aligned}}=\frac{15  G^4 m_1^9  \chi ^4 (3{}_{-2}A_{4,1}-{}_{-2}A_{4,3})}{b^6 m_2^2}\,,\\
&\qquad\quad=-\frac{ \pi  G^6 m_1^5 m_2^2 \sigma  \chi  \left(7 \sigma ^2 \left(9 \chi ^2{+}8\right)+33 \chi ^2{-}24\right)}{64 b^6}\,,\\
&{}_{-2}\Delta m_1^{4,\text{Kerr}}\Big|_{\text{aligned},\sigma\to1}=-\frac{\pi  G^6  m_1^5 m_2^2 \chi  \left(1+3 \chi ^2\right)}{2 b^6}\,,
\end{split}
\ee
where we have written $b = |\boldsymbol{b}|$ and $|\boldsymbol{S}|=\chi G m_1^2$. In the second line we inserted the LS coefficients of Table~\ref{tab:summary_coeffs} and the Teukolsky solutions in Eqs.(\ref{eq:c41and2}-\ref{eq:last-coeff}). In the last line we extracted the leading order non-relativistic limit ($\sigma\to1$) contributions.

Up to an overall factor (which can be attributed to the difference between unbound and bound orbits) the result in the last line of Eq.\eqref{eq:changemasskerrspin1gr} is consistent with the absorbed power computed in Eq.(35) of Ref.~\cite{Porto:2007qi}. 
The latter computation is based on the PN expansion, where the worldline absorptive degrees of freedom were matched to the Kerr absorption  cross   section in the polar limit. 
Therefore, the aligned spin observable is determined by polar scattering of gravitational waves from a Kerr BH, at least to this order in the PM and SME.

The second special case is given by the Schwarzschild limit ($\chi\to0$). The leading order mass change vanishes in this limit, ${}_{-2}\Delta m_1^{4,\text{Kerr}}\to0$, as none of the $c_{4,i}$ coefficients scale as  $\chi^{-4}$ (see Table~\ref{tab:summary_coeffs}). As discussed in Ref.~\cite{Poisson:2004cw}, taking the Schwarzschild limit of the Kerr result  is discontinuous since the mass change for Kerr [Eq.\eqref{eq:changemasskerrgravspin4}] and that for Schwarzschild [Eq.\eqref{eq:deltam1schgrav}] scale with different powers of $G$. 

We observed a similar behavior in Section~\ref{sec:ComptonsBHPT}, and argued that the correct Schwarzschild result is obtained by keeping contributions from transcendental functions of the Kerr spin parameter $\chi$, which appear at order $\epsilon^{2|s|+2}$ in the PM expansion of the absorptive factor ${}_{s}\eta_{\ell m}$. In the case of gravitational scattering, those transcendental-in-$\chi$ contributions are encapsulated by some of the  coefficients $c_{5,i}$ appearing at fifth order in the SME for the absorptive Compton amplitude given in Eq.\eqref{eq:ansatzA45}. We therefore need to compute the mass change for Kerr at this order to correctly approach the Schwarzschild case. Inserting the LS result of Eq.\eqref{eq:LSspin5gr} into  Eq.\eqref{eq:changeinmass} and evaluating the Fourier transform, we obtain the mass change for Kerr as
\begin{widetext}
\be
\begin{split}
{}_{-2}\Delta m_1^{5,\text{Kerr}} =&-\frac{45 |\boldsymbol{S}|}{i|\boldsymbol{b}|^{11} m_1^3 m_2^6 \sqrt{\sigma ^2-1}} \left(7 \boldsymbol{b}^2 m_1^2 m_2^2 \left(\boldsymbol{b}{\cdot}\boldsymbol{S}_{\perp}\right){}^2 \left(E^2{}_{-2}A_{5,8} \left(\boldsymbol{p}{\cdot}\boldsymbol{S}\right){}^2-E m_1 m_2 |\boldsymbol{S}|{}_{-2}A_{5,7} \boldsymbol{p}{\cdot}\boldsymbol{S}\right.\right.\\
&\left.\left.-2 m_1^2 m_2^2{}_{-2}A_{5,9} |\boldsymbol{S}_{\perp}|^2+m_1^2 m_2^2 |\boldsymbol{S}|^2{}_{-2}A_{5,6}\right)+\boldsymbol{b}^4 \left(5 E^4{}_{-2}A_{5,5} \left(\boldsymbol{p}{\cdot}\boldsymbol{S}\right){}^4+|\boldsymbol{S}| \left(E m_1^3 m_2^3{}_{-2}A_{5,7} \boldsymbol{p}{\cdot}\boldsymbol{S} |\boldsymbol{S}_{\perp}|^2\right.\right.\right.\\
&\left.\left.\left.-5 E^3 m_1 m_2{}_{-2}A_{5,4} \left(\boldsymbol{p}{\cdot}\boldsymbol{S}\right){}^3\right)+|\boldsymbol{S}|^2 \left(5 E^2 m_1^2 m_2^2 {}_{-2}A_{5,3} \left(\boldsymbol{p}{\cdot}\boldsymbol{S}\right){}^2-m_1^4 m_2^4{}_{-2}A_{5,6} |\boldsymbol{S}_{\perp}|^2\right)\right.\right.\\
&\left.\left.-E^2 m_1^2 m_2^2{}_{-2}A_{5,8} \left(\boldsymbol{p}{\cdot}\boldsymbol{S}\right){}^2 |\boldsymbol{S}_{\perp}|^2-5 E m_1^3 m_2^3 |\boldsymbol{S}|^3  \boldsymbol{p}{\cdot}\boldsymbol{S} {}_{-2}A_{5,2}+m_1^4 m_2^4{}_{-2}A_{5,9} |\boldsymbol{S}_{\perp}|^4\right.\right.\\
&\left.\left.+5 m_1^4 m_2^4 |\boldsymbol{S}|^4 {}_{-2}A_{5,1}\right)+21 m_1^4 m_2^4{}_{-2}A_{5,9} \left(\boldsymbol{b}{\cdot}\boldsymbol{S}_{\perp}\right){}^4\right)\,.
\end{split}
\ee
\end{widetext}
To obtain the Schwarzschild limit, we first take the align spin limit using the map in Eq.\eqref{eq:aligned}, and then take $\chi\to0$.
\begin{widetext}
\begin{align}
&{}_{-2}\Delta m_1^{5,\text{Kerr}}\Big|_{\text{aligned}}{ =} -\frac{45 G^5 m_1^{11} \chi ^5 \left(5 {}_{-2}A_{5,1}-{}_{-2}A_{5,6}+{}_{-2}A_{5,9}\right)}{ib^7 m_2^2 \sqrt{\sigma ^2-1}}\,,\\
&\quad\quad{=}\frac{45 \pi  G^7  m_1^6 m_2^2  \sqrt{\sigma ^2{-}1} \chi ^5}{i1024 b^7} \Big(105 \sigma ^4 c_{5,6}{-}154 \sigma ^2 c_{5,6}{-}84 \sigma ^2 c_{5,7}{+}\left(42 \sigma ^2{-}46\right) c_{5,4}{+}40 c_{5,2} {+}57 c_{5,6} {+}28 c_{5,7}{-}16 c_{5,8}\Big),\label{eq:nn}\\
&{}_{-2}\Delta m_1^{5,\text{Kerr}}\Big|_{\text{aligned},\chi\to0}{ =}\frac{5  \pi  G^7 m_1^6 m_2^2 \sqrt{\sigma ^2-1} \left(21 \sigma ^4-14 \sigma ^2+1\right)}{16 b^7}\label{eq:nnn}\,.
\end{align}
\end{widetext}
In the second line we inserted Teukolsky coefficients of Table~\ref{tab:summary_coeffs} into the LS coefficients. In the last line we approach the spinless limit ($\chi\to0$) following the arguments around Eq.\eqref{eq:smallspin}. 
The result [Eq.\eqref{eq:nnn}] is consistent with the Schwarzschild result [Eq.\eqref{eq:deltam1schgrav}]. The computation clarifies how the discontinuity in the Kerr and Schwarzschild absorptive observables can be addressed in the two body problem, an issue first raised in  Ref.~\cite{Poisson:2004cw}.

\subsubsection{Other spin-weight exchanges }

For completeness we also comment on the mass change for Kerr when other species of massless particles (spin-weight $s \neq -2$) mediate the interaction. The simplest case is the scalar ($s=0$) particle. The computation is very similar to the Schwarzschild case since the absorptive Compton amplitude for Kerr differs from that of Schwarzschild by an overall multiplicative factor of $\frac{\kappa+1}{2}$, yielding
\be
{}_0\Delta m_1^{\text{Kerr}}= \frac{G^2 (\kappa +1) m_1^2 Q_s^2 \sqrt{\sigma ^2-1}}{64 b^3 m_2^2} \,.
\ee
The Schwarzschild limit ($\kappa\to1$) recovers Eq.(3.36) of Ref.~\cite{Jones:2023ugm}. 

The mass change of Kerr due to $s=-1$ exchange is more interesting. The relevant Compton amplitude is given in Table~\ref{tab:summary_amplitudes_kerr}.
The LS [Eq.\eqref{LS}] evaluates to
\begin{align}
{}_{-1} \text{Im}[\mathcal{M}_4]^{2,\text{Kerr}} &=o_1 o_4 {}_{-1}A_{2,1}\,,\\
{}_{-1} \text{Im}[\mathcal{M}_4]^{3,\text{Kerr}} &=o_4 \Big(o_2^2{}_{-1}A_{2,4}+o_4^2 {}_{-1}A_{2,1}\nonumber\\
&\qquad\qquad+o_3 o_4 {}_{-1}A_{2,2}+o_3^2 A_{2,3}\Big)\,,
\end{align}
where once again LS coefficients are given in Table~\ref{tab:summary_coeffs}. The change in mass from Eq.\eqref{eq:changeinmass} is
\begin{widetext}
\begin{align}
{}_{-1}\Delta m_1^{2,\text{Kerr}} =&-\frac{3 E  |\boldsymbol{S}|{}_{-1}A_{2,1} \boldsymbol{ b}\cdot\left(\boldsymbol{p}\times \boldsymbol{S}\right)}{|\boldsymbol{b}|^5 m_2^3 \sqrt{\sigma ^2-1}}\\
{}_{-1}\Delta m_1^{3,\text{Kerr}} =& \frac{3  |\boldsymbol{S}|}{i\boldsymbol{ b}^7 m_1 m_2^4 \sqrt{\sigma ^2-1}}\Big(3 \boldsymbol{ b}^2 \left(E^2 \left(\boldsymbol{p}\cdot\boldsymbol{S}\right){}^2 {}_{-1}A_{3,1}-E m_1 m_2 |\boldsymbol{S}| \boldsymbol{p}\cdot\boldsymbol{S} {}_{-1}A_{3,2}+m_1^2 m_2^2 |\boldsymbol{S}|^2 {}_{-1}A_{3,1}\right)\nonumber\\
&\qquad\qquad\qquad\qquad\qquad+m_1^2 m_2^2 {}_{-1}A_{3,4} \left(5 \left(\boldsymbol{ b}.\boldsymbol{S}_{\perp}\right){}^2-\boldsymbol{ b}^2 \boldsymbol{S}_{\perp}^2\right)\Big) \,.
\end{align}
\end{widetext}
In the align spin configuration and in the Schwarzschild limit,
\begin{align}
{}_{-1}\Delta m_1^{2,\text{Kerr}} \Big|_{\text{aligned}}=&\frac{ G^3 m_1^3 Q_e^2 \sigma   \chi }{16 b^4}\to0\,\,\,\text{as}\,\chi\to0 \,,\\
{}_{-1}\Delta m_1^{3,\text{Kerr}} \Big|_{\text{aligned}}=&\frac{3 G^3 m_1^7  \chi ^3 (3 {}_{-1}A_{3,1}-{}_{-1}A_{3,4})}{ib^5 m_2^2 \sqrt{\sigma ^2-1}}\,,\\
{}_{-1}\Delta m_1^{3,\text{Kerr}} \Big|_{\text{aligned},\chi\to0}=&-\frac{3  G^4 m_1^4 Q_e^2 \sqrt{\sigma ^2-1} \left(5 \sigma ^2-1\right) }{32 b^5}\,,
\end{align}
where we inserted the LS and Teukolsky solutions given in Table~\ref{tab:summary_coeffs}. The last line recovers Eq.(3.38) of Ref.~\cite{Jones:2023ugm} for the change in mass of Schwarzschild BH due to the absorption of electromagnetic waves.

\section{Discussion}
\label{sec:conclusions}

In this paper we separated purely absorptive contributions in the elastic Compton amplitudes of massless linear perturbations of Kerr BH, and further used this amplitude to compute leading order mass change of black holes in scattering binary dynamics, based on the leading $\ell =|s|$ contributions to the partial wave solutions. While absorptive and conservative contributions can be completely separated at the leading PM order, this is not true for higher PM contributions; the purely conservative far zone contributions in Eq.\eqref{PM exp FZ}, the near-zone phase shift Eq.\eqref{PM exp NZ}, and the absorptive factor Eq.\eqref{PM exp eta} all mix at higher orders.
We leave identifying the iteration-type mixing terms and removing them for computation of two-body observables for future work, 
perhaps exploring on the gravitational self force approach to separate conservative from dissipative contributions via the time symmetric and asymmetric contributions to the Green's functions, respectively~\cite{Pound:2021qin}. 

While the work presented here avoided constructing an effective model parametrizing the absorptive degrees of freedom in the BH by directly using the absorptive contributions to the BHPT solutions, it is desirable to make   connections to the effective models. In particular, in subsection~\ref{sec:spectralIntegral} we showed that although  the imaginary part of the spectral integral in Eq.\eqref{eq:4pt} near the threshold   reproduces the imaginary Compton amplitudes, the real part of the spectral integral induces a na\"ive $\epsilon^{2s+1}\log(\epsilon)$ correction to the static Love numbers of Kerr. As $\delta^{\text{NZ}}$ in Eq.\eqref{PM exp} starts at $\epsilon^{2s+2}$, matching would require vanishing of the na\"ive log contribution. This implies that the spectral integral formula [Eq.\eqref{eq:4pt}] needs to be modified, most probably by the use of  a proper \textit{off-shell} parametrization of the spectral function in Eq.\eqref{eq:probab}, rather than the ``on-shell'' spectral density fixed from the absorption cross section. The situation is similar to the challenge of providing a microscopic description of the vanishing of static Love numbers. 
In this approach, the tidal operator arises as the leading term in the $\omega$ expansion of Eq.\eqref{eq:4pt}, where one integrates over the heavy states. The matching of these to the vanishing Love number in the BHPT computation  presents a non-trivial constraints on the spectral function for the BH.  
Construction of this spectral function is left for future exploration, including possible semi-classical extensions provided by Hawking particle exchanges~\cite{Goldberger:2020geb}, and connections to emergent BH macroscopic properties from microscopic dynamics~\cite{Raj:2023irr}.

Absorption can also induce changes in spin for macroscopic BHs. The study of BH spin-change and its connection to superradiance~\cite{Endlich:2016jgc,Porto:2007qi} is also an interesting avenue for future investigation.

\section*{Acknowledgments}  
We would like to thank  Rafael Aoude, Zvi Bern, Dimitrios Kosmopoulos, David Kosower,  Andres Luna, Nathan Moynihan, Rafael Porto, Radu Roiban, Riccardo Sturani, Fei Teng  and Raju Venugopalan for useful discussions.
The  work of Y.F.B.   has been supported by the European Research Council under Advanced Investigator Grant ERC–AdG–885414. This research used the   Black Hole Perturbation Toolkit~\cite{BHPToolkit}. Y.F.B. thanks the Galileo Galilei Institute for Theoretical Physics for the hospitality and the INFN for partial support during the completion of this work. Y.F.B. and J-W.K thank the hospitality of the Munich Institute for Astro-, Particle and BioPhysics (MIAPbP) and the organizers of the workshop ``EFT and Multi-Loop Methods for Advancing Precision in Collider and Gravitational Wave Physics'', where partial results reported in this paper were presented. This research was  supported in part by the MIAPbP which is funded by the Deutsche Forschungsgemeinschaft (DFG, German Research Foundation) under Germany´s Excellence Strategy – EXC-2094 – 390783311.  YTH is supported by MoST Grant 112-2811-M-002 -054 -MY2.

\appendix

\section{Review on Inelastic Scattering}
\label{app:cross_section}
In this appendix we review different approaches to study  inelastic scattering processes in quantum mechanics and general relativity, making emphasis on their similarities and differences.

\subsection{Inelastic scattering in  Quantum Mechanics}
For this subsection we  follow the discussion presented in Chapter XVIII of Ref.~\cite{Landau1981Quantum}. 
Consider a system of  quantum mechanical particles scattering one another. After the  scattering  process, the particles of the system can  either retain their initial internal states, or they can  change it; when the  former scenario happens, we say that an \textit{elastic interaction} took place, and in the latter case, the interaction is   called   \textit{inelastic}. Examples of inelastic interactions include ionization of atoms, nuclear disintegration,  excitation of atoms,  and  so on. When various of these physical processes happen simultaneously in a particles collision  experiment, we refer to each of them  as different scattering \textit{channels}. The wavefunction of the system of colliding particles is obtained by the direct sum of all terms corresponding to each possible  channel. 

It the presence of more than one scattering channel, the scattering-operator $\hat{S}$, controlling the unitary evolution of the system of  colliding particles will be given in general by a matrix  whose components are scattering  operators dictating the evolution of the particles in the different  channels.  The scattering operator for the elastic channel, also called the \textit{input channel},  corresponds to the diagonal elements of the scattering matrix, lets call it $\hat{S}_{ii}$.  The non-diagonal elements  $\hat{S}_{if}$ will control the evolution of particles  in other  channels, which we  refer also  as the \textit{reaction} channels, and label them with the suffix $f$. Notice that due to the existence of  different reaction channels, neither  $\hat{S}_{ii}$ nor $\hat{S}_{if}$ are unitary, but $\hat{S}$ is unitary\footnote{Unitarity of the scattering matrix is the  requirement that the flux in the ingoing waves must be equal to the sum of the fluxes in the outgoing waves in all scattering  channels.  This is then equal to require  the  sum of the probabilities of all processes (elastic and inelastic) that can occur in the  collision must be unity.}.

The wavefunction of the input channel has an asymptotic expansion consisting of the sum of an incident  plane wave and an elastically scattered outgoing wave. Let us consider for simplicity the system of interacting particles is composed  of only scalar particles. Then, the collision   happens in a plane. In such case, 
\begin{equation}
\psi_i=e^{ik_i z}+f_{ii}(\theta)\frac{e^{ik_i r}}{r}\,.
\end{equation}
Similarly, the  wave function for the other channels is represented by outgoing waves 
\begin{equation}
\psi_f = f_{fi}(\theta) \sqrt{\frac{m_f}{m_i}}\frac{e^{ik_i r}}{r}\,.
\end{equation}
Here $k_f$ is the momentum of the relative motion of the reaction products, and $m_i$ and $m_f$ are the reduced masses of the  initial and  final particles respectively. 

We can define the elements of the scattering matrix using the scattering amplitudes $\hat{f}$, in the different scattering channels as  
\begin{equation}\label{eq:scatteringmatrix}
\hat{S}_{fi} = \delta_{fi}\hat{1} +2 i \sqrt{k_i k_f} \hat{f}_{fi}
\end{equation}
Therefore, the  differential cross section for the elastic channel --- defined as the probability per unit time that a scatter particle passes a surface $r^2d\Omega$, normalized to the incident current density --- is given by
\begin{equation}
d\sigma_{\text{elastic}}:=d\sigma_{ii} = |f_{ii}|^2d\Omega = \frac{1}{4k_i^2}|1-\hat{S}_{ii}|^2d\Omega\,.
\end{equation}
In the same way, the   differential cross section for one of the   reaction channel $f$, is  
\begin{equation}\label{eq:reactcrosssect}
d\sigma_{fi} = |f_{fi}|^2
\frac{k_f}{k_i} d\Omega  =   \frac{1}{4k_i^2}|\hat{S}_{fi}|^2d\Omega\,.
\end{equation}
The total reaction differential cross section resulting from the sum over all possible non-elastic  channels is thus 
\begin{align}\label{eq:react}
 d\sigma_{\text{inelastic}} {:=} \sum_{f}d\sigma_{fi} { =}  \sum_{f} \frac{1}{4k_i^2}|\hat{S}_{fi}|^2d\Omega
 {=} \frac{1}{4k_i^2}[1{-}|\hat{S}_{ii}|^2]d\Omega\,,
\end{align}
where in the last equality we have use the unitarity condition of the total scattering matrix. Finally, the
 total cross section for  the collision is  given by the sum of the total elastic and total inelastic contributions
\begin{equation}
\sigma_{\text{total}} = \sigma_{\text{elastic}}+\sigma_{\text{inelastic}}\,,
\end{equation}
and is well known to be also obtained from the imaginary part of the forward limit of the total elastic cross section $\sigma_{\text{elastic}}$.

\subsection{Inelastic scattering in General Relativity}

In the previous subsection we learned that  both, the  total elastic cross section and the  total reaction (inelastic) cross section in a quantum mechanical scattering process involving different scattering channels are obtained purely from  the (non-unitary) diagonal element of the scattering matrix $\hat{S}_{ii}$. This statement holds also true  for scattering processes in general relativity. However, whereas, in  principle, in a quantum mechanical experiment, both, the diagonal and non-diagonal elements of the scattering matrix can be measured by an observer doing the experiment, in scattering processes involving BHs the non-diagonal elements $\hat{S}_{fi}$ are not available for observers sitting outside the BHs. 

In the wave scattering off Kerr process discussed in the main text, the   scattering amplitude for the elastic     (input ) channel in the presence of  absorptive (reaction) channels, 
analog to the the quantum mechanical amplitude   $f_{ii}$ in Eq.\eqref{eq:scatteringmatrix},
was  written in the partial wave basis in Eq.\eqref{Eq:fKerrGeneric}.  From such s amplitude, the total elastic cross section is obtained as 
\begin{equation}\label{eq:cross_sec_elas}
\begin{split}
&{}_s\sigma_{\text{elastic}} = \int d\Omega |{}_sf(\Omega)|^2\\
&\,{=}\frac{4\pi^2}{\omega^2}\sum_{P,\ell m}\Big|{}_s S_{\ell m}(\gamma,0; a\omega)\Big|^2\Big[1{+}{}_s \eta_{\ell m}^2 {-}2 {}_{s}\eta_{\ell m} \cos(4{}_s \delta^P_{\ell m})\Big]\,,
\end{split}
\end{equation}
were in the second line we have replaced the helicity preserving amplitude in Eq.\eqref{Eq:fKerrGeneric}, with the diagonal partial wave scattering element for the given spin-weight $s$,  ${}_s(\hat{S}_{ii})_{\ell m }:={}_s\mathsf{S}_{\ell m}^P$ as parameterize in Eq.\eqref{eq:connectionformula}, and we  made use  of completeness relations for the spin-weighted spheroidal harmonics to remove the integral and double sum. For $s=-2$, the elastic cross section in Eq.\eqref{eq:cross_sec_elas} receives an extra contribution from the helicity reversing amplitude Eq.\eqref{Eq:gKerrGeneric}.

Since in the BH scenario we cannot access the  individual non-diagonal reaction channels producing the individual scattering  amplitudes $f_{fi}$, the best we can do to account for unitarity is to sum over all such  non-diagonal scattering processes, accounting for the total flux of energy and angular momentum lost in these \textit{hidden} channels. 
Then, from the last equality in  Eq.\eqref{eq:react}, and using the partial wave elastic scattering matrix   ${}_s\mathsf{S}_{\ell m}^P$, 
the total absorption cross section for waves scattering off the BH  will be simply given by ~\cite{Macedo:2013afa}
\begin{equation}\label{eq:absoption_cross_Section_all}
    {}_s\sigma_{\text{inelastic}} = \frac{4\pi^2}{\omega^2}\sum_{\ell,m} \Big|{}_s S_{\ell m}(\gamma,0; a\omega)\Big|^2\big[1-{}_s \eta_{\ell m}^2\big]\,,
\end{equation}

Similarly to the quantum mechanical setup, the 
 total cross section in the wave scattering off Kerr process is  given by the sum of the total elastic and total inelastic cross sections 
\begin{equation}
\begin{split}
{}_s\sigma_{\text{total}} = &{}_s\sigma_{\text{elastic}} {+}{}_s\sigma_{\text{inelastic}} \\
=& \frac{8\pi^2}{\omega^2}\sum_{P,\ell m}\Big|{}_s S_{\ell m}(\gamma,0; a\omega)\Big|^2\Big[1-\eta_{\ell m}\cos(4{}_s \delta^P_{\ell m})\Big]\,,
\end{split}
\end{equation}
which by  the optical theorem can  also be  recovered from the forward limit of the total scattering amplitude in Eq.\eqref{Eq:fKerrGeneric}
\begin{equation}
{}_s\sigma_{\text{total}} = \frac{4\pi}{\omega} \text{Im}[{}_s f^{\text{Forward}}(\theta,\phi)]\,.
\end{equation}
At leading  order in the PM expansion, the inelastic cross section follows from the imaginary piece of the absorptive contribution to the  elastic amplitude in the forward limit, as mentioned around Eq.\eqref{eq:cross_section_LO_absoprtion}.

\section{Absorption factor from the  Nekrasov-Shatashvili function}\label{app:etafromF}
In this appendix  we provide all of the ingredients needed to obtain the explicit expressions for  the absorption  factors ${}_s\eta_{\ell m}$ used in the main body of this work, we provide them  for wave perturbations of generic spin-weight. We start from the  representation of the absorptive factor in the language of the  Nekrasov-Shatashvili (NS) function introduced in Ref.~\cite{Bautista:2023sdf}:
\begin{equation}\label{eq:eta_def}
    {}_s\eta_{\ell m}=\Bigg| \frac{1 {+} e^{-i \pi \bar{\nu}} \mathcal{K}}{1 {+} e^{i\pi\bar{\nu}} \frac{\cos(\pi(m_3 - \bar{\nu}))}{\cos(\pi(m_3{+}\bar{\nu}))} \mathcal{K}}  \Bigg| ~.
\end{equation}
The BH's tidal response function is is given by 
\begin{widetext}
\begin{equation}\label{eq:Kappa}
    \mathcal{K} = |L|^{-2 \bar{\nu}} \frac{ \Gamma(2 \bar{\nu}) \Gamma(2 \bar{\nu} {+}1) \Gamma\left(m_3-\bar{\nu} {+}\frac{1}{2}\right) \Gamma\left(m_2-\bar{\nu} {+}\frac{1}{2}\right) \Gamma\left(m_1-\bar{\nu} {+}\frac{1}{2}\right)}{ \Gamma(-2 \bar{\nu}) \Gamma(1-2 \bar{\nu}) \Gamma\left(m_3{+}\bar{\nu} {+}\frac{1}{2}\right) \Gamma\left(m_2{+}\bar{\nu} {+}\frac{1}{2}\right) \Gamma\left(m_1{+}\bar{\nu} {+}\frac{1}{2}\right)} e^{\partial_{\bar{\nu}} F}\,,
\end{equation}
\end{widetext}
 where the   dictionary of parameters is~\cite{Aminov:2020yma}
\begin{equation}\label{eq:dict}
\begin{aligned}
&m_1 {=} i \frac{m\chi-\epsilon}{\kappa} \,, \quad m_2 {=} -s {-} i \epsilon \, , \quad m_3 {=}  i \epsilon{-}s \,,\quad
 L {=} -2i \epsilon\kappa \,, \\
&\mathsf{u} = -{}_s\lambda_{\ell m} - s(s+1) + \epsilon (  i s\kappa- m \chi) 
 + \epsilon^2 (2 + \kappa ) \,,
\end{aligned}
\end{equation}
and  ${}_s\lambda_{\ell m}$ are the  spheroidal eigenvalue. We further  use the  Matone relation~\cite{Flume:2004rp,Matone:1995rx} to solve for the ``shifted-renormalized angular momentum'' $\bar{\nu}$, order by oder in the instanton expansion, $L$.
\begin{equation}\label{eq:Matone}
\mathsf{u} = \frac{1}{4}-\bar{\nu}^2+L \partial_L F \left(m_1,m_2,m_3,\bar{\nu},L\right) \,.
\end{equation}

To obtain the absorptive coefficients at leading PM order as needed in this paper, it is enough to provide the NS function up to second order in $L$. Recall the NS function   was provided up to order $L^9$ in  Ref.~\cite{Bautista:2023sdf}, and here we import such data up to the order of interest for this work
\begin{widetext}
\begin{equation}
\begin{split}
F=\frac{L m_3 \left(r^2-4 m_1 m_2\right)}{2 r^2}+\frac{L^2 }{64 r^6 \left(r^2+3\right)}&\Big(r^4 \left(r^2 \left(4 m_3^2-r^2-4\right)+4 m_2^2 \left(12 m_3^2+r^2\right)\right)+4 m_1^2 \left(r^4 \left(12 m_3^2+r^2\right)\right.\\
&\left.+4 m_2^2 \left(4 m_3^2 \left(5 r^2-12\right)+3 r^4\right)\right)\Big)+O\left(L^3\right)\,,
\end{split}
\end{equation}
we have further used  $r^2=1-4\bar{\nu}^2$ to simplify the expressions. In the same way, the Matone relation Eq.\eqref{eq:Matone} can be solved up to the same order to give
\begin{equation}
\begin{split}
\bar{\nu} =& -\frac{1}{2} \sqrt{1-4 \mathsf{u}}-\frac{L \left(m_3 \left(\mathsf{u}-m_1 m_2\right)\right)}{2 \left(\sqrt{1-4 \mathsf{u}} \mathsf{u}\right)}-\frac{L^2 }{8 \left((1-4 \mathsf{u})^{3/2} \mathsf{u}^3 (4 \mathsf{u}+3)\right)}\Big(\mathsf{u}^2 (4 \mathsf{u}-1) \left(\mathsf{u} \left(-m_2^2+\mathsf{u}+1\right)-m_1^2 \left(3 m_2^2+\mathsf{u}\right)\right)\\
&
-m_3^2 \left(\mathsf{u}^2 \left(3 m_2^2 (4 \mathsf{u}-1)+\mathsf{u} (12 \mathsf{u}+5)\right)+m_1^2 \left(m_2^2 \left(60 \mathsf{u}^2+5 \mathsf{u}-3\right)+3 (4 \mathsf{u}-1) \mathsf{u}^2\right)-2 m_1 m_2 (4 \mathsf{u}+3) (6 \mathsf{u}-1) \mathsf{u}\right)\Big)\\
&+O\left(L^3\right)
\end{split}
\end{equation}
Finally, the last ingredient for the computation is the  spheroidal eigenvalues ${}_s \lambda_{\ell m}$. Since there is no close form solution for these eigenvalues as these are the eigenvalues of the angular Confluent Heun differential equation, here we provide explicit analytical expressions  up to second order in the SME as obtained from the Black Hole Perturbation Toolkit~\cite{BHPToolkit}:  we have 
\begin{equation}
\begin{split}
{}_s \lambda_{\ell m}=&-\frac{2 a m \omega  \left(s^2+\ell ^2+\ell \right)}{\ell  (\ell +1)}-s (s+1)+\ell ^2+\ell+\\
&\frac{2 (a \omega) ^2 \left(m^2 \left(s^4 (5 \ell  (\ell +1)+3)-6 s^2 \ell ^2 (\ell +1)^2+\ell ^3 (\ell +1)^3\right)+\ell ^2 (\ell +1)^2 \left(-3 s^4+2 s^2 \ell  (\ell +1)+\ell ^3 (\ell +2)-\ell \right)\right)}{\ell ^3 (\ell +1)^3 (4 \ell  (\ell +1)-3)} \\
&+O\left((a\omega)^3\right)
\end{split}
\end{equation}

\end{widetext}
This completes the required ingredients  to compute the absorption factor in Eq.\eqref{eq:eta_def}.  Notice that since $\mathcal{K}$ scales as $(GM\omega)^{2|s|+1+O(G)}$, the absorption factor needs to be computed individually for each type of wave perturbation. For the different spin-weights we obtain the the coefficients entering into the PM expansion in Eq.\eqref{eq:beta-splittingeta}, as indicated in the main body.

\section{Polarisation vectors} \label{app:polvec}
From Eq.\eqref{eq:spinhel}, we construct the polarisation vectors as
\begin{widetext}
\begin{align}
\begin{aligned}
    \varepsilon_-^\mu &= \frac{[ \eta | \bar{\sigma}^\mu | \lambda \rangle}{\sqrt{2} [\eta \lambda]} = \frac{1}{\sqrt{2}} \left( 0 \,,\, \cos \theta \cos \phi + i \sin \phi \,,\, - i \cos \phi + \cos \theta \sin \phi \,,\, - \sin \theta \right) \,,
    \\ \varepsilon_+^\mu &= \frac{\langle \eta | {\sigma}^\mu | \lambda ]}{\sqrt{2} \langle \eta \lambda \rangle} = \frac{1}{\sqrt{2}} \left(0 \,,\, - \cos \theta \cos \phi + i \sin \phi \,,\, - i \cos \phi - \cos \theta \sin \phi \,,\,  \sin \theta \right) \,,
\end{aligned}
\end{align}
\end{widetext}
where we choose
\begin{align}
\begin{aligned}
    {} [ \eta | &= \left( e^{i \phi} \cos \theta /2 \,,\, \sin \theta / 2 \right) \,,
    \\ \langle \eta | &= \left( e^{i \phi} \sin \theta /2 \,,\, - \cos \theta / 2 \right) \,,
\end{aligned}
\end{align}
for the gauge spinors. Their limiting values as $\theta \to 0$ and $\theta \to \pi$ are
\begin{align}
    \varepsilon_-^\mu (\theta \to 0) &= + \frac{e^{i \phi}}{\sqrt{2}} \left( 0 \,,\, 1 \,,\, -i \,,\, 0 \right) \,,
    \\ \varepsilon_+^\mu (\theta \to 0) &= - \frac{e^{-i \phi}}{\sqrt{2}} \left( 0 \,,\, 1 \,,\, i \,,\, 0 \right) \,,
    \\ \varepsilon_-^\mu (\theta \to \pi) &= - \frac{e^{-i \phi}}{\sqrt{2}} \left( 0 \,,\, 1 \,,\, i \,,\, 0 \right) \,,
    \\ \varepsilon_+^\mu (\theta \to \pi) &= + \frac{e^{i \phi}}{\sqrt{2}} \left( 0 \,,\, 1 \,,\, - i \,,\, 0 \right) \,.
\end{align}
Therefore, to approach the correct forward limit, we must move on the same azimuthal slice $\phi = \text{const}$. If the incoming momentum is chosen to have $\phi = 0$, then the outgoing momentum must also have $\phi = 0$.

\section{One-loop integral coefficient extraction for eikonalised propagators \`a la Forde}\label{app:forde}
The leading singularity (LS) computation of Guevara based on the holomorphic classical limit~\cite{Guevara:2017csg} can be viewed as a variant of one-loop integral coefficient extraction explored by Forde~\cite{Forde:2007mi}.
As explained in section~\ref{app:tri_extract}, Forde's method computes the triangle coefficient by reading the residues of poles that appear when triple-cut conditions are imposed on the loop momentum, which is exactly how the triangle LS is computed.

The original argument by Forde relies on the parametrisation of the loop integrand introduced by Ossola, Papadopoulos and Pittau (OPP)~\cite{Ossola:2006us} where the propagators are chosen to be the Feynman type, 
\begin{align*}
    \frac{1}{D_j} &= \frac{1}{\ell^2 - m^2} \,. 
\end{align*}
The purpose of this section is to show that the method extends to integrals with eikonalised/linearised propagators,
\begin{align*}
    \frac{1}{D_j} &= \frac{1}{2 (\ell \cdot u)} \,.
\end{align*}
The argument naturally extends to cut conditions, since the cut conditions can be converted to a linear combination of propagators through the distributional identity $\frac{1}{x \pm i0^+} = \mathcal{P} \left(\frac{1}{x} \right) \mp i \delta (x)$. We mainly follow the presentation of the review Ref.~\cite{Ellis:2011cr}, where extension to non-renormalizable theories having higher-rank tensor integrals is new.

The key idea of OPP parametrisation is that there is a parametrisation of the one-loop integrand where the non-vanishing contribution can be solely attributed to the constant part of the numerator~\cite{Ossola:2006us}. Forde's method can be understood as extraction of the constant part using (multivariate) complex analysis, where the loop momentum is complexified and localised onto the relevant cut conditions~\cite{Forde:2007mi}. Therefore, it suffices to show that an OPP-like parametrisation of the one-loop integrand also exists for eikonalised/linearised propagators and non-renormalizable theories. Since we are only interested in the classical limit, we will limit the discussion to box-like and triangle-like integral coefficients.

A typical one-loop integrand takes the form
\begin{align}
    \int d^D \ell \frac{N (\ell)}{D_0 \cdots D_3} \,, \label{eq:1loop_gen}
\end{align}
where $N(\ell)$ is the numerator and $D_i$ are the inverse propagators. To simplify the analysis, we limit to $D = 4 - 2\epsilon$ and parametrise the inverse propagators as
\begin{align}
\begin{aligned}
    D_0 &= \ell^2 \,,
    \\ D_1 &= (\ell - q)^2 \,,
    \\ D_2 &= 2 (\ell \cdot u_1) \,,
    \\ D_3 &= 2 (\ell \cdot u_2) \,.
\end{aligned}
\end{align}
The physical space is defined as the space spanned by the external vectors appearing in the denominator, while the transverse space is defined as the remaining complementary space. For Eq.\eqref{eq:1loop_gen}, the physical space is three-dimensional and spanned by the vectors $\{ q^\mu , u_1^\mu , u_2^\mu \}$. The transverse space is $(1-2\epsilon)$-dimensional, and we span the space by two basis vectors $\{ n_4^\mu , n_{\epsilon}^\mu \}$. All basis vectors denoted by $n^\mu$ are unit vectors, i.e. $n_k^2 = 1$.

The one-loop integrand Eq.\eqref{eq:1loop_gen} can be algebraically decomposed into
\begin{align}
    \int_\ell \left[ \frac{d_{0123}(\ell)}{D_0 \cdots D_3} + \frac{c_{012}(\ell)}{D_0 D_1 D_2} + \frac{c_{013}(\ell)}{D_0 D_1 D_3} + \cdots \right] \,,
\end{align}
where the ellipsis denotes terms that do not contribute in the classical limit. We show that a parametrisation of the one-loop integrand exists such that only the constant piece $d_{0123} (0)$ ($c_{01j}(0)$) of the box (triangle) numerator $d_{0123} (\ell)$ ($c_{01j}(\ell)$) contributes to the integral.

\subsection{Quadruple cut and box coefficient}
We first analyse the box numerator $d_{0123} (\ell)$. For the physical space spanned by the vectors $v_j^\mu = \{ q^\mu , u_1^\mu , u_2^\mu \}$, we define the dual vectors $w_j^\mu = \{ \bar{q}^\mu , \bar{u}_1^\mu , \bar{u}_2^\mu \}$ such that $v_j \cdot w_k = \delta_{j,k}$.\footnote{Such a basis is known as the van Neerven-Vermaseren basis~\cite{vanNeerven:1983vr}.}

The loop momentum $\ell^\mu$ can be parametrised as
\begin{align} \label{eq:loopmom_box_param}
    \ell^\mu &= \sum_{j=1}^3 w_j^\mu (v_j \cdot \ell) + n_4^\mu (n_4 \cdot \ell) + n_{\epsilon}^\mu (n_\epsilon \cdot \ell) \,.
\end{align}
Inserting this parametrisation into $d_{0123} (\ell)$, we can redistribute the terms containing $(v_j \cdot \ell)$ to lower-point integrals (triangles, bubbles, etc.) by cancelling them against the denominator, e.g.\footnote{This procedure is known as the Passarino-Veltman reduction~\cite{Passarino:1978jh}.}
\begin{align}
    (v_1 \cdot \ell) := (q \cdot \ell) = \frac{q^2 + D_0 - D_1}{2} \,.
\end{align}

The resulting box numerator will be a polynomial of the scalars $(n_4 \cdot \ell)$ and $(n_\epsilon \cdot \ell)$. The polynomial dependence on $(n_4 \cdot \ell)$ can be reduced to linear dependence using the identity
\begin{align} \label{eq:box_loopmom_iden}
    (n_4 \cdot \ell)^2 &= - (n_\epsilon \cdot \ell)^2 + \cdots \,,
\end{align}
where ellipsis denotes terms polynomial in inverse propagators $D_i$. This relation is obtained by contracting Eq.\eqref{eq:loopmom_box_param} with $\ell^\mu$. 

The final form of the numerator $\tilde{d}_{0123} (\ell)$ after redistributing all terms that cancel against the denominator takes the general form\footnote{The linear dependence in $(n_\epsilon \cdot \ell)$ is absent since all external vectors are orthogonal to the $n_\epsilon^\mu$ direction.}
\begin{equation} \label{eq:box_integrand_param}
\begin{split}
    \tilde{d}_{0123} (\ell) &= \tilde{d}_{0123}^{(0)} + \sum_{j=0} \tilde{d}_{0123}^{(2j+1)} (n_4 \cdot \ell) (n_\epsilon \cdot \ell)^{2j}
    \\ & \phantom{=asdf} + \sum_{j=1} \tilde{d}_{0123}^{(2j)} (n_\epsilon \cdot \ell)^{2j} \,.
\end{split}
\end{equation}
The constant piece $\tilde{d}_{0123}^{(0)}$ is the only cut-constructible term that contributes to the box coefficient; the odd-power terms $\tilde{d}_{0123}^{(2j+1)}$ integrate to zero and the even-power terms $\tilde{d}_{0123}^{(2j>0)}$ become the rational part originating from dimensional regularisation $D = 4 - 2\epsilon$.

In $D = 4$ the quadruple cut conditions $D_{0,1,2,3} = 0$ have two solutions $\ell_\pm$, when viewed as a system of algebraic equations over the loop momentum $\ell^\mu$. The two solutions are exchanged under reflection by $n_4^\mu$, 
\begin{align}
    \ell^\mu &\to \ell^\mu - 2 n_4^\mu (n_4 \cdot \ell) \,,
\end{align}
since all cut conditions are invariant under the reflection. This means the wanted box coefficient can be obtained by averaging the cut numerator $\tilde{d}_{0123}(\ell)$ over the cut solutions,
\begin{align}
    \frac{\tilde{d}_{0123}(\ell_+) + \tilde{d}_{0123}(\ell_-)}{2} &= \tilde{d}_{0123}^{(0)} \,,
\end{align}
since $(n_4 \cdot \ell_+) = - (n_4 \cdot \ell_-)$.

\subsection{Triple cut and triangle coefficient} \label{app:tri_extract}
We analyse the triangle numerator $c_{012} (\ell)$. Now the physical space is spanned by the vectors $v_j^\mu = \{ q^\mu , u_1^\mu \}$ and their dual vectors are $w_j^\mu = \{ \bar{q}^\mu , \bar{u}_1^\mu \}$. For the transverse space we use basis vectors $\{ n_3^\mu , n_4^\mu , n_{\epsilon}^\mu \}$.

Similar to Eq.\eqref{eq:loopmom_box_param}, we parametrise $\ell^\mu$ as
\begin{align}
    \ell^\mu &= \sum_{j=1}^2 w_j^\mu (v_j \cdot \ell) + \sum_{k=3}^4 n_k^\mu (n_k \cdot \ell) + n_{\epsilon}^\mu (n_\epsilon \cdot \ell) \,.
\end{align}

Inserting this parametrisation into $c_{012}(\ell)$ and redistributing terms containing $(v_j \cdot \ell)$ to lower-point integrals, the resulting triangle numerator $\hat{c}_{012} (\ell)$ becomes a polynomial of the scalars
\begin{align}
    (n_3 \cdot \ell) \,,\, (n_4 \cdot \ell) \,,\, (n_{\epsilon} \cdot \ell) \,.
\end{align}
The terms in $\hat{c}_{012} (\ell)$ with odd-power dependence in any of the above scalars integrate to zero, but this is not the case for the even powers. Therefore, we use the triangle equivalent of Eq.\eqref{eq:box_loopmom_iden} to reorganise the numerator.
\begin{align} \label{eq:tri_loopmom_iden}
    (n_3 \cdot \ell)^2 + (n_4 \cdot \ell)^2 = - (n_\epsilon \cdot \ell)^2 + \cdots \,.
\end{align}
Similar to Eq.\eqref{eq:box_loopmom_iden}, the ellipsis denotes polynomial terms in $(v_j \cdot \ell)$ which (apart from the constant term) are redistributed to lower-point integrals.

We first reorganise the triangle numerator $\hat{c}_{012} (\ell)$ by converting all even powers of $(n_4 \cdot \ell)$ into $(n_3 \cdot \ell)$ using Eq.\eqref{eq:tri_loopmom_iden}.
\begin{align}
    \check{c}_{012} (\ell) &= \check{c}_{012}^{(0)} + \sum_{j,m=0} \check{c}_{012}^{(j,1,2m)} (n_3 \cdot \ell)^{j} (n_4 \cdot \ell) (n_\epsilon \cdot \ell)^{2m} \nonumber
    \\ &\phantom{=} + \sum_{j,m=0} \check{c}_{012}^{(j,0,2m)} (n_3 \cdot \ell)^{j} (n_\epsilon \cdot \ell)^{2m} \,.
\end{align}
It is obvious that all $\check{c}_{012}^{(j,1,2m)}$ terms vanish in the integral due to symmetry. On the other hand, the second line ($\check{c}_{012}^{(j,0,2m)}$ terms) of the triangle integrand do not vanish for even $j$. Therefore, we use Eq.\eqref{eq:tri_loopmom_iden} to recast the even terms as
\begin{align} \label{eq:tri_reparam_1}
    (n_3 \cdot \ell)^{2j} &\to \frac{(n_3 \cdot \ell)^{2j}}{2j} - \frac{2j-1}{2j} (n_3 \cdot \ell)^{2j-2}(n_4 \cdot \ell)^2 \,.
\end{align}

This combination integrates to zero, since the following integral is independent of $\phi$ due to rotational symmetry in the transverse space.
\begin{align}
    \int_\ell \frac{[(n_3 \cos \phi + n_4 \sin \phi) \cdot \ell]^{2j}}{D_0 D_1 D_2} \,.
\end{align}
Taking the double derivative 
$\frac{\partial^2}{\partial \phi^2}$ 
and setting $\phi = 0$, we obtain the RHS of Eq.\eqref{eq:tri_reparam_1} which vanishes under the integral.

The final form of the triangle numerator $\tilde{c}_{012} (\ell)$ is
\begin{align} \label{eq:tri_integrand_param}
    \tilde{c}_{012} (\ell) &= \tilde{c}_{012}^{(0)} + \sum_{j,m=0} \tilde{c}_{012}^{(j,1,2m)} (n_3 \cdot \ell)^j (n_4 \cdot \ell) (n_\epsilon \cdot \ell)^{2m} \nonumber
    \\ &\phantom{=} + \sum_{j,m=0} \tilde{c}_{012}^{(2j+1,0,2m)} (n_3 \cdot \ell)^{2j+1} (n_\epsilon \cdot \ell)^{2m} 
    \\ &\phantom{=} + \sum_{j=1,m=0} \tilde{c}_{012}^{(2j,0,2m)} (n_3 \cdot \ell)^{2j-2} (n_\epsilon \cdot \ell)^{2m} \nonumber
    \\ &\phantom{=asdfasdf} \times \left[ (n_3 \cdot \ell)^2 - (2j-1) (n_4 \cdot \ell)^2 \right] \,.\nonumber
\end{align}
The coefficient $\tilde{c}_{012}^{(0)}$ is cut-constructible and is the only term that contributes to the triangle coefficient; other terms with $m = 0$ vanish under the integral and the terms with $m \neq 0$ become the rational part.

In $D = 4$ the locus of triple cut conditions $D_{0,1,2} = 0$ is a complex 1-dimensional manifold, when viewed as a system of algebraic equations over the loop momentum $\ell^\mu$. Because of rotational symmetry in the transverse space, the triple cut solution $\ell^\mu$ has a fixed (Euclidean) norm when projected onto the transverse space. We parametrise the triple cut solution $\ell^\mu$ using the complex variable $T$ as
\begin{align}
    \ell^\mu &= V_P^\mu + L \left( T n_+^\mu + \frac{1}{T} n_-^\mu \right) \,, \label{eq:tri_loopmom_param}
    \\ n_\pm^\mu &= n_3^\mu \pm i n_4^\mu \,, 
\end{align}
where $V_P^\mu$ is a fixed vector in physical space. Since $n_\pm^\mu$ are null vectors, the projection of $\ell^\mu$ onto the transverse space has fixed norm $4 L^2$.

We now study the triangle numerator $\tilde{c}_{012}$ as a function of $T$. The terms contributing to the rational part, $m \neq 0$, are removed by the loop momentum parametrisation.
\begin{align}
    \tilde{c}_{012} (T) &= \tilde{c}_{012}^{(0)} + \sum_{j=0} i \tilde{c}_{012}^{(j,1,0)} L^{j+1} \left( T + \frac{1}{T} \right)^j \left( T - \frac{1}{T} \right) \nonumber
    \\ &\phantom{=} + \sum_{j=0} \tilde{c}_{012}^{(2j+1,0,0)} L^{2j+1} \left( T + \frac{1}{T} \right)^{2j+1}
    \\ &\phantom{=} + \sum_{j=1} \tilde{c}_{012}^{(2j,0,0)} \left( T + \frac{1}{T} \right)^{2j-2} \nonumber
    \\ &\phantom{=asdf} \times \left[ \left( T + \frac{1}{T} \right)^2 + (2j-1) \left( T - \frac{1}{T} \right)^2 \right] \,. \nonumber
\end{align}
As a Laurent expansion in $T$, the only non-vanishing coefficient of $T^0$ is $\tilde{c}_{012}^{(0)}$, which also is the only term that contributes to the triangle coefficient. The remaining task is to assure that quadruple cut contributions $\tilde{d}_{0123}$ do not contribute to the $T^0$ coefficient.

As a function of $T$, the quadruple cut contribution Eq.\eqref{eq:box_integrand_param} takes the form
\begin{align} \label{eq:box_integrand_tricut}
    \frac{\tilde{d}_{0123}}{D_3} &= \frac{\tilde{d}_{0123}^{(0)}}{D_3 (T)} + \tilde{d}_{0123}^{(1)} L \frac{(n_\perp \cdot n_+) T + \frac{(n_\perp \cdot n_-)}{T} }{D_3 (T)} \,.
\end{align}
We have restored the uncut propagator $D_3$ and renamed $n_4^\mu$ of Eq.\eqref{eq:box_integrand_param} to $n_\perp^\mu$, since it is not guaranteed that the basis vector of the transverse space are the same for both cuts. The $T$-dependence of the uncut propagator is given as
\begin{align}
    D_3 (T) &= (u_2 \cdot n_+) T + a_0 + (u_2 \cdot n_-) \frac{1}{T} \,,
\end{align}
where $a_0$ is some constant that does not matter.

We now investigate the limiting behaviour of Eq.\eqref{eq:box_integrand_tricut}.
\begin{align}
    \lim_{T \to 0} \frac{\tilde{d}_{0123}}{D_3} &= \tilde{d}_{0123}^{(1)} L \frac{(n_\perp \cdot n_-)}{(u_2 \cdot n_-)} \,,
    \\ \lim_{T \to \infty} \frac{\tilde{d}_{0123}}{D_3} &= \tilde{d}_{0123}^{(1)} L \frac{(n_\perp \cdot n_+)}{(u_2 \cdot n_+)} \,,
\end{align}
When we average over the two limiting values, they cancel against each other  due to the identity
\begin{align}
    (n_\perp \cdot n_+)(n_- \cdot u_2) + (n_\perp \cdot n_-)(n_+ \cdot u_2) \propto (n_\perp \cdot u_2) = 0 \,. \nonumber
\end{align}
The proportionality follows from the fact that $n_+^{(\mu} n_-^{\nu)}$ is proportional to the metric on the transverse space of the triple cut, and that the unit vector $n_\perp^\mu$ lives in this transverse space. $(n_\perp \cdot u_2) = 0$ follows from the definition of $n_\perp^\mu$ ($n_4^\mu$ in Eq.\eqref{eq:box_integrand_param}).

Based on the analyses, we conclude that the triangle coefficient $\tilde{c}_{012}^{(0)}$ can be obtained by studying the Laurent expansion of the triple cut integrand. Given a triple cut loop momentum parametrization  of the form Eq.\eqref{eq:tri_loopmom_param}, the triangle coefficient $\tilde{c}_{012}^{(0)}$ can be extracted from the triple cut integrand $c (T)$ as the difference between the residues of $\frac{c(T)}{T}$ at $T = 0$ and at $T = \infty$.
\begin{align}
    \tilde{c}_{012}^{(0)} &= \frac{1}{2} \left[ \text{Res}_{T=0} \frac{c(T)}{T} - \text{Res}_{T = \infty} \frac{c(T)}{T} \right] \,.
\end{align}

\bibliography{bibl}

\begin{thebibliography}{64}%
\makeatletter
\providecommand \@ifxundefined [1]{%
 \@ifx{#1\undefined}
}%
\providecommand \@ifnum [1]{%
 \ifnum #1\expandafter \@firstoftwo
 \else \expandafter \@secondoftwo
 \fi
}%
\providecommand \@ifx [1]{%
 \ifx #1\expandafter \@firstoftwo
 \else \expandafter \@secondoftwo
 \fi
}%
\providecommand \natexlab [1]{#1}%
\providecommand \enquote  [1]{``#1''}%
\providecommand \bibnamefont  [1]{#1}%
\providecommand \bibfnamefont [1]{#1}%
\providecommand \citenamefont [1]{#1}%
\providecommand \href@noop [0]{\@secondoftwo}%
\providecommand \href [0]{\begingroup \@sanitize@url \@href}%
\providecommand \@href[1]{\@@startlink{#1}\@@href}%
\providecommand \@@href[1]{\endgroup#1\@@endlink}%
\providecommand \@sanitize@url [0]{\catcode `\\12\catcode `\$12\catcode
  `\&12\catcode `\#12\catcode `\^12\catcode `\_12\catcode `\%12\relax}%
\providecommand \@@startlink[1]{}%
\providecommand \@@endlink[0]{}%
\providecommand \url  [0]{\begingroup\@sanitize@url \@url }%
\providecommand \@url [1]{\endgroup\@href {#1}{\urlprefix }}%
\providecommand \urlprefix  [0]{URL }%
\providecommand \Eprint [0]{\href }%
\providecommand \doibase [0]{http://dx.doi.org/}%
\providecommand \selectlanguage [0]{\@gobble}%
\providecommand \bibinfo  [0]{\@secondoftwo}%
\providecommand \bibfield  [0]{\@secondoftwo}%
\providecommand \translation [1]{[#1]}%
\providecommand \BibitemOpen [0]{}%
\providecommand \bibitemStop [0]{}%
\providecommand \bibitemNoStop [0]{.\EOS\space}%
\providecommand \EOS [0]{\spacefactor3000\relax}%
\providecommand \BibitemShut  [1]{\csname bibitem#1\endcsname}%
\let\auto@bib@innerbib\@empty
\bibitem [{\citenamefont {Abbott}\ \emph {et~al.}(2016)\citenamefont {Abbott}
  \emph {et~al.}}]{LIGOScientific:2016aoc}%
  \BibitemOpen
  \bibfield  {author} {\bibinfo {author} {\bibfnamefont {B.~P.}\ \bibnamefont
  {Abbott}} \emph {et~al.} (\bibinfo {collaboration} {LIGO Scientific,
  Virgo}),\ }\href {\doibase 10.1103/PhysRevLett.116.061102} {\bibfield
  {journal} {\bibinfo  {journal} {Phys. Rev. Lett.}\ }\textbf {\bibinfo
  {volume} {116}},\ \bibinfo {pages} {061102} (\bibinfo {year} {2016})},\
  \Eprint {http://arxiv.org/abs/1602.03837} {arXiv:1602.03837 [gr-qc]}
  \BibitemShut {NoStop}%
\bibitem [{\citenamefont {Goldberger}\ and\ \citenamefont
  {Rothstein}(2006{\natexlab{a}})}]{Goldberger:2004jt}%
  \BibitemOpen
  \bibfield  {author} {\bibinfo {author} {\bibfnamefont {W.~D.}\ \bibnamefont
  {Goldberger}}\ and\ \bibinfo {author} {\bibfnamefont {I.~Z.}\ \bibnamefont
  {Rothstein}},\ }\href {\doibase 10.1103/PhysRevD.73.104029} {\bibfield
  {journal} {\bibinfo  {journal} {Phys. Rev. D}\ }\textbf {\bibinfo {volume}
  {73}},\ \bibinfo {pages} {104029} (\bibinfo {year} {2006}{\natexlab{a}})},\
  \Eprint {http://arxiv.org/abs/hep-th/0409156} {arXiv:hep-th/0409156}
  \BibitemShut {NoStop}%
\bibitem [{\citenamefont {Porto}(2016)}]{Porto:2016pyg}%
  \BibitemOpen
  \bibfield  {author} {\bibinfo {author} {\bibfnamefont {R.~A.}\ \bibnamefont
  {Porto}},\ }\href {\doibase 10.1016/j.physrep.2016.04.003} {\bibfield
  {journal} {\bibinfo  {journal} {Phys. Rept.}\ }\textbf {\bibinfo {volume}
  {633}},\ \bibinfo {pages} {1} (\bibinfo {year} {2016})},\ \Eprint
  {http://arxiv.org/abs/1601.04914} {arXiv:1601.04914 [hep-th]} \BibitemShut
  {NoStop}%
\bibitem [{\citenamefont {Porto}(2008)}]{Porto:2007qi}%
  \BibitemOpen
  \bibfield  {author} {\bibinfo {author} {\bibfnamefont {R.~A.}\ \bibnamefont
  {Porto}},\ }\href {\doibase 10.1103/PhysRevD.77.064026} {\bibfield  {journal}
  {\bibinfo  {journal} {Phys. Rev. D}\ }\textbf {\bibinfo {volume} {77}},\
  \bibinfo {pages} {064026} (\bibinfo {year} {2008})},\ \Eprint
  {http://arxiv.org/abs/0710.5150} {arXiv:0710.5150 [hep-th]} \BibitemShut
  {NoStop}%
\bibitem [{\citenamefont {Goldberger}\ \emph {et~al.}(2021)\citenamefont
  {Goldberger}, \citenamefont {Li},\ and\ \citenamefont
  {Rothstein}}]{Goldberger:2020fot}%
  \BibitemOpen
  \bibfield  {author} {\bibinfo {author} {\bibfnamefont {W.~D.}\ \bibnamefont
  {Goldberger}}, \bibinfo {author} {\bibfnamefont {J.}~\bibnamefont {Li}}, \
  and\ \bibinfo {author} {\bibfnamefont {I.~Z.}\ \bibnamefont {Rothstein}},\
  }\href {\doibase 10.1007/JHEP06(2021)053} {\bibfield  {journal} {\bibinfo
  {journal} {JHEP}\ }\textbf {\bibinfo {volume} {06}},\ \bibinfo {pages} {053}
  (\bibinfo {year} {2021})},\ \Eprint {http://arxiv.org/abs/2012.14869}
  {arXiv:2012.14869 [hep-th]} \BibitemShut {NoStop}%
\bibitem [{\citenamefont {Goldberger}\ and\ \citenamefont
  {Rothstein}(2006{\natexlab{b}})}]{Goldberger:2005cd}%
  \BibitemOpen
  \bibfield  {author} {\bibinfo {author} {\bibfnamefont {W.~D.}\ \bibnamefont
  {Goldberger}}\ and\ \bibinfo {author} {\bibfnamefont {I.~Z.}\ \bibnamefont
  {Rothstein}},\ }\href {\doibase 10.1103/PhysRevD.73.104030} {\bibfield
  {journal} {\bibinfo  {journal} {Phys. Rev. D}\ }\textbf {\bibinfo {volume}
  {73}},\ \bibinfo {pages} {104030} (\bibinfo {year} {2006}{\natexlab{b}})},\
  \Eprint {http://arxiv.org/abs/hep-th/0511133} {arXiv:hep-th/0511133}
  \BibitemShut {NoStop}%
\bibitem [{\citenamefont {Goldberger}\ \emph {et~al.}(2014)\citenamefont
  {Goldberger}, \citenamefont {Ross},\ and\ \citenamefont
  {Rothstein}}]{Goldberger:2012kf}%
  \BibitemOpen
  \bibfield  {author} {\bibinfo {author} {\bibfnamefont {W.~D.}\ \bibnamefont
  {Goldberger}}, \bibinfo {author} {\bibfnamefont {A.}~\bibnamefont {Ross}}, \
  and\ \bibinfo {author} {\bibfnamefont {I.~Z.}\ \bibnamefont {Rothstein}},\
  }\href {\doibase 10.1103/PhysRevD.89.124033} {\bibfield  {journal} {\bibinfo
  {journal} {Phys. Rev. D}\ }\textbf {\bibinfo {volume} {89}},\ \bibinfo
  {pages} {124033} (\bibinfo {year} {2014})},\ \Eprint
  {http://arxiv.org/abs/1211.6095} {arXiv:1211.6095 [hep-th]} \BibitemShut
  {NoStop}%
\bibitem [{\citenamefont {Goldberger}\ and\ \citenamefont
  {Rothstein}(2020{\natexlab{a}})}]{Goldberger:2020wbx}%
  \BibitemOpen
  \bibfield  {author} {\bibinfo {author} {\bibfnamefont {W.~D.}\ \bibnamefont
  {Goldberger}}\ and\ \bibinfo {author} {\bibfnamefont {I.~Z.}\ \bibnamefont
  {Rothstein}},\ }\href {\doibase 10.1007/JHEP10(2020)026} {\bibfield
  {journal} {\bibinfo  {journal} {JHEP}\ }\textbf {\bibinfo {volume} {10}},\
  \bibinfo {pages} {026} (\bibinfo {year} {2020}{\natexlab{a}})},\ \Eprint
  {http://arxiv.org/abs/2007.00731} {arXiv:2007.00731 [hep-th]} \BibitemShut
  {NoStop}%
\bibitem [{\citenamefont {Saketh}\ \emph {et~al.}(2023)\citenamefont {Saketh},
  \citenamefont {Steinhoff}, \citenamefont {Vines},\ and\ \citenamefont
  {Buonanno}}]{Saketh:2022xjb}%
  \BibitemOpen
  \bibfield  {author} {\bibinfo {author} {\bibfnamefont {M.~V.~S.}\
  \bibnamefont {Saketh}}, \bibinfo {author} {\bibfnamefont {J.}~\bibnamefont
  {Steinhoff}}, \bibinfo {author} {\bibfnamefont {J.}~\bibnamefont {Vines}}, \
  and\ \bibinfo {author} {\bibfnamefont {A.}~\bibnamefont {Buonanno}},\ }\href
  {\doibase 10.1103/PhysRevD.107.084006} {\bibfield  {journal} {\bibinfo
  {journal} {Phys. Rev. D}\ }\textbf {\bibinfo {volume} {107}},\ \bibinfo
  {pages} {084006} (\bibinfo {year} {2023})},\ \Eprint
  {http://arxiv.org/abs/2212.13095} {arXiv:2212.13095 [gr-qc]} \BibitemShut
  {NoStop}%
\bibitem [{\citenamefont {Saketh}\ \emph
  {et~al.}(2024{\natexlab{a}})\citenamefont {Saketh}, \citenamefont {Zhou},
  \citenamefont {Ghosh}, \citenamefont {Steinhoff},\ and\ \citenamefont
  {Chatterjee}}]{Saketh:2024juq}%
  \BibitemOpen
  \bibfield  {author} {\bibinfo {author} {\bibfnamefont {M.~V.~S.}\
  \bibnamefont {Saketh}}, \bibinfo {author} {\bibfnamefont {Z.}~\bibnamefont
  {Zhou}}, \bibinfo {author} {\bibfnamefont {S.}~\bibnamefont {Ghosh}},
  \bibinfo {author} {\bibfnamefont {J.}~\bibnamefont {Steinhoff}}, \ and\
  \bibinfo {author} {\bibfnamefont {D.}~\bibnamefont {Chatterjee}},\
  }\href@noop {} {\  (\bibinfo {year} {2024}{\natexlab{a}})},\ \Eprint
  {http://arxiv.org/abs/2407.08327} {arXiv:2407.08327 [gr-qc]} \BibitemShut
  {NoStop}%
\bibitem [{\citenamefont {Ivanov}\ \emph {et~al.}(2024)\citenamefont {Ivanov},
  \citenamefont {Li}, \citenamefont {Parra-Martinez},\ and\ \citenamefont
  {Zhou}}]{Ivanov:2024sds}%
  \BibitemOpen
  \bibfield  {author} {\bibinfo {author} {\bibfnamefont {M.~M.}\ \bibnamefont
  {Ivanov}}, \bibinfo {author} {\bibfnamefont {Y.-Z.}\ \bibnamefont {Li}},
  \bibinfo {author} {\bibfnamefont {J.}~\bibnamefont {Parra-Martinez}}, \ and\
  \bibinfo {author} {\bibfnamefont {Z.}~\bibnamefont {Zhou}},\ }\href {\doibase
  10.1103/PhysRevLett.132.131401} {\bibfield  {journal} {\bibinfo  {journal}
  {Phys. Rev. Lett.}\ }\textbf {\bibinfo {volume} {132}},\ \bibinfo {pages}
  {131401} (\bibinfo {year} {2024})},\ \Eprint
  {http://arxiv.org/abs/2401.08752} {arXiv:2401.08752 [hep-th]} \BibitemShut
  {NoStop}%
\bibitem [{\citenamefont {Kim}\ and\ \citenamefont {Shim}(2021)}]{Kim:2020dif}%
  \BibitemOpen
  \bibfield  {author} {\bibinfo {author} {\bibfnamefont {J.-W.}\ \bibnamefont
  {Kim}}\ and\ \bibinfo {author} {\bibfnamefont {M.}~\bibnamefont {Shim}},\
  }\href {\doibase 10.1103/PhysRevD.104.046022} {\bibfield  {journal} {\bibinfo
   {journal} {Phys. Rev. D}\ }\textbf {\bibinfo {volume} {104}},\ \bibinfo
  {pages} {046022} (\bibinfo {year} {2021})},\ \Eprint
  {http://arxiv.org/abs/2011.03337} {arXiv:2011.03337 [hep-th]} \BibitemShut
  {NoStop}%
\bibitem [{\citenamefont {Aoude}\ and\ \citenamefont
  {Ochirov}(2023)}]{Aoude:2023fdm}%
  \BibitemOpen
  \bibfield  {author} {\bibinfo {author} {\bibfnamefont {R.}~\bibnamefont
  {Aoude}}\ and\ \bibinfo {author} {\bibfnamefont {A.}~\bibnamefont
  {Ochirov}},\ }\href {\doibase 10.1007/JHEP12(2023)103} {\bibfield  {journal}
  {\bibinfo  {journal} {JHEP}\ }\textbf {\bibinfo {volume} {12}},\ \bibinfo
  {pages} {103} (\bibinfo {year} {2023})},\ \Eprint
  {http://arxiv.org/abs/2307.07504} {arXiv:2307.07504 [hep-th]} \BibitemShut
  {NoStop}%
\bibitem [{\citenamefont {Jones}\ and\ \citenamefont
  {Ruf}(2024)}]{Jones:2023ugm}%
  \BibitemOpen
  \bibfield  {author} {\bibinfo {author} {\bibfnamefont {C.~R.~T.}\
  \bibnamefont {Jones}}\ and\ \bibinfo {author} {\bibfnamefont {M.~S.}\
  \bibnamefont {Ruf}},\ }\href {\doibase 10.1007/JHEP03(2024)015} {\bibfield
  {journal} {\bibinfo  {journal} {JHEP}\ }\textbf {\bibinfo {volume} {03}},\
  \bibinfo {pages} {015} (\bibinfo {year} {2024})},\ \Eprint
  {http://arxiv.org/abs/2310.00069} {arXiv:2310.00069 [hep-th]} \BibitemShut
  {NoStop}%
\bibitem [{\citenamefont {Poisson}(2004)}]{Poisson:2004cw}%
  \BibitemOpen
  \bibfield  {author} {\bibinfo {author} {\bibfnamefont {E.}~\bibnamefont
  {Poisson}},\ }\href {\doibase 10.1103/PhysRevD.70.084044} {\bibfield
  {journal} {\bibinfo  {journal} {Phys. Rev. D}\ }\textbf {\bibinfo {volume}
  {70}},\ \bibinfo {pages} {084044} (\bibinfo {year} {2004})},\ \Eprint
  {http://arxiv.org/abs/gr-qc/0407050} {arXiv:gr-qc/0407050} \BibitemShut
  {NoStop}%
\bibitem [{\citenamefont {Maselli}\ \emph {et~al.}(2018)\citenamefont
  {Maselli}, \citenamefont {Pani}, \citenamefont {Cardoso}, \citenamefont
  {Abdelsalhin}, \citenamefont {Gualtieri},\ and\ \citenamefont
  {Ferrari}}]{Maselli:2017cmm}%
  \BibitemOpen
  \bibfield  {author} {\bibinfo {author} {\bibfnamefont {A.}~\bibnamefont
  {Maselli}}, \bibinfo {author} {\bibfnamefont {P.}~\bibnamefont {Pani}},
  \bibinfo {author} {\bibfnamefont {V.}~\bibnamefont {Cardoso}}, \bibinfo
  {author} {\bibfnamefont {T.}~\bibnamefont {Abdelsalhin}}, \bibinfo {author}
  {\bibfnamefont {L.}~\bibnamefont {Gualtieri}}, \ and\ \bibinfo {author}
  {\bibfnamefont {V.}~\bibnamefont {Ferrari}},\ }\href {\doibase
  10.1103/PhysRevLett.120.081101} {\bibfield  {journal} {\bibinfo  {journal}
  {Phys. Rev. Lett.}\ }\textbf {\bibinfo {volume} {120}},\ \bibinfo {pages}
  {081101} (\bibinfo {year} {2018})},\ \Eprint
  {http://arxiv.org/abs/1703.10612} {arXiv:1703.10612 [gr-qc]} \BibitemShut
  {NoStop}%
\bibitem [{\citenamefont {Datta}\ \emph {et~al.}(2020)\citenamefont {Datta},
  \citenamefont {Brito}, \citenamefont {Bose}, \citenamefont {Pani},\ and\
  \citenamefont {Hughes}}]{Datta:2019epe}%
  \BibitemOpen
  \bibfield  {author} {\bibinfo {author} {\bibfnamefont {S.}~\bibnamefont
  {Datta}}, \bibinfo {author} {\bibfnamefont {R.}~\bibnamefont {Brito}},
  \bibinfo {author} {\bibfnamefont {S.}~\bibnamefont {Bose}}, \bibinfo {author}
  {\bibfnamefont {P.}~\bibnamefont {Pani}}, \ and\ \bibinfo {author}
  {\bibfnamefont {S.~A.}\ \bibnamefont {Hughes}},\ }\href {\doibase
  10.1103/PhysRevD.101.044004} {\bibfield  {journal} {\bibinfo  {journal}
  {Phys. Rev. D}\ }\textbf {\bibinfo {volume} {101}},\ \bibinfo {pages}
  {044004} (\bibinfo {year} {2020})},\ \Eprint
  {http://arxiv.org/abs/1910.07841} {arXiv:1910.07841 [gr-qc]} \BibitemShut
  {NoStop}%
\bibitem [{\citenamefont {Brown}\ \emph {et~al.}(2007)\citenamefont {Brown},
  \citenamefont {Fairhurst}, \citenamefont {Krishnan}, \citenamefont {Mercer},
  \citenamefont {Kopparapu}, \citenamefont {Santamaria},\ and\ \citenamefont
  {Whelan}}]{Brown:2007jx}%
  \BibitemOpen
  \bibfield  {author} {\bibinfo {author} {\bibfnamefont {D.}~\bibnamefont
  {Brown}}, \bibinfo {author} {\bibfnamefont {S.}~\bibnamefont {Fairhurst}},
  \bibinfo {author} {\bibfnamefont {B.}~\bibnamefont {Krishnan}}, \bibinfo
  {author} {\bibfnamefont {R.~A.}\ \bibnamefont {Mercer}}, \bibinfo {author}
  {\bibfnamefont {R.~K.}\ \bibnamefont {Kopparapu}}, \bibinfo {author}
  {\bibfnamefont {L.}~\bibnamefont {Santamaria}}, \ and\ \bibinfo {author}
  {\bibfnamefont {J.~T.}\ \bibnamefont {Whelan}},\ }\href@noop {} {\  (\bibinfo
  {year} {2007})},\ \Eprint {http://arxiv.org/abs/0709.0093} {arXiv:0709.0093
  [gr-qc]} \BibitemShut {NoStop}%
\bibitem [{\citenamefont {Tagoshi}\ \emph {et~al.}(1997)\citenamefont
  {Tagoshi}, \citenamefont {Mano},\ and\ \citenamefont
  {Takasugi}}]{Tagoshi:1997jy}%
  \BibitemOpen
  \bibfield  {author} {\bibinfo {author} {\bibfnamefont {H.}~\bibnamefont
  {Tagoshi}}, \bibinfo {author} {\bibfnamefont {S.}~\bibnamefont {Mano}}, \
  and\ \bibinfo {author} {\bibfnamefont {E.}~\bibnamefont {Takasugi}},\ }\href
  {\doibase 10.1143/PTP.98.829} {\bibfield  {journal} {\bibinfo  {journal}
  {Prog. Theor. Phys.}\ }\textbf {\bibinfo {volume} {98}},\ \bibinfo {pages}
  {829} (\bibinfo {year} {1997})},\ \Eprint
  {http://arxiv.org/abs/gr-qc/9711072} {arXiv:gr-qc/9711072} \BibitemShut
  {NoStop}%
\bibitem [{\citenamefont {Endlich}\ and\ \citenamefont
  {Penco}(2016)}]{Endlich:2015mke}%
  \BibitemOpen
  \bibfield  {author} {\bibinfo {author} {\bibfnamefont {S.}~\bibnamefont
  {Endlich}}\ and\ \bibinfo {author} {\bibfnamefont {R.}~\bibnamefont
  {Penco}},\ }\href {\doibase 10.1103/PhysRevD.93.064021} {\bibfield  {journal}
  {\bibinfo  {journal} {Phys. Rev. D}\ }\textbf {\bibinfo {volume} {93}},\
  \bibinfo {pages} {064021} (\bibinfo {year} {2016})},\ \Eprint
  {http://arxiv.org/abs/1510.08889} {arXiv:1510.08889 [gr-qc]} \BibitemShut
  {NoStop}%
\bibitem [{\citenamefont {Chatziioannou}\ \emph {et~al.}(2016)\citenamefont
  {Chatziioannou}, \citenamefont {Poisson},\ and\ \citenamefont
  {Yunes}}]{Chatziioannou:2016kem}%
  \BibitemOpen
  \bibfield  {author} {\bibinfo {author} {\bibfnamefont {K.}~\bibnamefont
  {Chatziioannou}}, \bibinfo {author} {\bibfnamefont {E.}~\bibnamefont
  {Poisson}}, \ and\ \bibinfo {author} {\bibfnamefont {N.}~\bibnamefont
  {Yunes}},\ }\href {\doibase 10.1103/PhysRevD.94.084043} {\bibfield  {journal}
  {\bibinfo  {journal} {Phys. Rev. D}\ }\textbf {\bibinfo {volume} {94}},\
  \bibinfo {pages} {084043} (\bibinfo {year} {2016})},\ \Eprint
  {http://arxiv.org/abs/1608.02899} {arXiv:1608.02899 [gr-qc]} \BibitemShut
  {NoStop}%
\bibitem [{\citenamefont {Taracchini}\ \emph {et~al.}(2013)\citenamefont
  {Taracchini}, \citenamefont {Buonanno}, \citenamefont {Hughes},\ and\
  \citenamefont {Khanna}}]{Taracchini:2013wfa}%
  \BibitemOpen
  \bibfield  {author} {\bibinfo {author} {\bibfnamefont {A.}~\bibnamefont
  {Taracchini}}, \bibinfo {author} {\bibfnamefont {A.}~\bibnamefont
  {Buonanno}}, \bibinfo {author} {\bibfnamefont {S.~A.}\ \bibnamefont
  {Hughes}}, \ and\ \bibinfo {author} {\bibfnamefont {G.}~\bibnamefont
  {Khanna}},\ }\href {\doibase 10.1103/PhysRevD.88.044001} {\bibfield
  {journal} {\bibinfo  {journal} {Phys. Rev. D}\ }\textbf {\bibinfo {volume}
  {88}},\ \bibinfo {pages} {044001} (\bibinfo {year} {2013})},\ \bibinfo {note}
  {[Erratum: Phys.Rev.D 88, 109903 (2013)]},\ \Eprint
  {http://arxiv.org/abs/1305.2184} {arXiv:1305.2184 [gr-qc]} \BibitemShut
  {NoStop}%
\bibitem [{\citenamefont {Chia}(2021)}]{Chia:2020yla}%
  \BibitemOpen
  \bibfield  {author} {\bibinfo {author} {\bibfnamefont {H.~S.}\ \bibnamefont
  {Chia}},\ }\href {\doibase 10.1103/PhysRevD.104.024013} {\bibfield  {journal}
  {\bibinfo  {journal} {Phys. Rev. D}\ }\textbf {\bibinfo {volume} {104}},\
  \bibinfo {pages} {024013} (\bibinfo {year} {2021})},\ \Eprint
  {http://arxiv.org/abs/2010.07300} {arXiv:2010.07300 [gr-qc]} \BibitemShut
  {NoStop}%
\bibitem [{\citenamefont {de~Cesare}\ and\ \citenamefont
  {Oliveri}(2023)}]{deCesare:2023rmg}%
  \BibitemOpen
  \bibfield  {author} {\bibinfo {author} {\bibfnamefont {M.}~\bibnamefont
  {de~Cesare}}\ and\ \bibinfo {author} {\bibfnamefont {R.}~\bibnamefont
  {Oliveri}},\ }\href {\doibase 10.1103/PhysRevD.108.044050} {\bibfield
  {journal} {\bibinfo  {journal} {Phys. Rev. D}\ }\textbf {\bibinfo {volume}
  {108}},\ \bibinfo {pages} {044050} (\bibinfo {year} {2023})},\ \Eprint
  {http://arxiv.org/abs/2305.04970} {arXiv:2305.04970 [gr-qc]} \BibitemShut
  {NoStop}%
\bibitem [{\citenamefont {Chen}\ \emph {et~al.}(2023)\citenamefont {Chen},
  \citenamefont {Hsieh}, \citenamefont {Huang},\ and\ \citenamefont
  {Kim}}]{Chen:2023qzo}%
  \BibitemOpen
  \bibfield  {author} {\bibinfo {author} {\bibfnamefont {Y.-J.}\ \bibnamefont
  {Chen}}, \bibinfo {author} {\bibfnamefont {T.}~\bibnamefont {Hsieh}},
  \bibinfo {author} {\bibfnamefont {Y.-T.}\ \bibnamefont {Huang}}, \ and\
  \bibinfo {author} {\bibfnamefont {J.-W.}\ \bibnamefont {Kim}},\ }\href@noop
  {} {\  (\bibinfo {year} {2023})},\ \Eprint {http://arxiv.org/abs/2312.04513}
  {arXiv:2312.04513 [hep-th]} \BibitemShut {NoStop}%
\bibitem [{\citenamefont {Vidal}\ \emph {et~al.}(2024)\citenamefont {Vidal},
  \citenamefont {Dantas}, \citenamefont {Sturani},\ and\ \citenamefont
  {Menezes}}]{Vidal:2024inh}%
  \BibitemOpen
  \bibfield  {author} {\bibinfo {author} {\bibfnamefont {G.}~\bibnamefont
  {Vidal}}, \bibinfo {author} {\bibfnamefont {G.~M.}\ \bibnamefont {Dantas}},
  \bibinfo {author} {\bibfnamefont {R.}~\bibnamefont {Sturani}}, \ and\
  \bibinfo {author} {\bibfnamefont {G.}~\bibnamefont {Menezes}},\ }\href@noop
  {} {\  (\bibinfo {year} {2024})},\ \Eprint {http://arxiv.org/abs/2410.23384}
  {arXiv:2410.23384 [gr-qc]} \BibitemShut {NoStop}%
\bibitem [{\citenamefont {Kosower}\ \emph {et~al.}(2019)\citenamefont
  {Kosower}, \citenamefont {Maybee},\ and\ \citenamefont
  {O'Connell}}]{Kosower:2018adc}%
  \BibitemOpen
  \bibfield  {author} {\bibinfo {author} {\bibfnamefont {D.~A.}\ \bibnamefont
  {Kosower}}, \bibinfo {author} {\bibfnamefont {B.}~\bibnamefont {Maybee}}, \
  and\ \bibinfo {author} {\bibfnamefont {D.}~\bibnamefont {O'Connell}},\ }\href
  {\doibase 10.1007/JHEP02(2019)137} {\bibfield  {journal} {\bibinfo  {journal}
  {JHEP}\ }\textbf {\bibinfo {volume} {02}},\ \bibinfo {pages} {137} (\bibinfo
  {year} {2019})},\ \Eprint {http://arxiv.org/abs/1811.10950} {arXiv:1811.10950
  [hep-th]} \BibitemShut {NoStop}%
\bibitem [{\citenamefont {Bautista}\ \emph
  {et~al.}(2023{\natexlab{a}})\citenamefont {Bautista}, \citenamefont
  {Guevara}, \citenamefont {Kavanagh},\ and\ \citenamefont
  {Vines}}]{Bautista:2022wjf}%
  \BibitemOpen
  \bibfield  {author} {\bibinfo {author} {\bibfnamefont {Y.~F.}\ \bibnamefont
  {Bautista}}, \bibinfo {author} {\bibfnamefont {A.}~\bibnamefont {Guevara}},
  \bibinfo {author} {\bibfnamefont {C.}~\bibnamefont {Kavanagh}}, \ and\
  \bibinfo {author} {\bibfnamefont {J.}~\bibnamefont {Vines}},\ }\href
  {\doibase 10.1007/JHEP05(2023)211} {\bibfield  {journal} {\bibinfo  {journal}
  {JHEP}\ }\textbf {\bibinfo {volume} {05}},\ \bibinfo {pages} {211} (\bibinfo
  {year} {2023}{\natexlab{a}})},\ \Eprint {http://arxiv.org/abs/2212.07965}
  {arXiv:2212.07965 [hep-th]} \BibitemShut {NoStop}%
\bibitem [{\citenamefont {Bautista}\ \emph
  {et~al.}(2024{\natexlab{a}})\citenamefont {Bautista}, \citenamefont {Khalil},
  \citenamefont {Sergola}, \citenamefont {Kavanagh},\ and\ \citenamefont
  {Vines}}]{Bautista:2024agp}%
  \BibitemOpen
  \bibfield  {author} {\bibinfo {author} {\bibfnamefont {Y.~F.}\ \bibnamefont
  {Bautista}}, \bibinfo {author} {\bibfnamefont {M.}~\bibnamefont {Khalil}},
  \bibinfo {author} {\bibfnamefont {M.}~\bibnamefont {Sergola}}, \bibinfo
  {author} {\bibfnamefont {C.}~\bibnamefont {Kavanagh}}, \ and\ \bibinfo
  {author} {\bibfnamefont {J.}~\bibnamefont {Vines}},\ }\href@noop {} {\
  (\bibinfo {year} {2024}{\natexlab{a}})},\ \Eprint
  {http://arxiv.org/abs/2408.01871} {arXiv:2408.01871 [gr-qc]} \BibitemShut
  {NoStop}%
\bibitem [{\citenamefont {Aoude}\ \emph {et~al.}(2024)\citenamefont {Aoude},
  \citenamefont {Cristofoli}, \citenamefont {Elkhidir},\ and\ \citenamefont
  {Sergola}}]{aoude2024inelasticcoupledchanneleikonalscattering}%
  \BibitemOpen
  \bibfield  {author} {\bibinfo {author} {\bibfnamefont {R.}~\bibnamefont
  {Aoude}}, \bibinfo {author} {\bibfnamefont {A.}~\bibnamefont {Cristofoli}},
  \bibinfo {author} {\bibfnamefont {A.}~\bibnamefont {Elkhidir}}, \ and\
  \bibinfo {author} {\bibfnamefont {M.}~\bibnamefont {Sergola}},\ }\href
  {https://arxiv.org/abs/2411.02294} {\enquote {\bibinfo {title} {Inelastic
  coupled-channel eikonal scattering},}\ } (\bibinfo {year} {2024}),\ \Eprint
  {http://arxiv.org/abs/2411.02294} {arXiv:2411.02294 [hep-th]} \BibitemShut
  {NoStop}%
\bibitem [{\citenamefont {Chandrasekhar}(1985)}]{Chandrasekhar:1985kt}%
  \BibitemOpen
  \bibfield  {author} {\bibinfo {author} {\bibfnamefont {S.}~\bibnamefont
  {Chandrasekhar}},\ }\href@noop {} {\emph {\bibinfo {title} {{The mathematical
  theory of black holes}}}}\ (\bibinfo {year} {1985})\BibitemShut {NoStop}%
\bibitem [{\citenamefont {Teukolsky}(1972)}]{PhysRevLett.29.1114}%
  \BibitemOpen
  \bibfield  {author} {\bibinfo {author} {\bibfnamefont {S.~A.}\ \bibnamefont
  {Teukolsky}},\ }\href {\doibase 10.1103/PhysRevLett.29.1114} {\bibfield
  {journal} {\bibinfo  {journal} {Phys. Rev. Lett.}\ }\textbf {\bibinfo
  {volume} {29}},\ \bibinfo {pages} {1114} (\bibinfo {year}
  {1972})}\BibitemShut {NoStop}%
\bibitem [{\citenamefont {Batic}\ and\ \citenamefont
  {Schmid}(2007)}]{Batic_2007}%
  \BibitemOpen
  \bibfield  {author} {\bibinfo {author} {\bibfnamefont {D.}~\bibnamefont
  {Batic}}\ and\ \bibinfo {author} {\bibfnamefont {H.}~\bibnamefont {Schmid}},\
  }\href {\doibase 10.1063/1.2720277} {\bibfield  {journal} {\bibinfo
  {journal} {Journal of Mathematical Physics}\ }\textbf {\bibinfo {volume}
  {48}} (\bibinfo {year} {2007}),\ 10.1063/1.2720277}\BibitemShut {NoStop}%
\bibitem [{\citenamefont {Bonelli}\ \emph {et~al.}(2023)\citenamefont
  {Bonelli}, \citenamefont {Iossa}, \citenamefont {Panea~Lichtig},\ and\
  \citenamefont {Tanzini}}]{Bonelli:2022ten}%
  \BibitemOpen
  \bibfield  {author} {\bibinfo {author} {\bibfnamefont {G.}~\bibnamefont
  {Bonelli}}, \bibinfo {author} {\bibfnamefont {C.}~\bibnamefont {Iossa}},
  \bibinfo {author} {\bibfnamefont {D.}~\bibnamefont {Panea~Lichtig}}, \ and\
  \bibinfo {author} {\bibfnamefont {A.}~\bibnamefont {Tanzini}},\ }\href
  {\doibase 10.1007/s00220-022-04497-5} {\bibfield  {journal} {\bibinfo
  {journal} {Commun. Math. Phys.}\ }\textbf {\bibinfo {volume} {397}},\
  \bibinfo {pages} {635} (\bibinfo {year} {2023})},\ \Eprint
  {http://arxiv.org/abs/2201.04491} {arXiv:2201.04491 [hep-th]} \BibitemShut
  {NoStop}%
\bibitem [{\citenamefont {Aminov}\ \emph {et~al.}(2022)\citenamefont {Aminov},
  \citenamefont {Grassi},\ and\ \citenamefont {Hatsuda}}]{Aminov:2020yma}%
  \BibitemOpen
  \bibfield  {author} {\bibinfo {author} {\bibfnamefont {G.}~\bibnamefont
  {Aminov}}, \bibinfo {author} {\bibfnamefont {A.}~\bibnamefont {Grassi}}, \
  and\ \bibinfo {author} {\bibfnamefont {Y.}~\bibnamefont {Hatsuda}},\ }\href
  {\doibase 10.1007/s00023-021-01137-x} {\bibfield  {journal} {\bibinfo
  {journal} {Annales Henri Poincare}\ }\textbf {\bibinfo {volume} {23}},\
  \bibinfo {pages} {1951} (\bibinfo {year} {2022})},\ \Eprint
  {http://arxiv.org/abs/2006.06111} {arXiv:2006.06111 [hep-th]} \BibitemShut
  {NoStop}%
\bibitem [{\citenamefont {Futterman}\ \emph {et~al.}(1988)\citenamefont
  {Futterman}, \citenamefont {Handler},\ and\ \citenamefont
  {Matzner}}]{futterman88}%
  \BibitemOpen
  \bibfield  {author} {\bibinfo {author} {\bibfnamefont {J.~A.~H.}\
  \bibnamefont {Futterman}}, \bibinfo {author} {\bibfnamefont {F.~A.}\
  \bibnamefont {Handler}}, \ and\ \bibinfo {author} {\bibfnamefont {R.~A.}\
  \bibnamefont {Matzner}},\ }\href {\doibase 10.1017/CBO9780511735615} {\emph
  {\bibinfo {title} {Scattering from Black Holes}}},\ Cambridge Monographs on
  Mathematical Physics\ (\bibinfo  {publisher} {Cambridge University Press},\
  \bibinfo {year} {1988})\BibitemShut {NoStop}%
\bibitem [{\citenamefont {Ivanov}\ and\ \citenamefont
  {Zhou}(2023)}]{Ivanov:2022qqt}%
  \BibitemOpen
  \bibfield  {author} {\bibinfo {author} {\bibfnamefont {M.~M.}\ \bibnamefont
  {Ivanov}}\ and\ \bibinfo {author} {\bibfnamefont {Z.}~\bibnamefont {Zhou}},\
  }\href {\doibase 10.1103/PhysRevLett.130.091403} {\bibfield  {journal}
  {\bibinfo  {journal} {Phys. Rev. Lett.}\ }\textbf {\bibinfo {volume} {130}},\
  \bibinfo {pages} {091403} (\bibinfo {year} {2023})},\ \Eprint
  {http://arxiv.org/abs/2209.14324} {arXiv:2209.14324 [hep-th]} \BibitemShut
  {NoStop}%
\bibitem [{\citenamefont {Bautista}\ \emph
  {et~al.}(2024{\natexlab{b}})\citenamefont {Bautista}, \citenamefont
  {Bonelli}, \citenamefont {Iossa}, \citenamefont {Tanzini},\ and\
  \citenamefont {Zhou}}]{Bautista:2023sdf}%
  \BibitemOpen
  \bibfield  {author} {\bibinfo {author} {\bibfnamefont {Y.~F.}\ \bibnamefont
  {Bautista}}, \bibinfo {author} {\bibfnamefont {G.}~\bibnamefont {Bonelli}},
  \bibinfo {author} {\bibfnamefont {C.}~\bibnamefont {Iossa}}, \bibinfo
  {author} {\bibfnamefont {A.}~\bibnamefont {Tanzini}}, \ and\ \bibinfo
  {author} {\bibfnamefont {Z.}~\bibnamefont {Zhou}},\ }\href {\doibase
  10.1103/PhysRevD.109.084071} {\bibfield  {journal} {\bibinfo  {journal}
  {Phys. Rev. D}\ }\textbf {\bibinfo {volume} {109}},\ \bibinfo {pages}
  {084071} (\bibinfo {year} {2024}{\natexlab{b}})},\ \Eprint
  {http://arxiv.org/abs/2312.05965} {arXiv:2312.05965 [hep-th]} \BibitemShut
  {NoStop}%
\bibitem [{\citenamefont {Saketh}\ \emph
  {et~al.}(2024{\natexlab{b}})\citenamefont {Saketh}, \citenamefont {Zhou},\
  and\ \citenamefont {Ivanov}}]{Saketh:2023bul}%
  \BibitemOpen
  \bibfield  {author} {\bibinfo {author} {\bibfnamefont {M.~V.~S.}\
  \bibnamefont {Saketh}}, \bibinfo {author} {\bibfnamefont {Z.}~\bibnamefont
  {Zhou}}, \ and\ \bibinfo {author} {\bibfnamefont {M.~M.}\ \bibnamefont
  {Ivanov}},\ }\href {\doibase 10.1103/PhysRevD.109.064058} {\bibfield
  {journal} {\bibinfo  {journal} {Phys. Rev. D}\ }\textbf {\bibinfo {volume}
  {109}},\ \bibinfo {pages} {064058} (\bibinfo {year} {2024}{\natexlab{b}})},\
  \Eprint {http://arxiv.org/abs/2307.10391} {arXiv:2307.10391 [hep-th]}
  \BibitemShut {NoStop}%
\bibitem [{\citenamefont {Landau}\ and\ \citenamefont
  {Lifshitz}(1981)}]{Landau1981Quantum}%
  \BibitemOpen
  \bibfield  {author} {\bibinfo {author} {\bibfnamefont {L.~D.}\ \bibnamefont
  {Landau}}\ and\ \bibinfo {author} {\bibfnamefont {L.~M.}\ \bibnamefont
  {Lifshitz}},\ }\href {http://www.worldcat.org/isbn/0750635398} {\emph
  {\bibinfo {title} {Quantum Mechanics Non-Relativistic Theory, Third Edition:
  Volume 3}}},\ \bibinfo {edition} {3rd}\ ed.\ (\bibinfo  {publisher}
  {Butterworth-Heinemann},\ \bibinfo {year} {1981})\BibitemShut {NoStop}%
\bibitem [{\citenamefont {Dolan}(2008)}]{Dolan:2008kf}%
  \BibitemOpen
  \bibfield  {author} {\bibinfo {author} {\bibfnamefont {S.~R.}\ \bibnamefont
  {Dolan}},\ }\href {\doibase 10.1088/0264-9381/25/23/235002} {\bibfield
  {journal} {\bibinfo  {journal} {Class. Quant. Grav.}\ }\textbf {\bibinfo
  {volume} {25}},\ \bibinfo {pages} {235002} (\bibinfo {year} {2008})},\
  \Eprint {http://arxiv.org/abs/0801.3805} {arXiv:0801.3805 [gr-qc]}
  \BibitemShut {NoStop}%
\bibitem [{\citenamefont {Bianchi}\ \emph {et~al.}(2022)\citenamefont
  {Bianchi}, \citenamefont {Consoli}, \citenamefont {Grillo},\ and\
  \citenamefont {Morales}}]{Bianchi:2021mft}%
  \BibitemOpen
  \bibfield  {author} {\bibinfo {author} {\bibfnamefont {M.}~\bibnamefont
  {Bianchi}}, \bibinfo {author} {\bibfnamefont {D.}~\bibnamefont {Consoli}},
  \bibinfo {author} {\bibfnamefont {A.}~\bibnamefont {Grillo}}, \ and\ \bibinfo
  {author} {\bibfnamefont {J.~F.}\ \bibnamefont {Morales}},\ }\href {\doibase
  10.1007/JHEP01(2022)024} {\bibfield  {journal} {\bibinfo  {journal} {JHEP}\
  }\textbf {\bibinfo {volume} {01}},\ \bibinfo {pages} {024} (\bibinfo {year}
  {2022})},\ \Eprint {http://arxiv.org/abs/2109.09804} {arXiv:2109.09804
  [hep-th]} \BibitemShut {NoStop}%
\bibitem [{\citenamefont {Starobinskil}\ and\ \citenamefont
  {Churilov}(1974)}]{Starobinskil:1974nkd}%
  \BibitemOpen
  \bibfield  {author} {\bibinfo {author} {\bibfnamefont {A.~A.}\ \bibnamefont
  {Starobinskil}}\ and\ \bibinfo {author} {\bibfnamefont {S.~M.}\ \bibnamefont
  {Churilov}},\ }\href@noop {} {\bibfield  {journal} {\bibinfo  {journal} {Sov.
  Phys. JETP}\ }\textbf {\bibinfo {volume} {65}},\ \bibinfo {pages} {1}
  (\bibinfo {year} {1974})}\BibitemShut {NoStop}%
\bibitem [{\citenamefont {Bautista}\ \emph
  {et~al.}(2023{\natexlab{b}})\citenamefont {Bautista}, \citenamefont
  {Guevara}, \citenamefont {Kavanagh},\ and\ \citenamefont
  {Vines}}]{Bautista:2021wfy}%
  \BibitemOpen
  \bibfield  {author} {\bibinfo {author} {\bibfnamefont {Y.~F.}\ \bibnamefont
  {Bautista}}, \bibinfo {author} {\bibfnamefont {A.}~\bibnamefont {Guevara}},
  \bibinfo {author} {\bibfnamefont {C.}~\bibnamefont {Kavanagh}}, \ and\
  \bibinfo {author} {\bibfnamefont {J.}~\bibnamefont {Vines}},\ }\href
  {\doibase 10.1007/JHEP03(2023)136} {\bibfield  {journal} {\bibinfo  {journal}
  {JHEP}\ }\textbf {\bibinfo {volume} {03}},\ \bibinfo {pages} {136} (\bibinfo
  {year} {2023}{\natexlab{b}})},\ \Eprint {http://arxiv.org/abs/2107.10179}
  {arXiv:2107.10179 [hep-th]} \BibitemShut {NoStop}%
\bibitem [{\citenamefont {De~Logi}\ and\ \citenamefont
  {Kov\'acs}(1977)}]{PhysRevD.16.237}%
  \BibitemOpen
  \bibfield  {author} {\bibinfo {author} {\bibfnamefont {W.~K.}\ \bibnamefont
  {De~Logi}}\ and\ \bibinfo {author} {\bibfnamefont {S.~J.}\ \bibnamefont
  {Kov\'acs}},\ }\href {\doibase 10.1103/PhysRevD.16.237} {\bibfield  {journal}
  {\bibinfo  {journal} {Phys. Rev. D}\ }\textbf {\bibinfo {volume} {16}},\
  \bibinfo {pages} {237} (\bibinfo {year} {1977})}\BibitemShut {NoStop}%
\bibitem [{\citenamefont {Barbieri}\ and\ \citenamefont
  {Guadagnini}(2005)}]{Barbieri:2005kp}%
  \BibitemOpen
  \bibfield  {author} {\bibinfo {author} {\bibfnamefont {A.}~\bibnamefont
  {Barbieri}}\ and\ \bibinfo {author} {\bibfnamefont {E.}~\bibnamefont
  {Guadagnini}},\ }\href {\doibase 10.1016/j.nuclphysb.2005.04.023} {\bibfield
  {journal} {\bibinfo  {journal} {Nucl. Phys. B}\ }\textbf {\bibinfo {volume}
  {719}},\ \bibinfo {pages} {53} (\bibinfo {year} {2005})},\ \Eprint
  {http://arxiv.org/abs/gr-qc/0504078} {arXiv:gr-qc/0504078} \BibitemShut
  {NoStop}%
\bibitem [{\citenamefont {Guevara}(2019)}]{Guevara:2017csg}%
  \BibitemOpen
  \bibfield  {author} {\bibinfo {author} {\bibfnamefont {A.}~\bibnamefont
  {Guevara}},\ }\href {\doibase 10.1007/JHEP04(2019)033} {\bibfield  {journal}
  {\bibinfo  {journal} {JHEP}\ }\textbf {\bibinfo {volume} {04}},\ \bibinfo
  {pages} {033} (\bibinfo {year} {2019})},\ \Eprint
  {http://arxiv.org/abs/1706.02314} {arXiv:1706.02314 [hep-th]} \BibitemShut
  {NoStop}%
\bibitem [{\citenamefont {Cachazo}\ and\ \citenamefont
  {Guevara}(2020)}]{Cachazo:2017jef}%
  \BibitemOpen
  \bibfield  {author} {\bibinfo {author} {\bibfnamefont {F.}~\bibnamefont
  {Cachazo}}\ and\ \bibinfo {author} {\bibfnamefont {A.}~\bibnamefont
  {Guevara}},\ }\href {\doibase 10.1007/JHEP02(2020)181} {\bibfield  {journal}
  {\bibinfo  {journal} {JHEP}\ }\textbf {\bibinfo {volume} {02}},\ \bibinfo
  {pages} {181} (\bibinfo {year} {2020})},\ \Eprint
  {http://arxiv.org/abs/1705.10262} {arXiv:1705.10262 [hep-th]} \BibitemShut
  {NoStop}%
\bibitem [{\citenamefont {Chung}\ \emph {et~al.}(2019)\citenamefont {Chung},
  \citenamefont {Huang}, \citenamefont {Kim},\ and\ \citenamefont
  {Lee}}]{Chung:2018kqs}%
  \BibitemOpen
  \bibfield  {author} {\bibinfo {author} {\bibfnamefont {M.-Z.}\ \bibnamefont
  {Chung}}, \bibinfo {author} {\bibfnamefont {Y.-T.}\ \bibnamefont {Huang}},
  \bibinfo {author} {\bibfnamefont {J.-W.}\ \bibnamefont {Kim}}, \ and\
  \bibinfo {author} {\bibfnamefont {S.}~\bibnamefont {Lee}},\ }\href {\doibase
  10.1007/JHEP04(2019)156} {\bibfield  {journal} {\bibinfo  {journal} {JHEP}\
  }\textbf {\bibinfo {volume} {04}},\ \bibinfo {pages} {156} (\bibinfo {year}
  {2019})},\ \Eprint {http://arxiv.org/abs/1812.08752} {arXiv:1812.08752
  [hep-th]} \BibitemShut {NoStop}%
\bibitem [{\citenamefont {Bautista}(2023)}]{Bautista:2023szu}%
  \BibitemOpen
  \bibfield  {author} {\bibinfo {author} {\bibfnamefont {Y.~F.}\ \bibnamefont
  {Bautista}},\ }\href {\doibase 10.1103/PhysRevD.108.084036} {\bibfield
  {journal} {\bibinfo  {journal} {Phys. Rev. D}\ }\textbf {\bibinfo {volume}
  {108}},\ \bibinfo {pages} {084036} (\bibinfo {year} {2023})},\ \Eprint
  {http://arxiv.org/abs/2304.04287} {arXiv:2304.04287 [hep-th]} \BibitemShut
  {NoStop}%
\bibitem [{\citenamefont {Bern}\ \emph {et~al.}(2021)\citenamefont {Bern},
  \citenamefont {Luna}, \citenamefont {Roiban}, \citenamefont {Shen},\ and\
  \citenamefont {Zeng}}]{Bern:2020buy}%
  \BibitemOpen
  \bibfield  {author} {\bibinfo {author} {\bibfnamefont {Z.}~\bibnamefont
  {Bern}}, \bibinfo {author} {\bibfnamefont {A.}~\bibnamefont {Luna}}, \bibinfo
  {author} {\bibfnamefont {R.}~\bibnamefont {Roiban}}, \bibinfo {author}
  {\bibfnamefont {C.-H.}\ \bibnamefont {Shen}}, \ and\ \bibinfo {author}
  {\bibfnamefont {M.}~\bibnamefont {Zeng}},\ }\href {\doibase
  10.1103/PhysRevD.104.065014} {\bibfield  {journal} {\bibinfo  {journal}
  {Phys. Rev. D}\ }\textbf {\bibinfo {volume} {104}},\ \bibinfo {pages}
  {065014} (\bibinfo {year} {2021})},\ \Eprint
  {http://arxiv.org/abs/2005.03071} {arXiv:2005.03071 [hep-th]} \BibitemShut
  {NoStop}%
\bibitem [{\citenamefont {Pound}\ and\ \citenamefont
  {Wardell}(2021)}]{Pound:2021qin}%
  \BibitemOpen
  \bibfield  {author} {\bibinfo {author} {\bibfnamefont {A.}~\bibnamefont
  {Pound}}\ and\ \bibinfo {author} {\bibfnamefont {B.}~\bibnamefont
  {Wardell}},\ }\href {\doibase 10.1007/978-981-15-4702-7\_38-1} {\  (\bibinfo
  {year} {2021}),\ 10.1007/978-981-15-4702-7\_38-1},\ \Eprint
  {http://arxiv.org/abs/2101.04592} {arXiv:2101.04592 [gr-qc]} \BibitemShut
  {NoStop}%
\bibitem [{\citenamefont {Goldberger}\ and\ \citenamefont
  {Rothstein}(2020{\natexlab{b}})}]{Goldberger:2020geb}%
  \BibitemOpen
  \bibfield  {author} {\bibinfo {author} {\bibfnamefont {W.~D.}\ \bibnamefont
  {Goldberger}}\ and\ \bibinfo {author} {\bibfnamefont {I.~Z.}\ \bibnamefont
  {Rothstein}},\ }\href {\doibase 10.1103/PhysRevLett.125.211301} {\bibfield
  {journal} {\bibinfo  {journal} {Phys. Rev. Lett.}\ }\textbf {\bibinfo
  {volume} {125}},\ \bibinfo {pages} {211301} (\bibinfo {year}
  {2020}{\natexlab{b}})},\ \Eprint {http://arxiv.org/abs/2007.00726}
  {arXiv:2007.00726 [hep-th]} \BibitemShut {NoStop}%
\bibitem [{\citenamefont {Raj}\ and\ \citenamefont
  {Venugopalan}(2024)}]{Raj:2023irr}%
  \BibitemOpen
  \bibfield  {author} {\bibinfo {author} {\bibfnamefont {H.}~\bibnamefont
  {Raj}}\ and\ \bibinfo {author} {\bibfnamefont {R.}~\bibnamefont
  {Venugopalan}},\ }\href {\doibase 10.1103/PhysRevD.109.044064} {\bibfield
  {journal} {\bibinfo  {journal} {Phys. Rev. D}\ }\textbf {\bibinfo {volume}
  {109}},\ \bibinfo {pages} {044064} (\bibinfo {year} {2024})},\ \Eprint
  {http://arxiv.org/abs/2311.03463} {arXiv:2311.03463 [hep-th]} \BibitemShut
  {NoStop}%
\bibitem [{\citenamefont {Endlich}\ and\ \citenamefont
  {Penco}(2017)}]{Endlich:2016jgc}%
  \BibitemOpen
  \bibfield  {author} {\bibinfo {author} {\bibfnamefont {S.}~\bibnamefont
  {Endlich}}\ and\ \bibinfo {author} {\bibfnamefont {R.}~\bibnamefont
  {Penco}},\ }\href {\doibase 10.1007/JHEP05(2017)052} {\bibfield  {journal}
  {\bibinfo  {journal} {JHEP}\ }\textbf {\bibinfo {volume} {05}},\ \bibinfo
  {pages} {052} (\bibinfo {year} {2017})},\ \Eprint
  {http://arxiv.org/abs/1609.06723} {arXiv:1609.06723 [hep-th]} \BibitemShut
  {NoStop}%
\bibitem [{BHP()}]{BHPToolkit}%
  \BibitemOpen
  \href@noop {} {\enquote {\bibinfo {title} {{Black Hole Perturbation
  Toolkit}},}\ }\bibinfo {howpublished}
  {(\href{http://bhptoolkit.org/}{bhptoolkit.org})}\BibitemShut {NoStop}%
\bibitem [{\citenamefont {Macedo}\ \emph {et~al.}(2013)\citenamefont {Macedo},
  \citenamefont {Leite}, \citenamefont {Oliveira}, \citenamefont {Dolan},\ and\
  \citenamefont {Crispino}}]{Macedo:2013afa}%
  \BibitemOpen
  \bibfield  {author} {\bibinfo {author} {\bibfnamefont {C.~F.~B.}\
  \bibnamefont {Macedo}}, \bibinfo {author} {\bibfnamefont {L.~C.~S.}\
  \bibnamefont {Leite}}, \bibinfo {author} {\bibfnamefont {E.~S.}\ \bibnamefont
  {Oliveira}}, \bibinfo {author} {\bibfnamefont {S.~R.}\ \bibnamefont {Dolan}},
  \ and\ \bibinfo {author} {\bibfnamefont {L.~C.~B.}\ \bibnamefont
  {Crispino}},\ }\href {\doibase 10.1103/PhysRevD.88.064033} {\bibfield
  {journal} {\bibinfo  {journal} {Phys. Rev. D}\ }\textbf {\bibinfo {volume}
  {88}},\ \bibinfo {pages} {064033} (\bibinfo {year} {2013})},\ \Eprint
  {http://arxiv.org/abs/1308.0018} {arXiv:1308.0018 [gr-qc]} \BibitemShut
  {NoStop}%
\bibitem [{\citenamefont {Flume}\ \emph {et~al.}(2004)\citenamefont {Flume},
  \citenamefont {Fucito}, \citenamefont {Morales},\ and\ \citenamefont
  {Poghossian}}]{Flume:2004rp}%
  \BibitemOpen
  \bibfield  {author} {\bibinfo {author} {\bibfnamefont {R.}~\bibnamefont
  {Flume}}, \bibinfo {author} {\bibfnamefont {F.}~\bibnamefont {Fucito}},
  \bibinfo {author} {\bibfnamefont {J.~F.}\ \bibnamefont {Morales}}, \ and\
  \bibinfo {author} {\bibfnamefont {R.}~\bibnamefont {Poghossian}},\ }\href
  {\doibase 10.1088/1126-6708/2004/04/008} {\bibfield  {journal} {\bibinfo
  {journal} {JHEP}\ }\textbf {\bibinfo {volume} {04}},\ \bibinfo {pages} {008}
  (\bibinfo {year} {2004})},\ \Eprint {http://arxiv.org/abs/hep-th/0403057}
  {arXiv:hep-th/0403057} \BibitemShut {NoStop}%
\bibitem [{\citenamefont {Matone}(1995)}]{Matone:1995rx}%
  \BibitemOpen
  \bibfield  {author} {\bibinfo {author} {\bibfnamefont {M.}~\bibnamefont
  {Matone}},\ }\href {\doibase 10.1016/0370-2693(95)00920-G} {\bibfield
  {journal} {\bibinfo  {journal} {Phys. Lett. B}\ }\textbf {\bibinfo {volume}
  {357}},\ \bibinfo {pages} {342} (\bibinfo {year} {1995})},\ \Eprint
  {http://arxiv.org/abs/hep-th/9506102} {arXiv:hep-th/9506102} \BibitemShut
  {NoStop}%
\bibitem [{\citenamefont {Forde}(2007)}]{Forde:2007mi}%
  \BibitemOpen
  \bibfield  {author} {\bibinfo {author} {\bibfnamefont {D.}~\bibnamefont
  {Forde}},\ }\href {\doibase 10.1103/PhysRevD.75.125019} {\bibfield  {journal}
  {\bibinfo  {journal} {Phys. Rev. D}\ }\textbf {\bibinfo {volume} {75}},\
  \bibinfo {pages} {125019} (\bibinfo {year} {2007})},\ \Eprint
  {http://arxiv.org/abs/0704.1835} {arXiv:0704.1835 [hep-ph]} \BibitemShut
  {NoStop}%
\bibitem [{\citenamefont {Ossola}\ \emph {et~al.}(2007)\citenamefont {Ossola},
  \citenamefont {Papadopoulos},\ and\ \citenamefont {Pittau}}]{Ossola:2006us}%
  \BibitemOpen
  \bibfield  {author} {\bibinfo {author} {\bibfnamefont {G.}~\bibnamefont
  {Ossola}}, \bibinfo {author} {\bibfnamefont {C.~G.}\ \bibnamefont
  {Papadopoulos}}, \ and\ \bibinfo {author} {\bibfnamefont {R.}~\bibnamefont
  {Pittau}},\ }\href {\doibase 10.1016/j.nuclphysb.2006.11.012} {\bibfield
  {journal} {\bibinfo  {journal} {Nucl. Phys. B}\ }\textbf {\bibinfo {volume}
  {763}},\ \bibinfo {pages} {147} (\bibinfo {year} {2007})},\ \Eprint
  {http://arxiv.org/abs/hep-ph/0609007} {arXiv:hep-ph/0609007} \BibitemShut
  {NoStop}%
\bibitem [{\citenamefont {Ellis}\ \emph {et~al.}(2012)\citenamefont {Ellis},
  \citenamefont {Kunszt}, \citenamefont {Melnikov},\ and\ \citenamefont
  {Zanderighi}}]{Ellis:2011cr}%
  \BibitemOpen
  \bibfield  {author} {\bibinfo {author} {\bibfnamefont {R.~K.}\ \bibnamefont
  {Ellis}}, \bibinfo {author} {\bibfnamefont {Z.}~\bibnamefont {Kunszt}},
  \bibinfo {author} {\bibfnamefont {K.}~\bibnamefont {Melnikov}}, \ and\
  \bibinfo {author} {\bibfnamefont {G.}~\bibnamefont {Zanderighi}},\ }\href
  {\doibase 10.1016/j.physrep.2012.01.008} {\bibfield  {journal} {\bibinfo
  {journal} {Phys. Rept.}\ }\textbf {\bibinfo {volume} {518}},\ \bibinfo
  {pages} {141} (\bibinfo {year} {2012})},\ \Eprint
  {http://arxiv.org/abs/1105.4319} {arXiv:1105.4319 [hep-ph]} \BibitemShut
  {NoStop}%
\bibitem [{\citenamefont {van Neerven}\ and\ \citenamefont
  {Vermaseren}(1984)}]{vanNeerven:1983vr}%
  \BibitemOpen
  \bibfield  {author} {\bibinfo {author} {\bibfnamefont {W.~L.}\ \bibnamefont
  {van Neerven}}\ and\ \bibinfo {author} {\bibfnamefont {J.~A.~M.}\
  \bibnamefont {Vermaseren}},\ }\href {\doibase 10.1016/0370-2693(84)90237-5}
  {\bibfield  {journal} {\bibinfo  {journal} {Phys. Lett. B}\ }\textbf
  {\bibinfo {volume} {137}},\ \bibinfo {pages} {241} (\bibinfo {year}
  {1984})}\BibitemShut {NoStop}%
\bibitem [{\citenamefont {Passarino}\ and\ \citenamefont
  {Veltman}(1979)}]{Passarino:1978jh}%
  \BibitemOpen
  \bibfield  {author} {\bibinfo {author} {\bibfnamefont {G.}~\bibnamefont
  {Passarino}}\ and\ \bibinfo {author} {\bibfnamefont {M.~J.~G.}\ \bibnamefont
  {Veltman}},\ }\href {\doibase 10.1016/0550-3213(79)90234-7} {\bibfield
  {journal} {\bibinfo  {journal} {Nucl. Phys. B}\ }\textbf {\bibinfo {volume}
  {160}},\ \bibinfo {pages} {151} (\bibinfo {year} {1979})}\BibitemShut
  {NoStop}%
\end{thebibliography}%

\end{document}